\documentclass[a4paper,11pt]{article}
\pdfoutput=1 

\usepackage{jheppub} 

\usepackage[T1]{fontenc} 

\usepackage{hyperref}
\hypersetup{
colorlinks=true,
linkcolor=blue,
citecolor=blue}
\usepackage{url}
\usepackage{caption}
\usepackage{graphicx}
\usepackage{float}
\usepackage{subcaption}
\usepackage{multirow}
\usepackage{makecell}
\usepackage{cases}

\usepackage{natbib}

\usepackage{newtxtext,newtxmath}

\usepackage{pifont}

\usepackage{amsfonts,amsmath,amssymb}

\usepackage{multicol}

\DeclareMathAlphabet{\mathcal}{OMS}{cmsy}{m}{n}

\allowdisplaybreaks[3]

\def \bal#1\eal  {\begin{align} #1 \end{align}}
\def\({\left(}
\def\){\right)}
\def\[{\left[}
\def\]{\right]}

\newcommand{\be} {\begin{equation}}
\newcommand{\ee} {\end{equation}}

\newcommand{\vnba}{\bar{\vmathbb{1}}} \newcommand{\vnbb}{\bar{\vmathbb{2}}} \newcommand{\vnbc}{\bar{\vmathbb{3}}} \newcommand{\vnbd}{\bar{\vmathbb{4}}}

\title{{\huge \boldmath Positivity Bounds on parity-violating scalar-tensor EFTs}}

\author[a,b]{\large Hao Xu}
\author[a,b]{\large , Dong-Yu Hong}
\author[a,b]{\large , Zhuo-Hui Wang}
\author[a,b]{\large and Shuang-Yong Zhou}

\affiliation[a]{Interdisciplinary Center for Theoretical Study, University of Science and Technology of China,
Hefei, Anhui 230026, China}
\affiliation[b]{Peng Huanwu Center for Fundamental Theory, Hefei, Anhui 230026, China}

\emailAdd{haoxu@mail.ustc.edu.cn}
\emailAdd{principle@mail.ustc.edu.cn}
\emailAdd{wzh33@mail.ustc.edu.cn}
\emailAdd{zhoushy@ustc.edu.cn}

\abstract{
Using dispersion relations of the scattering amplitudes and semi-definite programming, we calculate causality bounds on the Wilson coefficients in scalar-tensor effective field theories that include parity-violating operators. Particular attention has been paid to the dynamical-Chern-Simons (dCS) and scalar-Gauss-Bonnet (sGB) couplings, along with higher order coefficients, and the interplay between them. For the leading terms, the bounds on the parity-conserving and -violating coefficients are simply projections of the complex coefficients. Some parity-violating coefficients are found to be upper bounded by the parity-conserving counterparts, or the higher order parity-conserving coefficients. While the observational constraints on parity-violating coefficients are weaker than the parity-conserving counterparts, the causality bounds are of comparable strength and thus may play a more prominent role in constraining strong gravity effects in upcoming observations.

}

\begin{document}

\hfill{ {\small USTC-ICTS/PCFT-24-36} }

\maketitle
\flushbottom

\section{Introduction and summary}

General relativity has been instrumental in understanding gravitational dynamics and has been thoroughly tested in weak-field regimes through numerous experiments~\cite{Will:2014kxa}.
However, there remains the possibility that deviations from general relativity could emerge in strong-field regimes, such as those involving binary black holes and neutron stars \cite{Berti:2015itd}. Recent gravitational wave observations from LIGO and Virgo have opened a new avenue for probing these phenomena in such extreme environments~\cite{LIGOScientific:2016aoc,LIGOScientific:2016dsl}.
One possibility of the strong-regime deviations involves introducing additional degrees of freedom, and scalar-tensor theories, due to its simplicity, represent a significant class of such kind. Indeed, pure gravitational EFTs that give rise to significant deviations from general relativity seem to either violate causality or be non-observable in the planned gravitational wave experiments \cite{deRham:2021bll}.
Increasing attention has also been directed toward models in which the scalar field is non-minimally coupled. When the model respects parity, it often involves the Gauss-Bonnet invariant. The scalar-Gauss-Bonnet (sGB) models are particularly notable for admitting hairy black hole solutions~\cite{Kanti:1995vq,Yagi:2011xp,Sotiriou:2013qea,Sotiriou:2014pfa},
as well as exhibiting the phenomenon of spontaneous scalarization~\cite{Doneva:2017bvd,Silva:2017uqg,Antoniou:2017acq}, and are undergoing rigorous tests through various astrophysical observations.
On the other hand, the dynamical Chern-Simons (dCS) theory~\cite{Jackiw:2003pm,Alexander:2009tp} is characterized by the presence of the Pontryagin density, which is the parity-violating counterpart of the Gauss-Bonnet topological invariant, and
has various origins from more fundamental theories~\cite{Alvarez-Gaume:1983ihn, Alexander:2021ssr, Alexander:2022cow}. The Pontryagin coupling in the dCS model has also been shown to give rise to hairy black hoes~\cite{Grumiller:2007rv,Alexander:2009tp,Yunes:2009hc,Yagi:2012ya,Konno:2014qua,McNees:2015srl,Cardenas-Avendano:2018ocb}.
Extensive phenomenological studies on gravitational waves and cosmology within the dCS framework have been conducted~\cite{Lue:1998mq,Alexander:2009tp,Yunes:2010yf,Gluscevic:2010vv,Kawai:2017kqt,Nair:2019iur,Li:2022grj}.
Novel features of the parity-violating dCS theory include gravitational wave birefringence~\cite{Alexander:2004wk,Takahashi:2009wc,Wang:2012fi,Wang:2021gqm}
and distinct parity-violating patterns in the cosmic microwave background~\cite{QUaD:2008ado,Sorbo:2011rz,Shiraishi:2013kxa,Philcox:2023ffy}. The dCS couplings have been constrained by current gravitational wave observations  ~\cite{Silva:2020acr,Silva:2022srr,Califano:2023aji}.
Various other parity-violating gravitational models have been shown to give rise to interesting effects in astrophysics and cosmology ~\cite{Nieh:1981ww,Hohmann:2020dgy,Li:2020xjt,Iosifidis:2020dck,Cai:2021uup, Li:2022vtn}.

On the other hand, positivity or causality bounds (see~\cite{deRham:2022hpx} for a review), rooted in fundamental principles of quantum field theory such as unitarity, analyticity and crossing symmetry, offer a powerful tool for constraining the parameter spaces of these gravitational effective field theories (EFTs). These bounds originate from connections between UV theories and low-energy EFTs in the form of dispersion relations, which pass down the UV unitarity constraints to the low-energy Wilson coefficients.
In the simplest case of the forward 2-to-2 scattering in a single scalar EFT, this method leads to the requirement of the $s^2$ coefficient being positive~\cite{Adams:2006sv}. In the multi-field generalization of the forward positivity bounds, a geometric representation in terms of convex cones or spectrahedra naturally emerge~\cite{Bellazzini:2014waa,Zhang:2020jyn,Li:2021lpe}. These positivity bounds have been used to constrain the parameter space of the Standard Model EFT (SMEFT) (see, {\it e.g.,}
~\cite{Zhang:2018shp,Bellazzini:2018paj,Remmen:2019cyz,Remmen:2020vts,Bonnefoy:2020yee,Fuks:2020ujk,Gu:2020ldn,Davighi:2021osh,Li:2022rag,Ghosh:2022qqq,Chen:2023bhu,Gu:2023emi}.).
Various scalar positivity bounds away from the forward limit were formulated in~\cite{deRham:2017avq,Arkani-Hamed:2020blm,Bellazzini:2020cot},
along with extensions to particles with spins~\cite{Bellazzini:2016xrt,deRham:2017zjm, deRham:2018qqo}.
When full crossing symmetry is imposed, that is, when the two-channel dispersion relations are supplemented by $s$-$t$ null constraints, two-sided positivity bounds on ratios of EFT coefficients can be obtained~\cite{Tolley:2020gtv,Caron-Huot:2020cmc}. The triple-symmetric positivity bounds can be easily generalized to models with multiple degrees of freedom~\cite{Du:2021byy, Bern:2021ppb,Chowdhury:2021ynh,Caron-Huot:2022ugt,Chiang:2022jep,Henriksson:2022oeu} and massive cases~\cite{Xu:2023lpq, Bertucci:2024qzt}.
When non-positive parts of UV partial unitarity are used, upper bounds on the EFT coefficients can also be obtained~\cite{Caron-Huot:2020cmc,Chiang:2022jep,Chiang:2022ltp,Hong:2024fbl}.
Alternative approaches to obtain the same optimal bounds have subsequently been proposed, which include the use of fully crossing-symmetric dispersion relations~\cite{Sinha:2020win} and the double moment formalism~\cite{Chiang:2021ziz}.

In the presence of gravity, new subtleties arise due to $t$-channel graviton exchange, which causes a divergence in the forward limit of the dispersion relations, preventing a Taylor expansion of the dispersion relations in terms of $t$. Unlike in theories with spin less than 2, the presence of the spin-2 $t$-channel pole leads to negative lower bounds on the $s^2$ coefficients, which are suppressed by the Planck scale~\cite{Alberte:2020jsk,Tokuda:2020mlf,Alberte:2020bdz}. Nevertheless, the $s$-expanded dispersive relations can still be treated as one-parameter families of sum rules with respect to $t$, allowing for functional optimization via continuous decision functions of $t$~\cite{Caron-Huot:2021rmr}. This method has been applied to constrain EFTs of modified Einstein gravity~\cite{Caron-Huot:2022ugt,Henriksson:2022oeu,Caron-Huot:2022jli}.
Fully crossing-symmetric positivity bounds have also been employed to constrain scalar-tensor theories when parity is respected~\cite{Hong:2023zgm}. Several other positivity bounds on scalar-tensor theories, which either do not rely on full crossing symmetry or neglect the effects of the $t$-channel pole, can be found in~\cite{Melville:2019wyy,Tokuda:2020mlf,deRham:2021fpu,Herrero-Valea:2021dry,Serra:2022pzl,Hertzberg:2022bsb,Bellazzini:2022wzv,Bellazzini:2024dco}. Constraints can also be derived from causality conditions within gravitational EFTs~\cite{Camanho:2014apa,Camanho:2016opx,deRham:2020zyh,Chen:2021bvg, Goon:2016une,Hinterbichler:2017qyt,AccettulliHuber:2020oou,Bellazzini:2021shn,Melville:2024zjq,Cassem:2024djm}. Similar complementary bounds can also be derived in the primal bootstrap approach, for example,~\cite{Paulos:2017fhb, Guerrieri:2020bto,Guerrieri:2021ivu,EliasMiro:2022xaa,Haring:2022sdp,Haring:2023zwu,Bhat:2024agd} (see \cite{Kruczenski:2022lot} for a review).
Other applications of positivity/causality bounds in the gravitational setup can be found in, for example,~\cite{Bellazzini:2015cra,Cheung:2016yqr,Bonifacio:2016wcb,Bellazzini:2017fep,Bonifacio:2018vzv,deRham:2019ctd,Alberte:2019xfh, Chen:2019qvr,Huang:2020nqy,Wang:2020xlt,Herrero-Valea:2020wxz,Arkani-Hamed:2021ajd,Aoki:2023khq,CarrilloGonzalez:2023emp,deBoe:2024gpf,Eichhorn:2024wba,Alviani:2024sxx,Caron-Huot:2024tsk,Caron-Huot:2024lbf,Beadle:2024hqg}.

In this paper, we extend the framework of causality bounds to place constraints on gravitational EFT couplings in the parity-violating theories. Assuming the theory is weakly coupled below the EFT cutoff, we adopt a tree-level approximation for the low-energy amplitudes. We begin by deriving two-channel dispersion relations and additionally imposing the missing $s$-$t$ crossing symmetry. This leads to a series of sum rules that contain the momentum transfer $t$, as the $t$-channel graviton exchange prevents further expansion in terms of $t$.
These sum rules are then used to quickly power-count the Wilson coefficients, {\it i.e.,} estimate the sizes of these coefficients, which helps the numerical evaluation later. After that, $t$-channel-resolving functional semi-definite optimization is applied to numerically obtain causality bounds on various EFT coefficients, the dependence of the bounds on various factors are investigated, and some results are examined semi-analytically. For example, we can intuitively extract the scalings of the bounds with respect to the lowest UV states from the sum rules and find that they match the numerical results very well. We also briefly discuss the observational implications of these causality bounds. The main difference from a parity-conserving theory is that parity-violating operators now contribute to the imaginary part of the scattering amplitude through contractions of the Levi-Civita symbol with external polarizations,
and thus the dimensions of the semi-definite matrices to be evaluated are larger, significantly increasing the computational complexity. The main results of this paper include:

\begin{itemize}

\item We first calculate the causality bounds involving the dynamical-Chern-Simons couplings and the scalar-Gauss-Bonnet couplings,  where the Pontryagin (topological) invariant and the Gauss-Bonnet invariant are respectively given by
\be
\tilde{\mathcal{R}}^{(2)} =  R_{\mu \nu \rho \sigma} \epsilon^{\mu \nu \alpha \beta} R_{\alpha \beta}{}^{\ \rho \sigma}/2,~~~~~~~~\mathcal{G} = R_{\mu\nu\rho\sigma} R^{\mu\nu\rho\sigma} - 4R_{\mu\nu} R^{\mu\nu} + R^2.
\ee
For the causality bounds on the leading-order scalar-tensor couplings,  $\beta_1 \phi\mathcal{G}/2$ and $\breve{\beta}_1\phi\tilde{\mathcal{R}}^{(2)}/2$, we find that the bounds can be essentially formulated as constraints on $|\bar{\beta}_1|$, where
\be
\bar{\beta}_1 \equiv \beta_1 + i\breve{\beta}_1 = |\bar{\beta}_1| e^{i\varphi_1} ,
\ee
as the bounds are insensitive to the phase $\varphi_1 = \arg \bar{\beta}_1$. Consequently, the causality bounds on the parity-conserving coupling $\beta_1$ and the parity-violating coupling $\breve{\beta}_1$ can be obtained through simple projections from the bounds on the complex parameter $\bar{\beta}_1$; see Figure \ref{b1g0}. This accidental degeneracy will be lifted if $\mathcal{G}$ and $\tilde{\mathcal{R}}^{(2)}$ were coupled to different scalars. On the other hand, fixing higher-order coefficients (to smaller values) can significantly reduce the bounds on $\beta_1$ and $\breve{\beta}_1$. All of these can be intuitively understood semi-analytically. In particular, the insensitivity of the bounds on $\varphi_1$ stems from the helicity structures inherent in the relevant sum rules.

\item Generally, the causality bounds can not be formulated as bounds only on the moduli of the complex parameters. For example, for the Lagrangian terms $\beta_2\phi^2\mathcal{G}/4+\breve{\beta_2}\phi^2\tilde{\mathcal{R}}^{(2)}/4$ and $\gamma_2\nabla_\mu\phi\nabla^\mu\phi\mathcal{R}^{(2)}/2+\breve{\gamma}_2\nabla_\mu\phi\nabla^\mu\phi\tilde{\mathcal{R}}^{(2)}/2$, the causality bounds on $|\bar{\beta}_2|=|\beta_2 + i\breve{\beta}_2|$ and $|\bar{\gamma}_2|=|\gamma_2 + i\breve{\gamma}_2|$ depend on the ratio $\beta_2/\breve{\beta}_2$ and $\gamma_2/\breve{\gamma}_2$ respectively, in the presence of a non-vanishing $\bar{\beta_1}$ (see Figure \ref{fixb1}). For a vanishing $\bar{\beta}_1$, $|\bar{\beta}_2|$ and $|\bar{\gamma}_2|$ become again insensitive to complex angles, for the same reason as that of the insensitivity of $|\bar{\beta}_1|$ with respect to its angle.

\item We find that some parity-violating couplings are upper-bounded by their parity-conserving counterparts. For example, at order $\mathcal{O}((s,t,u)^4)$ in the amplitude, which involves the Lagrangian terms $\alpha_4(\mathcal{R}^{(2)})^2/4$, $\alpha_4^\prime(\tilde{\mathcal{R}}^{(2)})^2/4$ and $\breve{\alpha}_4\mathcal{R}^{(2)}\tilde{\mathcal{R}}^{(2)}/4$, the parity-violating coefficient $\breve{\alpha}_4$ is upper bounded by the parity-conserving coefficients $\alpha_4$ and ${\alpha}_4^\prime$~(see Figure \ref{quartic}). This fact may be used to constrain the parity-violating effects based on the observational constraints on the parity-conserving effects, which would be especially useful if the parity-conserving effects are easier to observe. Of course, importantly, it does not work the opposite way, and we can always consistently set all the parity-violating coefficients to zero, without introducing any constraints on the parity-conserving coefficients. Also, some parity-violating coefficients are upper bounded by higher order coefficients. For example, $\breve{\beta}_1$ and $\breve{\gamma}_0$ are upper bounded by $\alpha_4$ and ${\alpha}_4^\prime$~(see Figure \ref{quartic}). However, in this case, the parity-conserving ones, ${\beta}_1$ and ${\gamma}_0$, are also upper bounded by the same higher order coefficients.

\item The current observational constraints on the dCS coupling $\breve{\beta}_1$ are significantly weaker than those on the sGB coupling ${\beta}_1$, even though the theoretical causality bounds for both are of comparable strength (see Figure \ref{obscmp}). Consequently, causality bounds on parity-violating coefficients are expected to play a more prominent role in constraining strong gravity effects in upcoming observations.

\end{itemize}

The paper is organized as follows. In Section \ref{sec2}, we derive the dispersive sum rules needed for obtaining the causality bounds. In Section \ref{sec3}, we specify the general optimization framework for numerically computing the causality bounds in the presence of parity-violating operators. In Section \ref{sec4}, we calculate the tree level scattering amplitudes in a general parity-violating gravitational EFT, and perform the power-counting of the Wilson coefficients using the dispersion relations. In Section \ref{sec5}, we first demonstrate how to practically compute the bounds with a simple example, then present various numerical results, along with some numerical and semi-analytical analyses, and briefly confront the bounds with observations. In Appendix \ref{numde}, we provide some numerical details about computing the causality bounds, and in Appendix \ref{secsr}, we explicitly list the relevant sum rules utilized in this paper.

\section{Dispersion relations}\label{sec2}

Effective field theory is a powerful tool to parameterize our ignorance of high energy physics from the low energy point of view.
The causality/positivity bounds tell us that the Wilson coefficients of an EFT can not take arbitrary values in order to avoid running into contradictions with fundamental S-matrix principles such as unitarity and analyticity. In this section, we shall derive the dispersion relations/sum rules for general scattering amplitudes with bosonic spins, which form the basis for obtaining causality bounds on general scalar-tensor theories with a parity-violating sector.

\begin{figure}[ht]
\centering
\begin{subfigure}{0.49\linewidth}
    \centering
    \includegraphics[width=1\linewidth]{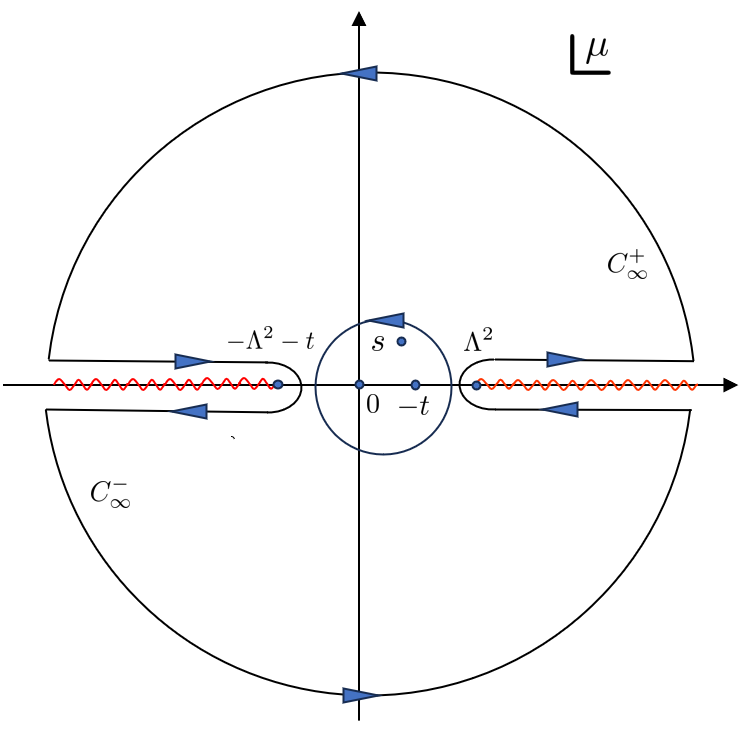}
    \end{subfigure}
\caption{ Analytical structure of $\mathcal{M}^{\vmathbb{1234}}(\mu,t)/(\mu-s)$ on the $\mu$ (center-of-mass energy squared) complex plane, where $\Lambda$ is the EFT cutoff, and the blue dots on the real $\mu$ axis denote poles of the amplitude within the EFT. The $C^\pm_\infty$ are semi-circle contours at complex $\mu$ infinity. }
\label{fig:contour}
\end{figure}

The analyticity assumption suggests that, for fixed $t$, a scattering amplitude $\mathcal{M}^{\vmathbb{1234}}(s,t)$ ($\vmathbb{i}$ labeling a state with helicity $h_i$) is analytic on the $s$ complex plane except for the poles and branch cuts on the real $s$ axis. In addition we shall assume that the EFT is weakly coupled in the IR so that the leading tree-level approximation can be applied. Thus the only singularities below the cutoff, {\it i.e.}, within the range $-\Lambda^2-t<s<\Lambda^2$, are the poles located on the real $s$ axis, which come from the exchange diagrams.
This is illustated in Fig \ref{fig:contour}
for $\mathcal{M}^{\vmathbb{1234}}(\mu,t)/(\mu-s)$ on the $\mu$ (center-of-mass energy squared) complex plane, with $|s|<\Lambda^2$ and $0<-t<\Lambda^2$. Performing the corresponding contour integral around the branch cuts and complex infinity and using the residue theorem, we get
\begin{align}\label{conint}
&\sum_{\text {EFT poles }}\operatorname{Res} \frac{\mathcal{M}^{\vmathbb{1 2 3 4}}(\mu, t)}{\mu-s} \notag \\
= & \int_{\Lambda^2}^{+\infty} \frac{\mathrm{d} \mu}{\pi} \frac{\operatorname{Disc} \mathcal{M}^{\vmathbb{1 2 3 4}}(\mu, t)}{\mu-s}+\int_{-\infty}^{-t-\Lambda^2} \frac{\mathrm{d} \mu}{\pi} \frac{\operatorname{Disc} \mathcal{M}^{\vmathbb{1 2 3 4}}(\mu, t)}{\mu-s}
 +\int_{C_{\infty}^{ \pm}} \frac{\mathrm{d} \mu}{2 \pi i} \frac{\mathcal{M}^{\vmathbb{1 2 3 4}}(\mu, t)}{\mu-s}  \notag\\
= & \int_{\Lambda^2}^{+\infty} \frac{\mathrm{d} \mu}{\pi}\left(\frac{\operatorname{Disc} {\mathcal{M}}^{\vmathbb{1 2 3 4}}(\mu, t)}{\mu-s}+\frac{\operatorname{Disc} \mathcal{M}^{\vmathbb{1 4 32}}(\mu, t)}{\mu-u}\right)
 +\int_{C_{\infty}^{ \pm}} \frac{\mathrm{d} \mu}{2 \pi i} \frac{\mathcal{M}^{\vmathbb{1 2 3 4}}(\mu, t)}{\mu-s},
\end{align}
where "EFT poles" denote the amplitude poles at $\mu=0,-t$ as well as the pole at $\mu=s$ due to the $1/(\mu-s)$ factor.
The discontinuity is defined as $\operatorname{Disc}\mathcal{M}^{\vmathbb{1 2 3 4}}(\mu, t) \equiv (\mathcal{M}^{\vmathbb{1 2 3 4}}(s+i\, \epsilon, t)$$-\mathcal{M}^{\vmathbb{1 2 3 4}}(s-i\,\epsilon, t))/2i$. In the last line we have adjusted the integration variable and used the $s$-$u$ crossing symmetries of the amplitudes to make the expression formally $s$-$u$ symmetric. However, the above expression in general diverges in the UV parts of the integration. The standard approach to overcome this is to perform some "subtractions", which essentially takes into account the polynomial boundedness of the UV amplitudes.
If the scattering amplitude has the asymptotical behavior $\mathcal{M}(s,t)/s^N=0$ as $|s|\rightarrow\infty$, then an $N\text{-}{\text{th}}$ subtraction is sufficient. A dispersion relation with $N\text{-}{\text{th}}$ subtractions can then be written as (see, for example, \cite{Hong:2023zgm})
\begin{align}\label{eq:Nsub}
& \sum_{\text {EFT poles }} \operatorname{Res} \frac{\mathcal{M}^{\vmathbb{1234}}(\mu, t)}{\mu-s}=\sum_{m=0}^{N-1} b_{(N) m}^{\vmathbb{1 2 3 4}}(t) s^m \notag\\
& \quad+\int_{\Lambda^2}^{+\infty} \frac{\mathrm{d} \mu}{\pi}\left(\frac{\left(s-\mu_s\right)^N}{\left(\mu-\mu_s\right)^N} \frac{\operatorname{Disc} \mathcal{M}^{\vmathbb{1234}}(\mu, t)}{\mu-s}+\frac{\left(u-\mu_u\right)^N}{\left(\mu-\mu_u\right)^N} \frac{\operatorname{Disc}\mathcal{M}^{\vmathbb{14 3 2}}(\mu, t)}{\mu-u}\right)
\end{align}
where $\mu_s$ and $\mu_u$ are subtraction points for $s$ and $u$ channels, which can be taken differently in general. Often, an $N=2$ subtraction is sufficient, especially for non-gravitational theories with a mass gap.
For massless gravitational theories, due to the graviton $t$ channel pole, we generally expect the following Regge behavior \cite{Alberte:2021dnj, Caron-Huot:2021rmr}
\begin{eqnarray}\label{regge}
\left\{
\begin{aligned}
  &\lim_{|s|\rightarrow\infty}\mathcal{M}(s,t)/s^2=0,\quad & & t<0, \\
  &\lim_{|s|\rightarrow\infty}\mathcal{M}(s,t)/s^3=0,\quad & & 0\leq t\leq \xi,
\end{aligned}
\right.
\end{eqnarray}
with $\xi$ being a small positive number. In other words, in the physical region $t<0$, a twice subtraction is sufficient for gravitational amplitudes at least in the weakly coupled case ~\cite{Caron-Huot:2022ugt}. However, as $t\rightarrow 0^-$, the existence of the graviton $t$-channel pole implies that the dispersion relation diverges in this limit, and some further subtractions are needed. Nevertheless, the Regge behavior for $0\leq t\leq \xi$ in Eq.~(\ref{regge}) implies that a thrice subtraction can safely remove divergent $t$-channel pole. In the following, we shall consider the dispersion relations in the physical region $t\leq 0$, and take $N=2$ for $t<0$ and $N=3$ for the range $t\leq 0$.

Now, let us consider the unitarity conditions of the UV amplitude. A general 2$\rightarrow$2 scattering amplitude with spins can be decomposed into partial waves in the helicity basis using the Wigner d-matrices
\begin{equation}\label{par}
  \mathcal{M}^{\vmathbb{1234}}(s,t)=16\pi\sum_l^{} (2l+1)d^l_{h_{12},h_{43}}({\theta})A_l^{\vmathbb{1234}}(s)
\end{equation}
where the scattering angle $\theta$ is defined as $\cos \theta \equiv 1+2t/s$, $A_l^{\vmathbb{1234}}(s)$ are the $l$-th partial wave amplitude and we have defined $h_{ij}\equiv h_i-h_j$. Conservation of the angular momentum in the scattering implies unitarity for partial wave amplitudes $A_l^{\vmathbb{1234}}(s)$:
\begin{equation}\label{gopt}
  i(A_l^{\vnbc\vnbd\vnba\vnbb}(s)^*-A_l^{\vmathbb{1234}}(s))=\sum_X A_{l,s}^{\vmathbb{12}\rightarrow X}A_{l,s}^{*\vnbc\vnbd\rightarrow X},
\end{equation}
where $\bar{\vmathbb{i}}$ denotes particle $\vmathbb{i}$ carrying helicity $-h_i$, $X$ is the intermediate state spanned by the complete basis in the Hilbert space, and $A_{l,s}^{\vmathbb{12}\rightarrow X}$ indicates the partial wave amplitude with respect to the process $\vmathbb{12}\rightarrow X$ with the center of mass energy square being $s$. With the help of the Hermitian analyticity of S-matrix $(A_l^{\vnbc\vnbd\vnba\vnbb}(s+i\epsilon))^*=A_l^{\vmathbb{1234}}(s-i\epsilon)$, we have
\begin{equation}\label{dissum}
  \operatorname{Disc}A_l^{\vmathbb{1234}}(s)=\sum_X c_{l,s}^{\vmathbb{12}}c_{l,s}^{*\vnbc\vnbd}
\end{equation}
where $c_{l,s}^{\vmathbb{12}}\equiv A_{l,s}^{\vmathbb{12}\rightarrow X}/\sqrt{2}$.
Then we can cast the dispersive relation (\ref{eq:Nsub}) into the form
\begin{align}\label{eq:Nsubex}
& \sum_{\text {EFT poles }} \operatorname{Res} \frac{\mathcal{M}^{\vmathbb{1234}}(\mu, t)}{\mu-s}=\sum_{m=0}^{N-1} b_{(N) m}^{\vmathbb{1 2 3 4}}(t) s^m \notag\\
& \quad+16\pi\sum_{l,X}(2l+1)\int_{\Lambda^2}^{+\infty} \frac{\mathrm{d} \mu}{\pi}\left(\frac{\left(s-\mu_s\right)^N}{\left(\mu-\mu_s\right)^N} \frac{d^{l,\mu,t}_{h_{12},h_{43}}c_{l,\mu}^{\vmathbb{12}}c_{l,\mu}^{*\vnbc\vnbd}}{\mu-s}+\frac{\left(u-\mu_u\right)^N}{\left(\mu-\mu_u\right)^N} \frac{d^{l,\mu,t}_{h_{14},h_{23}}c_{l,\mu}^{\vmathbb{14}}c_{l,\mu}^{*\vnbc\vnbb}}{\mu-u}\right)
\end{align}
where we have used the abbreviation $d^{l,\mu,t}_{h_{12},h_{43}}\equiv d^l_{h_{12},h_{43}}({\arccos}(1+2t/\mu))$, and $\{b_{(N) m}^{\vmathbb{1 2 3 4}}(t)\}$ denote the remnant, unspecified functions after the $N$-th subtraction. We emphasize that for an appropriate $t$ and $N$ this dispersion relation is now finite on both sides of the equality, which allows us to connect the Wilson coefficients in the IR to some integrals of the UV amplitudes. (In the presence of the spin-2 $t$-channel pole, one cannot simply double taylor-expand both sides in terms of $s$ and $t$, and then match the taylor coefficients, unlike the non-gravitational theories.)

Before extracting these final sum rules connection the IR and the UV, we shall first take into account the $s\text{-}t$ crossing symmetry of the amplitudes, which plays an important role in enhancing causality bounds~ \cite{Tolley:2020gtv,Caron-Huot:2020cmc}. Note that, to obtain Eq.~\eqref{eq:Nsubex}, we only used the $s\text{-}u$ crossing symmetry.
For this purpose, let us focus on scalar-tensor theories. The general amplitudes for scalar-tensor theories can be grouped into two kinds, one with the full $stu$ symmetries and the other only possessing one of the $su$, $st$ or $tu$ crossing symmetries \cite{Hong:2023zgm}. For the latter kind, the extra $s\text{-}t$ crossing ``symmetries'' merely link different amplitudes, rather than actually impose symmetries on the amplitudes.
The triple crossing symmetric amplitudes are of the form $\mathcal{M}^{\vmathbb{1111}}$ or $\mathcal{M}^{\vmathbb{1222}}$, which contain $\mathcal{M}^{0000}(s,t,u)$, $\mathcal{M}^{+++0}(s,t,u)$, $\mathcal{M}^{+++0}(s,t,u)$ and $\mathcal{M}^{++++}(s,t,u)$, where $\pm$ is a shorthand for $\pm 2$. The independent amplitudes with only one {\it bona fide} crossing symmetry are of the form $\mathcal{M}^{\vmathbb{1123}}$, which include $\mathcal{M}^{++00}(s,t,u)$, $\mathcal{M}^{+-00}(s,t,u)$, $\mathcal{M}^{++-0}(s,t,u)$ and $\mathcal{M}^{++--}(s,t,u)$. The remaining amplitudes with other helicities can be obtained from these amplitudes via crossing and the relation $\mathcal{M}^{\overline{\vmathbb{1234}}}(s,t,u)=(\mathcal{M}^{\vmathbb{1234}}(s^*,t^*,u^*))^*$.
To make use of the information of the $s\text{-}t$ crossing symmetries, we parameterize the left hand side of Eq.~\eqref{eq:Nsubex} in the following form
\begin{equation}\label{para}
  \sum_{\text {EFT poles }} \operatorname{Res} \frac{\mathcal{M}^{\vmathbb{1234}}(\mu, t)}{\mu-s}=a_{2,-1}^{\vmathbb{1234}}\frac{s^2}{t}+\sum_{k,n\geq 0}a_{k,n}^{\vmathbb{1234}}s^k t^n.
\end{equation}
As the left hand of Eq.~\eqref{eq:Nsubex} can be reliably computed within the low energy EFT. The coefficients $a_{k,n}^{\vmathbb{1234}}$ can be expressed in terms of the EFT couplings. To utilize the $s\text{-}t$ crossing symmetries, we have $\mathcal{M}^{\vmathbb{1234}}(s, t)=\mathcal{M}^{\vmathbb{1324}}(t, s)$, which implies
\begin{equation}
a_{k,n}^{\vmathbb{1234}}=a_{n,k}^{\vmathbb{1324}} \text{{~ , for $k,n\geq 0$}} .
\end{equation}
The possible $s^2/t$ term comes from the graviton $t$-channel exchange.

Away from the forward with $t<0$, we can make use of twice subtracted dispersion relations (Eq.~(\ref{eq:Nsubex}) with $N=2$). Choosing the subtraction points at $\mu_s=0,\,\mu_u=0$, we get
\begin{align}\label{twsub}
\!\!\!\sum_{\text {EFT poles }}\!\!\!\!\operatorname{Res} \frac{\mathcal{M}^{\vmathbb{1234}}(\mu, t)}{\mu-s}=\!\sum_{m=0}^{1}\! b_{(2) m}^{\vmathbb{1 2 3 4}}(t) s^m +\left\langle\frac{s^2d^{l,\mu,t}_{h_{12},h_{34}}c_{l,\mu}^{\vmathbb{12}}c_{l,\mu}^{*\vnbc\vnbd}}{\mu^2(\mu-s)}
  +\frac{(-s-t)^2d^{l,\mu,t}_{h_{14},h_{23}}c_{l,\mu}^{\vmathbb{14}}c_{l,\mu}^{*\vnbc\vnbb}}{\mu^2(\mu+s+t)}\right\rangle ,
\end{align}
where we have defined the shorthand
\begin{equation}\label{abbr}
  \Big\langle\cdots\Big\rangle : = 16\pi\sum_{l,X}(2l+1)\int_{\Lambda^2}^{+\infty}\frac{\mathrm{d\,\mu}}{\pi}(\cdots)
\end{equation}
Note that in the weakly coupled case the EFT amplitude only contains simple poles, so the left hand side of \eqref{twsub} is analytic in the neighborhood of $s=0$. Therefore we are able to taylor-expand Eq.~(\ref{twsub}) on both sides with respect to $s$ and match the expansion coefficients:
\begin{equation}\label{twmat}
  \delta_{k,2}a_{k,-1}^{\vmathbb{1234}}\frac{1}{t}+\sum_{n=0} a_{k,n}^{\vmathbb{1234}}t^n=\left\langle \frac{\partial_s^k}{k!} \left(\frac{s^2d^{l,\mu,t}_{h_{12},h_{43}}c_{l,\mu}^{\vmathbb{12}}c_{l,\mu}^{*\vnbc\vnbd}}{\mu^2(\mu-s)}
  +\frac{(-s-t)^2d^{l,\mu,t}_{h_{14},h_{23}}c_{l,\mu}^{\vmathbb{14}}c_{l,\mu}^{*\vnbc\vnbb}}{\mu^2(\mu+s+t)}\right)\Bigg{|}_{s\rightarrow0}\right\rangle
\end{equation}
where $k$ is fixed and $n$ counts the power of $t$. When $\mathcal{M}^{\vmathbb{1234}}(s,t)$ is $su$-symmetric, $b_{(2) 1}^{\vmathbb{1 2 3 4}}(t)=0$ and $k$ can take $k\geq 1$. For the non-$su$-symmetric cases, generally the linear $s$ term does not vanish, so we require $k\geq 2$.

The drawback of the dispersion relation (\ref{twmat}) is that on the left hand side the summation over $n$ is from $n=0$ to infinity, meaning that this dispersion relation relies on very high order EFT coefficients. In practice, we are usually concerned with a few low order coefficients.
To isolate these low order coefficients, we use the $st$-crossed dispersion relations of $\mathcal{M}^{\vmathbb{1234}}$ and make three subtractions. Similar to the twice subtracted dispersion relation in Eq.~(\ref{twsub}), following~\cite{Caron-Huot:2022ugt,Hong:2023zgm}, the thrice subtracted dispersion relation with the crossed channel has the form
\begin{align}\label{thsub}
\sum_{\text {EFT poles }}\!\! \!\operatorname{Res} \frac{\mathcal{M}^{\vmathbb{1324}}(\mu, t)}{\mu-s}=\sum_{m=0}^{2} b_{(3) m}^{\vmathbb{1 3 2 4}}(t) s^m +\left\langle\frac{s^3d^{l,\mu,t}_{h_{13},h_{42}}c_{l,\mu}^{\vmathbb{13}}c_{l,\mu}^{*\vnbb\vnbd}}{\mu^3(\mu-s)}
  +\frac{(-s)^3d^{l,\mu,t}_{h_{14},h_{32}}c_{l,\,u}^{\vmathbb{14}}c_{l,\mu}^{*\vnbb\vnbc}}{(\mu+t)^3(\mu+s+t)}\right\rangle
\end{align}
where we have chosen $\mu_s=0$ and $\mu_u=-t$. For this thrice subtracted dispersion relation, we can safely expand around $t=0$, as the $1/t$ pole has been eliminated. Matching the $t^n$ coefficients from the both sides, we get
\begin{equation}\label{thmat}
  \sum_{k\geq3} a_{k,n}^{\vmathbb{1234}}s^k=\left\langle \frac{\partial_t^n}{n!} \left(\frac{s^3d^{l,\mu,t}_{h_{13},h_{42}}c_{l,\mu}^{\vmathbb{13}}c_{l,\mu}^{*\vnbb\vnbd}}{\mu^3(\mu-s)}
  +\frac{(-s)^3d^{l,\mu,t}_{h_{14},h_{32}}c_{l,\mu}^{\vmathbb{14}}c_{l,\mu}^{*\vnbb\vnbc}}{(\mu+t)^3(\mu+s+t)}\right)\Bigg{|}_{t\rightarrow0}\right\rangle
\end{equation}
with $n\geq 0$. Swapping $s$ with $t$ and combining Eq.~(\ref{thmat}) and Eq.~(\ref{twmat}), we get the final dispersion relations/sum rules we will use later
\begin{equation}\label{disf}
  \delta_{k,2}a_{k,-1}^{\vmathbb{1234}}\frac{1}{t}+\sum_{n=0}^2 a_{k,n}^{\vmathbb{1234}}t^n=\left\langle F_{k,l}^{\vmathbb{1234}}(\mu,t)    \right\rangle
\end{equation}
with
\begin{equation}\label{Fdef}
\begin{aligned}
F_{k,l}^{\vmathbb{1234}}(\mu,t)\equiv& \frac{\partial_s^k}{k!} \left(\frac{s^2d^{l,\mu,t}_{h_{12},h_{43}}c_{l,\mu}^{\vmathbb{12}}c_{l,\mu}^{*\vnbc\vnbd}}{\mu^2(\mu-s)}
  +\frac{(-s-t)^2d^{l,\mu,t}_{h_{14},h_{23}}c_{l,\mu}^{\vmathbb{14}}c_{l,\mu}^{*\vnbc\vnbb}}{\mu^2(\mu+s+t)}\right)\Bigg{|}_{s\rightarrow0}\\
  &-\frac{\partial_t^k}{k!} \left(\frac{s^3d^{l,\mu,t}_{h_{13},h_{42}}c_{l,\mu}^{\vmathbb{13}}c_{l,\mu}^{*\vnbb\vnbd}}{\mu^3(\mu-s)}
  +\frac{(-s)^3d^{l,\mu,t}_{h_{14},h_{32}}c_{l,\mu}^{\vmathbb{14}}c_{l,\mu}^{*\vnbb\vnbc}}{(\mu+t)^3(\mu+s+t)}\right)\Bigg{|}_{t\rightarrow0,s\rightarrow t}
\end{aligned}
\end{equation}
Fully utilizing these dispersion relations is in principle sufficient to extract the optimal causality bounds. In practice, for an easier numerical implementation, we also use extra forward-limit sum rules that can be easily derived from the above dispersion relations \cite{Hong:2023zgm}. These forward limit dispersion relations (with $k\geq3$) we will use are given by
\begin{equation}\label{fwddis}
  \left\langle \frac{\partial_t^n}{n!} F_{k,l}^{\vmathbb{1234}}(\mu,0)    \right\rangle=
  \begin{cases}
    a_{k,n}^{\vmathbb{1234}}, & 0\leq n\leq 2 \\
    0, & n\geq 3.
  \end{cases}
\end{equation}

\section{Optimization framework}\label{sec3}

In this section we will outline the main procedures to compute the optimal causality bounds utilizing the dispersion relations established previously, for which we will closely follow the method pioneered by Ref.~\cite{Caron-Huot:2021rmr}. As this has been discussed in details for parity-conserving scalar-tensor theories in Ref \cite{Hong:2023zgm}, we will be relatively briefly in this section, mainly highlighting the differences in  parity-violating theories. The detailed numerical implementation along with some convergence tests is discussed in Appendix \ref{numde}.

We will mainly use the $st$-enhanced sum rules (\ref{disf}) to obtain the optimal bounds for the Wilson coefficients by construct suitable semi-definite programs (SDPs). A sum rule of the form (\ref{disf}) is essentially one-parameter ($t$) family of constraints. To efficiently extract the information of these constraints, it is natural to introduce decision variables that are themselves functionals \cite{Caron-Huot:2022ugt}. That is, we integrate a sum rule of the form (\ref{disf}) against weight functions $f_k^{\vmathbb{1234}}(p)$ to get a weighted sum rule
\begin{equation}\label{sumint}
\begin{aligned}
  \sum_{\vmathbb{1234},k}\int_{0}^{\Lambda}\mathrm{d}p\,&f_k^{\vmathbb{1234}}(p)\(\delta_{k,2}a_{k,-1}^{\vmathbb{1234}}\(-\frac{1}{p^2}\)+\sum_{n=0}^2 a_{k,n}^{\vmathbb{1234}}(-p^2)^n\)\\
  =&\left\langle\sum_{\vmathbb{1234},k}\int_{0}^{\Lambda}\mathrm{d}p\, f_k^{\vmathbb{1234}}(p) F_{k,l}^{\vmathbb{1234}}(\mu,-p^2)    \right\rangle,
  \end{aligned}
\end{equation}
where real number $p$ is defined as $t\equiv-p^2$, and use this weighted sum rule in the numerical SDP. Obviously, in this setup, there are an infinite number of the decision variables $f_k^{\vmathbb{1234}}(p)$, and thus the problem is transferred to how to sample $f_k^{\vmathbb{1234}}(p)$ appropriately. In Eq.~(\ref{sumint}), we have summed over all possible asymptotic states and over various $k$, in order to obtain the optimal bounds.
Additionally, one also needs to efficiently sample $l$ and $\mu$ in $F_{k,l}^{\vmathbb{1234}}(\mu,-p^2)$, which in principle can also take an infinite number of values. For this enormous task, several approximations have to be implemented to simplify the optimization procedure. Essentially, different regions of the $\mu$-$l$ parameter space need to be handled differently, and a viable truncation scheme of the $f_k^{\vmathbb{1234}}(p)$ functional spaces is in need. The detailed approximation approaches are described in Appendix \ref{numde}.

Now, let us construct the SDP problem. Note that, by appropriately choosing a set of the weight functions $f_k^{\vmathbb{1234}}(p)$, it is possible that we can make the real part of the weight integrated sum rule (\ref{sumint}) positive
\begin{equation}\label{sumpos}
 \mathrm{Re}\sum_{\vmathbb{1234},k}\int_{0}^{\Lambda}\mathrm{d}p\,f_k^{\vmathbb{1234}}(p)\(\delta_{k,2}a_{k,-1}^{\vmathbb{1234}}\(-\frac{1}{p^2}\)+\sum_{n=0}^2 a_{k,n}^{\vmathbb{1234}}(-p^2)^n\)\geq 0.
\end{equation}
This is possible if $f_k^{\vmathbb{1234}}(p)$ can be chosen such that the real part of the expression inside $\langle~\rangle$ in Eq.~\eqref{sumint} is positive:
\begin{equation}
\mathrm{Re}\sum_{\vmathbb{1234},k}\int_{0}^{\Lambda}\mathrm{d}p\, f_k^{\vmathbb{1234}}(p) F_{k,l}^{\vmathbb{1234}}(\mu,-p^2)  \geq 0
\end{equation}
Since $a_{k,n}^{\vmathbb{1234}}$ are combinations of the Wilson coefficients, this means that we can get a causality bound on the Wilson coefficients. The difference from the parity-conserving case is that now we need to take the real part in the evaluation when the theory may violate parity. Taking the real part is already sufficient as we can always choose $f_k^{\vmathbb{1234}}(p)$ to swap the real and imaginary part.

Note that $F_{k,l}^{\vmathbb{1234}}(\mu,-p^2)$ contain numerous unknown partial wave amplitudes $c_{l,\mu}^{\vmathbb{12}}$ along with their conjugates. To formulate conditions without these partial wave amplitudes, we can identify the real part of weight-integrated $F_{k,l}^{\vmathbb{1234}}(\mu,-p^2)$ as a quadratic form of $\mathrm{Re}c_{l,s}^{\vmathbb{12}}$ and $\mathrm{Im}c_{l,s}^{\vmathbb{12}}$:
\begin{equation}\label{quad}
 \mathrm{Re}\int_{0}^{\Lambda}\mathrm{d}p\, f_{k}^{\vmathbb{1234}}(p)  F_{k,l}^{\vmathbb{1234}}(\mu,-p^2)=\mathbf{c}_{l,\mu}\mathrm{Re}\(\int_{0}^{\Lambda}\mathrm{d}p\, f_{k}^{\vmathbb{1234}} (p)\mathbf{B}_{k,l,\mu}^{\vmathbb{1234}}(p)\)\mathbf{c}^{T}_{l,\mu}
\end{equation}
where the vector is defined as $\mathbf{c}_{l,\mu}\equiv(c^{00}_{l,\mu},\mathrm{Re}\,c^{+0}_{l,\mu},\mathrm{Re}\,c^{++}_{l,\mu},c^{+-}_{l,\mu},\mathrm{Im}\,c^{+0}_{l,\mu},\mathrm{Im}\,c^{++}_{l,\mu})$, running over all independent $c_{l,s}^{\vmathbb{12}}$,
and $\mathbf{B}_{k,l,\mu}^{\vmathbb{1234}}$ is in general a $6\times6$ complex matrix with respect to these entries. Note that $c^{00}_{l,\mu}$ and $c^{+-}_{l,\mu}$ are self conjugate and automatically real. The independent $c_{l,s}^{\vmathbb{12}}$ can be inferred by the following identities:
\begin{align}\label{cid}
  c_{l,\mu}^{\vmathbb{12}}&=(-1)^l c_{l,\mu}^{\vmathbb{21}}  \\
  \label{cid2}
  c_{l,\mu}^{\vmathbb{12}}&=  c_{l,\mu}^{*\bar{\vmathbb{2}}\bar{\vmathbb{1}}}.
\end{align}
Eq.~\eqref{cid} follows from the bose symmetry and the symmetries of the Wigner d-matrices, and Eq.~\eqref{cid2} can be inferred from on-shell 3-point amplitudes $\mathcal{M}^{1^{h_1},2^{h_2},3^{X}}$ in terms of little group scalings. The positivity of Eq.~(\ref{sumpos}) is then readily formulated as the positivity of $\mathbf{B}_{k,l,\mu}^{\vmathbb{1234}}(p)$:
\begin{equation}\label{sdcond1}
  \bar{\mathbf{B}}_{l,\mu}\equiv\sum_{\vmathbb{1234},k}\mathrm{Re}\int_{0}^{1}\mathrm{d}p\, f_k^{\vmathbb{1234}}(p)\mathbf{B}_{k,l,\mu}^{\vmathbb{1234}}(p) \succeq 0\quad \text{for all }\mu\geq 1,\, l\geq0
\end{equation}
or more explicitly
\begin{equation}\label{sdcond2}
  \sum_{\vmathbb{1234},k}\!\int_{0}^{1}\!\!\mathrm{d}p\,
  \(\mathrm{Re}f_k^{\vmathbb{1234}}(p)\mathrm{Re}\mathbf{B}_{k,l,\mu}^{\vmathbb{1234}}(p)-\mathrm{Im}f_k^{\vmathbb{1234}}(p)\mathrm{Im}\mathbf{B}_{k,l,\mu}^{\vmathbb{1234}}(p)\) \succeq 0\quad\text{for all}~ \mu\geq 1,\, l\geq0
\end{equation}
where $\succeq$ denotes the positive semidefiniteness of the matrix
and we have set $\Lambda=1$ for simplicity. For odd $l$, further simplification can be made by using the identity (\ref{cid}), which reduces the $\mathbf{c}_{l,\mu}$ vector to $\mathbf{c}_{l,\mu}=(\mathrm{Re}\,c^{+0}_{l,\mu},c^{+-}_{l,\mu},\mathrm{Im}\,c^{+0}_{l,\mu})$.

\begin{figure}[ht]
    \centering\!\!\!
    \begin{subfigure}{0.43\linewidth}
    \centering
    \includegraphics[width=0.99\linewidth]{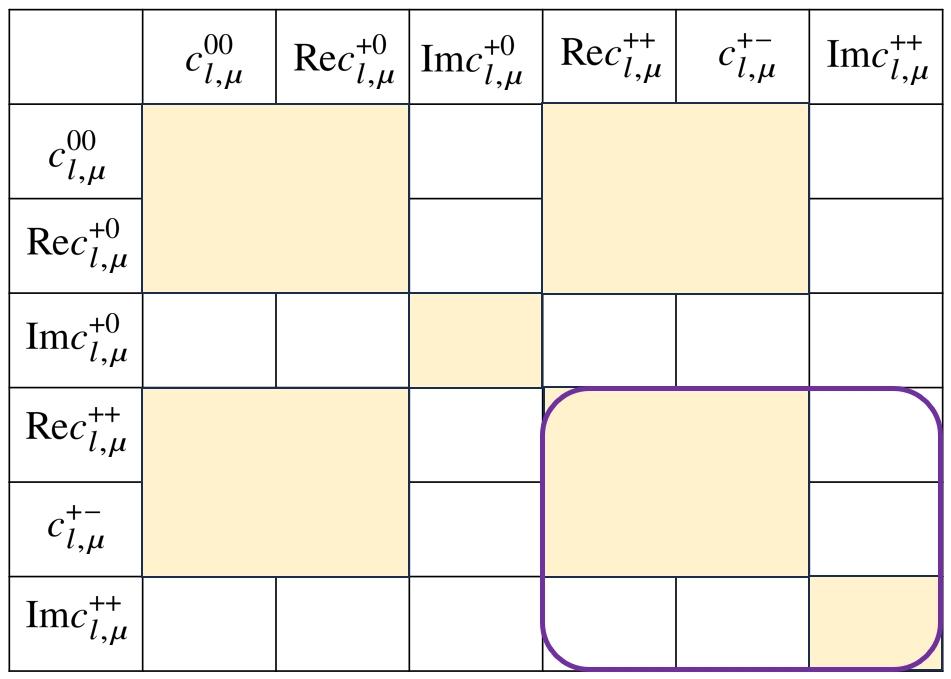}
    \end{subfigure}
\caption{Structure of the $\bar{\mathbf{B}}_{l,\mu}$ matrix with even $l$ (see Eq.~\eqref{sdcond1})) in a parity-violating theory.
If parity-conserving, the ${\rm Im}c_{l,\mu}^{\vmathbb{12}}$ columns and rows are effectively absent. The yellow regions denote the parity-conserving components and the purple rectangle denotes the pure gravitational sector.
}
\label{blockmat}
\end{figure}

Note that the matrix $\bar{\mathbf{B}}_{l,\mu}$ is significantly larger than the corresponding matrix in a parity-conserving theory \cite{Hong:2023zgm}, where the ${\rm Im}c_{l,\mu}^{\vmathbb{12}}$ columns and rows are absent; see Figure \ref{blockmat}. (Strictly speaking, in the parity-conserving case, one can effectively lump the imaginary parts into the corresponding real parts.) For example, the number of independent components in the $6\times 6$ symmetric matrix $\bar{\mathbf{B}}_{l,\mu}$ are $21$ for even $l$, while for the parity-conserving case there are only 13 independent components.
The enlargement in the dimensionality of the $\bar{\mathbf{B}}_{l,\mu}$ matrices significantly increases the computational costs. With the same number of $\mu$-$l$ semi-definite constraints implemented, the parity-violating case usually takes about one order of magnitude longer than the similar parity-conserving case to run the code.
The details of the numerical setup is provided in Appendix \ref{numde}. To verify the numerical validity, the convergence tests for various SDPs are also shown.

In order to make the SDP numerically solvable, we also need to specify the truncated forms of weight functions $f_k^{\vmathbb{1234}}(p)$. We will choose the following trunacated polynomials
\begin{equation}\label{eq:fk}
  f_k^{\vmathbb{1234}}(p)=\sum_{i=i_{\mathrm{min}}}^{i_\mathrm{max}}f_{k,i}^{\vmathbb{1234}}(1-p)^2p^i.
\end{equation}
where $i$ is integer while $i_{\rm min}$ depends on the dispersion relation $F^{\vmathbb{1234}}_{k,\ell}$. The reason for the form of this function and the choices of $i_{\rm min}$ and $i_{\rm max}$ are explained in Appendix \ref{numde}. Generally, the weight functions $f_k^{\vmathbb{1234}}(p)$ are complex, but for the sum rules $F_{k,l}^{\vmathbb{1234}}(\mu,-p^2)$ with only real parts such as $F_{k,l}^{0000}(\mu,-p^2)$, we should choose real $f_k^{\vmathbb{1234}}(p)$. To numerically implement the semi-definite conditions of $\bar{\mathbf{B}}_{l,\mu}$, we can use the widely used SDPB package~\cite{Simmons-Duffin:2015qma}.

\section{Scalar-tensor theories with parity violation}\label{sec4}

In this section we turn to the Lagrangian and scattering amplitudes of the scalar-tensor theories in the presence of parity-violating operators, and perform power-counting for the Wilson coefficients via the method of the dispersion relations, which was introduced in Ref \cite{Hong:2023zgm}.

\subsection{Lagrangian and 2-to-2 amplitudes}

We consider a 4D scalar-tensor theory where the mass of the scalar is negligible, compared to the EFT cutoff. Although our causality bounds directly constrain the amplitude coefficients, the Wilson coefficients of the Lagrangian are widely used in many circumstances, so we start with the Lagrangian.
The lowest order Lagrangian terms in a general scalar-tensor theory with parity violation are given by \cite{Li:2023wdz}
\begin{align}\label{stlag}
  S&=\int \mathrm{d}x^4 \sqrt{-g}\,\(\mathcal{L}_{\text{even}}+ \mathcal{L}_{\text{scalar}}+\mathcal{L}_{\text{odd}}  \), \\
  \mathcal{L}_{\text{even}}&\equiv\frac{M_P^2}{2}R+\frac{\beta_1}{2!}\phi\mathcal{G}+\frac{\beta_2}{4}\phi^2\mathcal{G}
  +\frac{\gamma_1}{3!}\phi\mathcal{R}^{(3)} +\frac{\gamma_2}{2}\nabla_\mu\phi\nabla^\mu\phi\mathcal{R}^{(2)} \notag\\
   &\quad+\frac{4\gamma_3}{3}\phi\nabla_\mu\nabla_\rho\phi\nabla_\nu\phi\nabla_\sigma\phi R^{\mu\nu\rho\sigma}
   +\frac{\gamma_0}{3!}\mathcal{R}^{(3)}+\frac{\alpha_4}{4}(\mathcal{R}^{(2)})^2 +\frac{\alpha^\prime_4}{4}(\tilde{\mathcal{R}}^{(2)})^2 ,\\
  \mathcal{L}_{\text{scalar}}&\equiv -\frac{1}{2}\nabla_\mu\phi\nabla^\mu\phi -\frac{\lambda_3}{3!}\phi^3-\frac{\lambda_4}{4!}\phi^4
 +\frac{\alpha}{2}\(\nabla_\mu\phi\nabla^\mu\phi\)^2 +\frac{\gamma_4}{3}\nabla_\mu\phi\nabla^\mu\phi\nabla_\rho\nabla_\sigma\phi\nabla^\rho\nabla^\sigma\phi   ,\\
  \mathcal{L}_{\text{odd}}&\equiv  \frac{\breve{\beta}_1}{2!}\phi\tilde{\mathcal{R}}^{(2)}+\frac{\breve{\beta}_2}{4}\phi^2\tilde{\mathcal{R}}^{(2)}
  +\frac{\breve{\gamma}_1}{3!}\phi\tilde{\mathcal{R}}^{(3)} +\frac{\breve{\gamma}_2}{2}\nabla_\mu\phi\nabla^\mu\phi\tilde{\mathcal{R}}^{(2)} \notag\\
   &\quad+\frac{4\breve{\gamma}_3}{3}\phi\nabla_\mu\nabla_\rho\phi\nabla_\nu\phi\nabla_\sigma\phi \tilde{R}^{\mu\nu\rho\sigma}
   +\frac{\breve{\gamma}_0}{3!}\tilde{\mathcal{R}}^{(3)}+\frac{\breve{\alpha}_4}{4}\mathcal{R}^{(2)}\tilde{\mathcal{R}}^{(2)},
\end{align}
where we have defined
\begin{equation}
\begin{aligned}
\mathcal{R}^{(2)}&=R_{\mu \nu \rho \sigma} R^{\mu \nu \rho \sigma}, \quad \tilde{\mathcal{R}}^{(2)}=R_{\mu \nu \rho \sigma} \tilde{R}^{\mu \nu \rho \sigma}, \quad \tilde{R}_{\mu \nu \rho \sigma} \equiv \frac{1}{2} \epsilon_{\mu \nu}{ }^{\alpha \beta} R_{\alpha \beta \rho \sigma} \\
\mathcal{R}^{(3)}&=R_{\mu \nu}{ }^{\rho \sigma} R_{\rho \sigma}{ }^{\alpha \beta} R_{\alpha \beta}{ }^{\mu \nu}, \quad \tilde{\mathcal{R}}^{(3)}=R_{\mu \nu}{ }^{\rho \sigma} R_{\rho \sigma}{ }^{\alpha \beta} \tilde{R}_{\alpha \beta}{ }^{\mu \nu} \,
\end{aligned}
\end{equation}
and $\mathcal{G}=R_{\mu\nu\rho\sigma}R^{\mu\nu\rho\sigma}-4R_{\mu\nu}R^{\mu\nu}+R^2$ is the Gauss-Bonnet invariant. $\tilde{\mathcal{R}}^{(2)}$ is often referred to as Pontryagin density. For a shift-symmetric scalar, the terms without a derivative on $\phi$ are absent except for the $\phi \mathcal{G}$ term, as $\mathcal{G}$ is a total derivative. Performing a standard perturbative calculation with the Feynman rules obtained from the above Lagrangian for the lowest orders and parametrizing the higher order terms, the independent tree-level scattering amplitudes for various external helicities are given by
\begin{align}\label{amp}
\mathcal{M}^{0000} & =-\lambda_3^2\left(\frac{1}{s}+\frac{1}{t}+\frac{1}{u}\right)-\lambda_4+\frac{1}{M_P^2}\left(\frac{s u}{t}+\frac{s t}{u}+\frac{u t}{s}\right)+\alpha x+\gamma_4 y+\sum_{\substack{n \geq 0, m \geq 0,\\ m+n \geq 2}} g_{m, n}^S x^n y^m, \\
\mathcal{M}^{++--} & =\frac{1}{M_P^2} \frac{s^3}{t u}-\frac{\beta_1^2+\breve{\beta}_1^2}{M_P^4} s^3+\frac{2\left(\alpha_4+\alpha_4^{\prime}\right)}{M_P^4} s^4+\frac{\gamma_0^2+\breve{\gamma}_0^2}{M_P^6} s^3 t u+\sum_{\substack{n \geq 4, m \geq 0,\\ m+n \geq 5}} g_{n, m}^{T_1} s^n(t u)^m, \\
\mathcal{M}^{+++-} & =\frac{\bar{\gamma}_0}{M_P^4} y+\sum_{n \geq 0, m \geq 2} \bar{g}_{m, n}^{T_2} x^n y^m \\
\mathcal{M}^{++++} & =\frac{10 \bar{\gamma}_0 -3 \bar{\beta}_1^2}{M_P^4} y+\frac{\alpha_4-\alpha_4^{\prime}+i \breve{\alpha}_4}{M_P^4} x^2+\(\frac{\bar{\gamma}_0^2}{M_P^6}+\bar{c}_{1, 1}^{T_3}\)x y+\bar{g}_{2,0}^{T_3} y^2+\!\!\sum_{\substack{n \geq 0, m \geq 0,\\ m+n \geq 3}}\!\! \bar{g}_{m, n}^{T_3} x^n y^m \\
\mathcal{M}^{+++0} & =\frac{\bar{\gamma}_1}{M_P^3} y+\sum_{\substack{n \geq 0, m \geq 1,\\ m+n \geq 2}} \bar{g}_{m, n}^{M_1} x^n y^m \\
\mathcal{M}^{++0-} & =\frac{\bar{\beta}_1}{M_P^3} s^2-\frac{\bar{\gamma}_0\left(\beta_1-i \breve{\beta}_1\right)}{M_P^5} s^2 t u+\sum_{n \geq 3, m \geq 1} \bar{g}_{n, m}^{M_2} s^n(t u)^m, \\
\mathcal{M}^{++00} & =\frac{\lambda_3 \bar{\beta}_1}{M_P^2} s+\frac{\bar{\beta}_2}{M_P^2} s^2+\frac{\bar{\gamma}_0}{M_P^4} s t u+\frac{\bar{\gamma}_2 M_P^2+\bar{\beta}_1^2}{M_P^4} s^3+\bar{g}_{2,1}^{M_3} s^2 t u+\sum_{\substack{n \geq 2, m \geq 0 \\
m+n \geq 4}} \bar{g}_{n, m}^{M_3} s^n(t u)^m, \\
\mathcal{M}^{+-00} & =\frac{1}{M_P^2} \frac{t u}{s}+\frac{\beta_1^2+\breve{\beta}_1^2}{M_P^4} y+\sum_{n \geq 0, m \geq 2} g_{n, m}^{M_4} s^n(t u)^m \\
\mathcal{M}^{+000} & =\frac{\bar{\beta}_1}{2 M_P^3} x+\frac{\bar{\gamma}_3}{M_P} y+\sum_{\substack{n \geq 0, m \geq 1,\\ m+n \geq 2}} \bar{g}_{m, n}^{M_5} x^n y^m
\end{align}
where we have introduced $x\equiv s^2+t^2+u^2,\,y\equiv stu$, and the complex Wilson coefficients are defined with $\bar{g}\equiv g +i\,\breve{g}$. Other amplitudes can be obtained by crossing and conjugation. The $\mathcal{O}(x y)$ term in $\mathcal{M}^{++++}$ consists of two parts, one from double 3-vertex insertions connected with a propagator, the remaining $\bar{c}_{1, 1}^{T_3} x y$ term from 4-point contact interactions of dimension-10 curvature operators of the form $\nabla^2 (\mathcal{R}^{(2)})^2$ and $\nabla^2 (\mathcal{R}^{(2)}\tilde{\mathcal{R}}^{(2)})$. If we are agnostic about them, we may replace the coefficient in front of $xy$ with a generic one $\bar{g}^{T_3}_{1,1}$, and adjust the decision variables such that it does not contribute in the optimization. In Section \ref{secho}, we will also discuss the scenario when $\bar{c}_{1, 1}^{T_3} x y$ is specified to exact values.  For the real amplitudes such as $\mathcal{M}^{0000}$, $\mathcal{M}^{+-00}$ and $\mathcal{M}^{++--}$, we have taken the corresponding amplitude coefficients to be real.

Let us illustrate how the parity-violating operators generate the imaginary parts of the scattering amplitudes. Take the $\breve{\beta}_2$ term of $\mathcal{M}^{++00}$ for example, which comes from second-order metric perturbations of the term $\breve{\beta}_2 \phi^2 \tilde{\mathcal{R}}^{(2)}/4$:
\begin{equation}\label{tb2h2}
  \phi^2 \tilde{\mathcal{R}}^{(2)}\Big|_{\mathcal{O}(h^2)}\supset\(\frac{2}{M_P^2}\)^2\phi^2\(-\varepsilon^{\mu\nu\rho\sigma}h_{\alpha\mu,\nu}{}^{,\beta}
  h_{\beta\rho,\sigma}{}^{,\alpha}+\varepsilon^{\mu\nu\rho\sigma}h_{\alpha\mu,\nu}{}^{,\beta}h^{\alpha}{}_{\rho,,\sigma\beta}\)
\end{equation}
where $\varepsilon^{\mu\nu\rho\sigma}$ is the Levi-Civita symbol. From this, we can read the associated $\phi\phi h h$ Feynman rule as
\begin{equation}\label{frtb2}
  i\,2\cdot\frac{\breve{\beta}_2}{4}\(\frac{2}{M_P^2}\)^2\varepsilon^{\mu\nu\rho\sigma}p_{1\nu}p_1{}^\beta p_{2\sigma}p_{2\beta}\eta{}_{\alpha\mu,ab} \eta^{\alpha}{}_{\rho,cd}+\{(p_1,{a,b})\Leftrightarrow(p_2,(c,d))\}
\end{equation}
where $\eta_{\mu\nu,ab}\equiv(\eta_{\mu a}\eta_{\nu b}+\eta_{\mu b}\eta_{\nu a})/2$ with the Latin indices labeling external polarizations. The scattering amplitude can be obtained by contracting the above vertex with external polarization tensors $\tilde{\epsilon}_{i+}^{\mu\nu}\equiv\epsilon_{i+}^{\mu}\epsilon_{i+}^{\nu},~
  \epsilon_{1+}^{\mu}=(0,1,-i,0)/\sqrt{2},~\epsilon_{2+}^{\mu}=(0,1,i,0)/\sqrt{2}$ and using the associated momenta in the center of mass frame $p_{1\mu}=E(1,0,0,1),~p_{2\mu}=E(1,0,0,-1)$, which is given by
  \begin{align}\label{tb2amp}
  i\mathcal{M}^{++00}&\supset i\,\frac{\breve{\beta}_2}{2}\(\frac{2}{M_P^2}\)^2 \varepsilon^{\mu\nu\rho\sigma}p_{1\nu}p_1{}^\beta p_{2\sigma}p_{2\beta}\eta{}_{\alpha\mu,ab} \eta^{\alpha}{}_{\rho,cd}\tilde{\epsilon}^{ab}_{1+}\tilde{\epsilon}^{cd}_{2+}\cdot2
  =i\(i\frac{\breve{\beta}_2}{M_P^2}s^2\)
\end{align}
From this example, we see that the imaginary part of the amplitude comes from the contraction of the Levi-Civita symbol, which originates from a parity-violating operator, with external polarization tensors.
In contrast, the amplitudes calculated from the usual parity-conserving operators do not contain such kind of contractions, and are always real at tree-level.

\subsection{Power-counting via dispersion relations}
\label{powercounting}

The causality bounds essentially provide a way to power-count the Wilson coefficients, that is, to determine the typical allowed values for the Wilson coefficients. This is automatically done once the upper and lower bound are numerically obtained. However, before that, one can inspect the dispersion relations (Appendix \ref{secsr}) to estimate the typical sizes of the coefficients \cite{Hong:2023zgm}, which can also act as a sanity check for the final numerical bounds.

In the absence of gravitational interactions, these estimates are consistent with naive dimensional analysis.
For a purely gravitational EFT, the Lagrangian operators generally scale as $\mathcal{O}_R\sim M_P^2\Lambda^2 (\nabla/\Lambda)^{N_\nabla}(R/\Lambda^2)^{N_R}$, with $N_\nabla$ and $N_R$ being the number of covariant derivatives and curvature tensors respectively \cite{Caron-Huot:2022ugt}.
For a scalar-tensor theory, it is found that a Lagrangian operator scales as
\begin{equation}\label{stsca}
  \mathcal{O}_{\phi R}\sim M_P^2\Lambda^2 \(\frac{\nabla}{\Lambda}\)^{N_\nabla}\(\frac{R}{\Lambda^2}\)^{N_R}\(\frac{\phi}{M_P}\)^{N_\phi}\(\frac{M_P}{\Lambda}\)^{\tilde{N}_\phi}
\end{equation}
where $\tilde{N}_\phi$ is determined by the number of $c_{l,\mu}^{00}$ in the most constraining sum rule \cite{Hong:2023zgm}. For the lowest orders, $\tilde{N}_\phi$ happens to be $\lfloor N_\phi / 2\rfloor$, where $\left\lfloor~ \right\rfloor$ is to take the flooring integer.

To see how these scalings come about, we introduce the dimensionless versions of the dispersive sum rules (\ref{disf}):
\begin{equation}\label{disfdl}
  \delta_{k,2}a_{k,-1}^{\vmathbb{1234}}\frac{1}{\Lambda^2\hat{t}}+\sum_{n=0}^2 \Lambda^{2n}a_{k,n}^{\vmathbb{1234}}\hat{t}^n=\frac{1}{\Lambda^{2k}}\boldsymbol{\big{\langle}} \hat{F}_{k,l}^{\vmathbb{1234}}(\hat{\mu},\hat{t})\boldsymbol{\big{\rangle}}
\end{equation}
where the hatted variables $\hat{z}\equiv z/\Lambda^2,\, z=\mu,\,s,\,t,\,u$, are dimensionless, and we have defined $\hat{F}_{k,l}^{\vmathbb{1234}}(\hat{\mu},\hat{t})/\Lambda^{2k}\equiv F_{k,l}^{\vmathbb{1234}}(\mu,t)$. The bold angle bracket
\begin{equation}
\boldsymbol{\langle}\cdots\boldsymbol{\rangle}\equiv16\pi\sum_{l,X}(2l+1)\int_{1}^{+\infty}\mathrm{d\,\hat{\mu}}/\pi(\cdots)
\end{equation}
is the dimensionless version of the abbreviation $\langle\cdots\rangle$. With these done, we next estimate the scaling behavior of $c_{l,\mu}^{\vmathbb{12}}$ by balancing the scaling hierarchies on both sides of the sum rules (\ref{disfdl}). For example, the sum rule with $\hat{F}_{2,l}^{++--}(\hat{\mu},\hat{t})$ can be explicitly written as
\begin{equation}\label{scaegf}
  -\frac{1}{M_P^2\Lambda^2\hat{t}}=\frac{1}{\Lambda^4}\boldsymbol{\Big{\langle}}c_1|c_{l,\mu}^{++}|^2+c_2|c_{l,\mu}^{+-}|^2\boldsymbol{\Big{\rangle}}~~~ \Rightarrow~~~
    -\frac{\Lambda^2}{M_P^2}=\hat{t}\boldsymbol{\Big{\langle}}c_1|c_{l,\mu}^{++}|^2+c_2|c_{l,\mu}^{+-}|^2\boldsymbol{\Big{\rangle}},
\end{equation}
where $c_1$ and $c_2$ are dimensionless numbers for each $\mu,\ell$ and $\hat{t}$. This sum rule is valid for $|t|\lesssim \Lambda^2$, with the most constraining power coming from $|t|\sim\Lambda^2$ or $|\hat{t}|\sim 1$. As a quick estimate, we can take $|\hat{t}|\sim 1$. Another reason for this choice is that for $|\hat{t}|\sim 1$ the sum rules converge more quickly than $|t|\sim 0$, which is a result of the Regge behavior of the amplitudes. This fact makes the estimation here more accurate. Since all the other dimensionless quantities are more or less $\mathcal{O}(1)$, to balance the equation, we need
\begin{equation}\label{corres}
  c_{l,\mu}^{++},\,c_{l,\mu}^{+-},\,c_{l,\mu}^{-+},c_{l,\mu}^{--}\Leftrightarrow \mathcal{O}\(\frac{\Lambda}{M_P}\),
\end{equation}
where the notation $\Leftrightarrow$ means that the contribution of some partial amplitude in the dispersion relations is of some order. We have also estimated $c_{l,\mu}^{-+},c_{l,\mu}^{--}$ above because they are related to $c_{l,\mu}^{++},c_{l,\mu}^{+-}$ via Eq.~(\ref{cid}) and Eq.~(\ref{cid2}). (Note that the $c_{l,\mu}^{\vmathbb{12}}$ functions are in general complex; by Eq.~\eqref{corres}, we mean that the real and imaginary part are of $\mathcal{O}\({\Lambda}/{M_P}\)$.) With these correspondences established, we can further estimate other Wilson coefficients from other dispersion relations.

Let us look at the sum rule with $\hat{F}_{3,l}^{++--}(\hat{\mu},\hat{t})$, which goes like
\begin{equation}\label{gb}
  -\frac{|\bar{\beta}_1|^2}{M_P^4}-\frac{|\bar{\gamma}_0|^2}{M_P^6}\Lambda^4\hat{t}^2=\frac{1}{\Lambda^6}\boldsymbol{\Big{\langle}} \hat{F}_{3,l}^{++--}(\hat{\mu},\hat{t}) \boldsymbol{\Big{\rangle}}.
\end{equation}
In order to balance this equation, considering $c_{l,\mu}^{++}$, $c_{l,\mu}^{+-}\Leftrightarrow\mathcal{O}\({\Lambda}/{M_P}\)$, the relevant Wilson coefficients should be typically around
\begin{equation}\label{eq:gbscal}
  \beta_1^2,\,\breve{\beta}_1^2\sim\frac{M_P^2}{\Lambda^4},\quad\gamma_0^2,\,\breve{\gamma}_0^2\sim\frac{M_P^4}{\Lambda^8},
\end{equation}
We may get a different scaling for these coefficients from other sum rules, in which case we should adopt the most constraining one, since fine tuned cancellations may happen in the dispersion relations.

For Wilson coefficients residing in dispersion relations with mixed helicities, we also need to establish the scaling hierarchies of $c_{l,\mu}^{+0},\,c_{l,\mu}^{-0}$ and $c_{l,\mu}^{00}$. To this end, note that, for example, the forward limit sum rules of $\hat{F}_{1,l}^{+0-0}(\hat{\mu},0)$ is
\begin{equation}
  -\frac{\Lambda^2}{M_P^2}=\boldsymbol{\Big{\langle}}c_3|c_{l,\mu}^{+0}|^2\boldsymbol{\Big{\rangle}},
\end{equation}
where $c_3$ is a dimensionless number for $\mu$ and $\ell$. Following a similar argument that leads to Eq.~\eqref{corres}, we can extract the following scaling correspondences
\begin{equation}\label{cp0cor}
  c_{l,\mu}^{+0},\,c_{l,\mu}^{-0},\,c_{l,\mu}^{0+},\,c_{l,\mu}^{0-}\Leftrightarrow \mathcal{O}\(\frac{\Lambda}{M_P}\).
\end{equation}
However, from the sum rules with pure helicity-0 states, there are two distinct scenarios for the scaling hierarchy of $c_{l,\mu}^{00}$, corresponding to whether the scalar is more of matter or gravitational interactions. For the former, we expect that $a_{k,n}^{0000}\sim1/\Lambda^{2k+2n}$, which implies $c_{l,\mu}^{00}\Leftrightarrow \mathcal{O}(1)$. In this scenario, from the sum rule with $\hat{F}_{1,l}^{0000}(\hat{\mu},\hat{t})$, we have
\begin{equation}
   -\frac{1}{M_P^2}+2\alpha\Lambda^2 \hat{t}-\gamma_4\Lambda^4 \hat{t}^2=\frac{1}{\Lambda^2}\boldsymbol{\Big{\langle}}c_4|c_{l,\mu}^{00}|^2\boldsymbol{\Big{\rangle}}
\end{equation}
If $M_P\gg\Lambda$, the scalar self couplings scale as  $\alpha\sim1/\Lambda^4,\,\gamma_4\sim1/\Lambda^6$. On the other hand, if the interactions involved with the scalar field are more of the gravitational nature, which is phenomenologically more interesting, the $1/M_P^2$ term on the left side of the above equation is comparable with the remaining terms. This leads to a suppressed scalar partial wave function
\begin{equation}
c_{l,\mu}^{00}\Leftrightarrow \mathcal{O}\(\Lambda/M_P\) .
\end{equation}
In the following of this paper, we will focus on this scaling behavior. In this scaling scheme, the scalar Wilson coefficients scale as
\begin{equation}\label{sccp}
  \alpha\sim \frac{1}{M_P^2\Lambda^2},\quad \gamma_4\sim\frac{1}{M_P^2\Lambda^4}.
\end{equation}
Then, from the dispersion relation with $\hat{F}_{1,l}^{+++0}(\hat{\mu},\hat{t})$, we have
\begin{equation}
   -\frac{\bar{\gamma}_1}{M_P^3}\Lambda^4\hat{t}^2=\frac{1}{\Lambda^2}\boldsymbol{\Big{\langle}}(\cdots)c_{l,\mu}^{++}c_{l,\mu}^{+0}\boldsymbol{\Big{\rangle}}
\end{equation}
Balancing the scaling hierarchies on both sides, we have
\begin{align}\label{g1sca}
  \gamma_1\sim \frac{M_P}{\Lambda^4},~~~~
  \breve{\gamma}_1\sim \frac{M_P}{\Lambda^4}
\end{align}
Following similar procedures, the typical sizes of the other Wilson coefficients explicitly listed in the Lagrangian \eqref{stlag} can be inferred to be
\begin{equation}\label{allsca}
  {\gamma}_2,\,\breve{\gamma}_2\sim\frac{1}{\Lambda^4}, \quad\gamma_3,\,\breve{\gamma}_3\sim\frac{1}{M_P\Lambda^4}, \quad\beta_2,\,\breve{\beta}_2\sim\frac{1}{\Lambda^2},
  \quad\alpha_4,\alpha_4^\prime,\breve{\alpha}_4\sim\frac{M_P^2}{\Lambda^6}
\end{equation}
As we will see later, our numerical results obtained will be consistent with these analyses in this subsection.

\section{Bounds on parity-violating theories}\label{sec5}

In this section, we will numerically implement the SDPs to calculate the optimal causality bounds in scalar-tensor theories with parity violation. We first calculate the bounds on cubic gravitational couplings, which also serves as an explicit example to demonstrate the construction of the SDP. We will also look into some main features of these leading causality bounds in a limit that can be handled analytically, and investigate how the causality bounds depend on the UV spin $l$ and mass scale $\mu$, uncovering some spin selection rules. The causality bounds for high order EFT couplings along with the interplay between different coefficients are discussed in the remaining subsections. Finally, we will briefly confront the causality bounds on the sGB/dCS coefficients with the current observational constraints.

\subsection{Bounds on cubic gravitational couplings}

Let us first look at the bounds on the leading, cubic gravitational EFT couplings
\begin{equation}\label{mod}
  \mathcal{L}\supset \frac{\beta_1}{2!}\phi\mathcal{G}+\frac{\breve{\beta}_1}{2!}\phi\tilde{\mathcal{R}}^{(2)}+\frac{\gamma_0}{3!}{\mathcal{R}}^{(3)}+
  \frac{\breve{\gamma}_0}{3!}{\tilde{\mathcal{R}}}^{(3)}.
\end{equation}
These terms give rise to 3-vertices in the Feynman rules, so these coefficients can be mixed in the scattering amplitudes. The bounds on the first term, which gives rise to hairy black holes, have been discussed in \cite{Hong:2023zgm}. The second term, the dynamical-Chern-Simons coupling, violates parity~\cite{Alexander:2009tp}, and the leading pure gravitational terms are also included for completeness.

\subsubsection{The SDP setup}

To illustrate how to practically implement the SDP proposed in the last section, we consider four sum rules with opposite helicities (\ref{pmpm1}), (\ref{pmpm2}), (\ref{ppmm2}) and (\ref{ppmm3}) and two sum rules with all plus helicities (\ref{pppp1}) and (\ref{pppp2}). These sum rules contain the coefficients of Eq.~\eqref{mod} that we wish to optimize over. This limited choice of sum rules is for illustration purposes, not for obtaining the best bounds. If we want to get better bounds, more sum rules including those without the target coefficients have to be taken into account, since they give rise to more constraints, largely from the crossing symmetry. Substituting the explicit expressions into Eq.~(\ref{sumpos}), the positivity condition with the mentioned sum rules taken into account can be written as
\begin{align}\label{sdeg}
  &\frac{1}{M_P^2}\int_{0}^{1}\mathrm{d}p\,f_2^{++--}(p)\frac{1}{p^2}\notag\\
  \quad&+\int_{0}^{1}\mathrm{d}p\, \Big[\frac{\beta_1^2}{M_P^2}\Big(-f_3^{++--}(p)
  +\frac{3}{M_P^2}\big(\mathrm{Re}f_1^{++++}(p)(-p^2)^2+\mathrm{Re}f_2^{++++}(p)(-p^2)\big)\Big)\notag\\
  \quad& -\frac{\breve{\beta}_1^2}{M_P^2}\Big(f_3^{++--}(p)+\frac{3}{M_P^2}\(\mathrm{Re}f_1^{++++}(p)(-p^2)^2+\mathrm{Re}f_2^{++++}(p)(-p^2)\)\Big)
  \notag\\
  \quad&-\frac{6\beta_1\breve{\beta}_1}{M_P^4}\(\mathrm{Im}f_1^{++++}(p)(-p^2)^2+\mathrm{Im}f_2^{++++}(p)(-p^2)\)
  -\frac{\gamma_0^2}{M_P^6}f_3^{++--}(p)(-p^2)^2 \notag\\
  \quad&-\frac{\breve{\gamma}_0^2}{M_P^6}f_3^{++--}(p)(-p^2)^2 +\frac{12}{M_P^4}\(\(\alpha_4-\alpha_4^\prime\)\mathrm{Re}f_2^{++++}(p)-\breve{\alpha}_4\mathrm{Im}f_2^{++++}(p)\)(-p^2)^2 \notag\\
  \quad&-\frac{10\gamma_0}{M_P^4}\(\mathrm{Re}f_1^{++++}(p)(-p^2)^2+\mathrm{Re}f_2^{++++}(p)(-p^2)\)\notag\\
  \quad&+\frac{10\breve{\gamma}_0}{M_P^4}\(\mathrm{Im}f_1^{++++}(p)(-p^2)^2+\mathrm{Im}f_2^{++++}(p)(-p^2)\)\Big]\geq 0,
\end{align}
where we have set the EFT cutoff $\Lambda=1$ for simplicity.  This inequality constrains the EFT couplings if the decision variables, {\it i.e.}, the weight functions, are fixed. (For a quicker and more stable numerical evaluation, the forward limit sum rules can also be used; see Appendix \ref{numde} for more details.) How are the weight functions fixed then? This is via the dispersion relations (\ref{disf}): the best choices of the weight functions are fixed by satisfying the $\bar{\mathbf{B}}_{l,\mu}\succeq 0$ conditions (see Eqs.~(\ref{quad}) and (\ref{sdcond1})).

The $t$-channel graviton exchange gives rise to an IR divergence, but with an appropriate choice of the weight functions, only a mild logarithmic divergence occurs. In our case the divergence comes from the $\int_{0}^{1}\mathrm{d}p\,f_2^{++--}(p)/p^2$ (see the first line of Eq.~\eqref{sdeg}), because, as it turns out, the basis functions of $f_2^{++--}(p)$ must start with $p(1-p)^2$ \cite{Caron-Huot:2022ugt}.
To regulate the logarithmic divergence, we introduce an IR cutoff $m_{\mathrm{IR}}$, which leads to the divergent term $f_{2,-1}^{++++}\log(\Lambda/m_{\mathrm{IR}})/M_P^2$ in Eq.~\eqref{sdeg}.
For the coefficients that we are agnostic about, we will impose the constraints on the relevant weight functions such that they disappear in Eq.~(\ref{sumpos}) after integrating over $p$. For example, if we wish to be agnostic about the quartic curvature coefficients $\alpha_4$, $\alpha_4^\prime$ and $\breve{\alpha}_4$, then we can supplement the following constraints to the SDP
\begin{equation}
  \int_{0}^{1}\mathrm{d}p\,\mathrm{Re}f_2^{++++}(-p^2)^2=0,\quad
  \int_{0}^{1}\mathrm{d}p\,\mathrm{Im}f_2^{++++}(-p^2)^2=0 .
\end{equation}

The target coefficients $\beta_1$ and $\breve{\beta}_1$ along with $\gamma_0$ and $\breve{\gamma}_0$ can be combined into the complex coefficients $\bar{\beta}_1$ and $\bar{\gamma}_0$, which we will parametrize as
\begin{equation}\label{complex}
  \bar{\beta}_1=|\bar{\beta}_1|e^{i\varphi_1}, \quad \bar{\gamma}_0=|\bar{\gamma}_0|e^{i\varphi_2},
\end{equation}
or
\begin{equation}\label{excpl}
\begin{aligned}
  \beta_1&=|\bar{\beta}_1|\cos\varphi_1,&\,\breve{\beta}_1&=|\bar{\beta}_1|\sin\varphi_1, \\
  \gamma_0&=|\bar{\gamma}_0|\cos\varphi_2,&\,\breve{\gamma}_0&=|\bar{\gamma}_0|\sin\varphi_2.
\end{aligned}
\end{equation}
Note that the complex phase $\varphi_1$ and $\varphi_2$ will be relevant in the sum rules with mixing terms such as $\bar{\beta}_1^2$ and $\bar{\beta}_1^*\bar{\gamma}_0$, which are contained in $\langle F_{3,l}^{++00} \rangle$ and $\langle F_{3,l}^{++0-} \rangle$ respectively. Also, we shall often make use of the angular optimization for a 2D parameter space. For example, we can use the following parametrization for $|\bar{\beta}_1|$ and $|\bar{\gamma}_0|$:
\begin{equation}\label{angular}
  \frac{\Lambda^2|\bar{\beta}_1|}{M_P}=r\cos\theta,\quad \frac{\Lambda^4|\bar{\gamma}_0|}{M_P^2}=r\sin\theta
\end{equation}
With this setup, for fixed $\varphi_1$ and $\varphi_2$ , the Eq.~(\ref{sdeg}) is reduced to
\begin{equation}\label{sdang}
  f_{2,1}^{++--}\log\(\frac{\Lambda}{m_{\mathrm{IR}}}\)+\(A\cos^2\theta+B\cos\theta\sin\theta+C\sin^2\theta\)r^2+(\cdots) r \geq0,
\end{equation}
where $A,\,B$ and $C$ are linear combinations of the weight function coefficients of $\mathrm{Re}f_{k,i}^{\vmathbb{1234}}$ and $\mathrm{Im}f_{k,i}^{\vmathbb{1234}}$. In this case, we get a quadratic inequality with respect to $r$. This kind of optimization objective cannot be directly solved by the SDPB package, and a recursive solving procedure is required. Nevertheless, for phenomenological interesting cases, we may assume that $\log(\Lambda/m_{\mathrm{IR}})$ is relatively large so that the linear terms in $\breve{\beta}_1$ and $\breve{\gamma}_0$ are suppressed. In this case we can simply optimizing over $r^2$, which is directly solvable in SDPB. A sharp bound for a fixed $\theta$ can then be obtained by normalizing the coefficient in front of $r^2$ to $-1$ and minimizing $f_{2,1}^{++--}$. Explicitly, this optimization scheme can be written as follows
\begin{eqnarray}\label{kbcons}
\left\{
\begin{aligned}
  &\text{Minimize:}\quad f_{2,1}^{++--} \notag \\
    &\text{Subject to:}\notag \\
  &\quad \bar{\mathbf{B}}_{l,\mu} \succeq 0\quad \text{for all }\mu\geq 1,\, l\geq0 \notag \\
  &\quad A\cos^2\theta+B\cos\theta\sin\theta+C\sin^2\theta=-1,\notag \\
  &\quad \int_{0}^{1}\mathrm{d}p\,\mathrm{Re}f_2^{++++}(-p^2)^2=0,\quad
  \int_{0}^{1}\mathrm{d}p\,\mathrm{Im}f_2^{++++}(-p^2)^2=0,\,\text{etc.}
\end{aligned}
\right.
\end{eqnarray}
for a given value of $\theta$. This gives a bound on $r^2$.
An alternative choice of the normalization condition is setting $f_{2,1}^{++--}$ to unity, in which case we maximize $A\cos^2\theta+B\cos\theta\sin\theta+C\sin^2\theta$.

The conditions $\bar{\mathbf{B}}_{l,\mu}\succeq 0$ should be imposed for all $\mu$ and $l$. To improve numerical efficiency, we separate them into even and odd $l$ groups.
Still, the SDPB package can only deal with polynomial matrix programs with a single continuous decision variable, so further approximations are needed.
Following \cite{Caron-Huot:2021rmr, Caron-Huot:2022ugt}, we divide the $\mu$-$l$ constraint space into five regions: finite $\mu$ and $l$, large $\mu$ and finite $l$, finite $\mu$ and large $l$, large $\mu$ and $l$ with finite $b$, large $\mu$ and $l$ with large $b$, where $b\equiv 2l/\sqrt{\mu}$ is essentially the impact parameter in the scattering process. It is important to utilize the constraints in most of these regions (see Appendix \ref{numde}).
Now, scanning over many $\theta$, we can plot 2D causality bounds on the $\bar{\beta}_1^2$-$\bar{\gamma}_0^2$ plane. The results for this demonstrating example can be found in Figure \ref{figdis} with the black lines. Note that in the figures we plot tilded coefficients, which are dimensionless and for example defined as
\begin{equation}
\tilde{\bar{\beta}}_1 \equiv \frac{\bar{\beta}_1 \Lambda^2}{M_P\sqrt{\log(\Lambda/m_{\mathrm{IR}})}},~~~~~~ \tilde{\bar{\gamma}}_0 \equiv \frac{\bar{\gamma}_0 \Lambda^4}{ M_P^2\sqrt{\log(\Lambda/m_{\mathrm{IR}})}} .
\end{equation}

\begin{figure}[ht]
    \centering\!\!\!
    \begin{subfigure}{0.49\linewidth}
    \centering
    \includegraphics[width=1\linewidth]{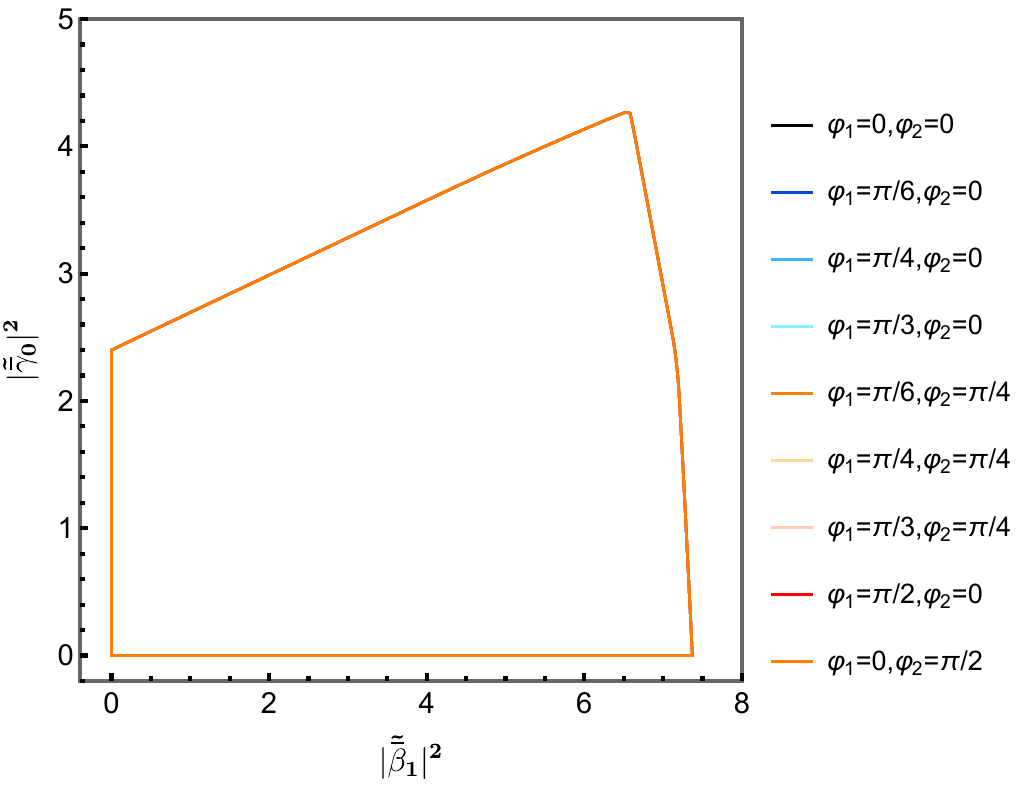}
    \end{subfigure}
    \begin{subfigure}{0.49\linewidth}
    \centering
    \includegraphics[width=1.04\linewidth]{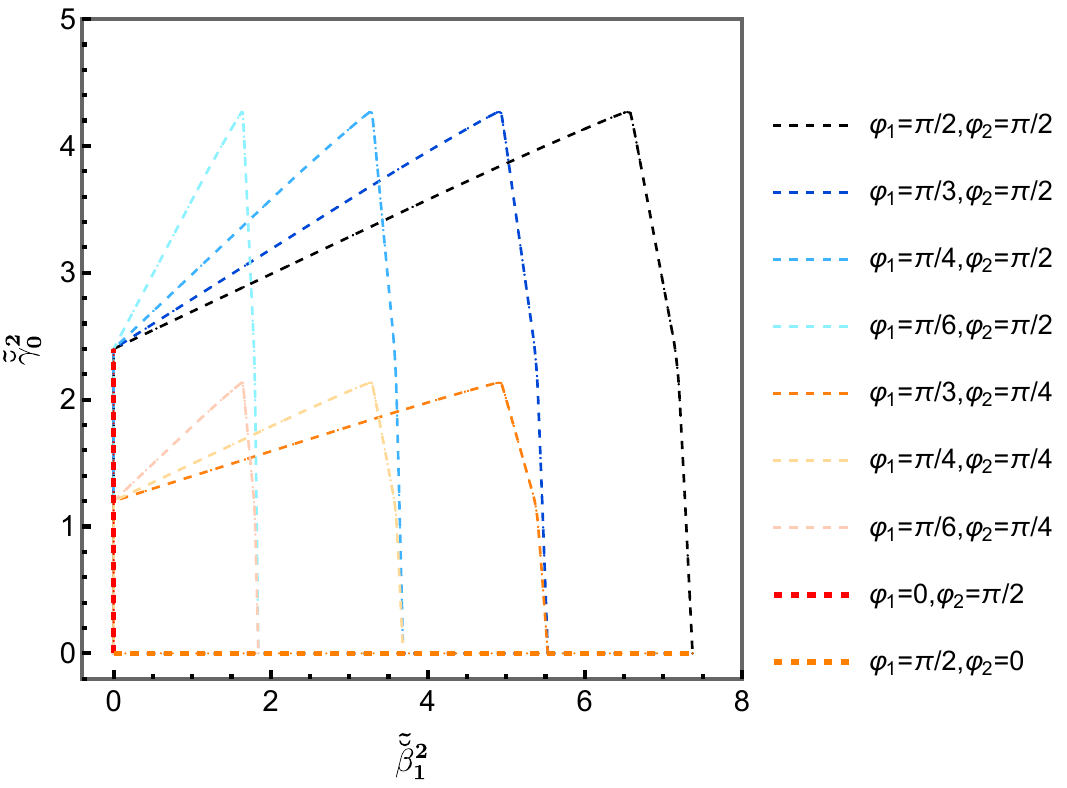}
    \end{subfigure}
\caption{ Causality constraints on the |$\tilde{\beta}_1|^2$-$|\tilde{\gamma}_0|^2$ ({\it left}) and $\tilde{\breve{\beta}}_1^2$-$\tilde{\breve{\gamma}}_0^2$ ({\it right}) plane, where ${\bar{\beta}}_1=|{\bar{\beta}}_1|e^{i\varphi_1}\equiv \beta_1+i \breve{\beta}_1$, ${\bar{\gamma}}_0=|{\bar{\gamma}}_0|e^{i\varphi_2}\equiv \gamma_0+i\breve{\gamma}_0$ and the tilded quantities are defined as $\tilde{\bar{\beta}}_1 \equiv {\bar{\beta}_1 \Lambda^2}/({M_P\sqrt{\log(\Lambda/m_{\mathrm{IR}})}}),~ \tilde{\bar{\gamma}}_0 \equiv {\bar{\gamma}_0 \Lambda^4}/({ M_P^2\sqrt{\log(\Lambda/m_{\mathrm{IR}})}}) $. The bounds in the modular length space $|\tilde{\bar{\beta}}_1|^2$-$|\tilde{\bar{\gamma}}_0|^2$ are nearly the same for different $\varphi_{1,2}$. The parameter space allowed by the causality constraints on $\tilde{\beta}_1^2$-$\tilde{\gamma}_0^2$ and $\tilde{\breve{\beta}}_1^2$-$\tilde{\breve{\gamma}}_0^2$ are  (almost) dual to each other, that is, one can obtain the parity-violating coefficients' region from the parity-conserving one via $\varphi_{1,2}\rightarrow\pi/2-\varphi_{1,2}$ and vice versa. }
\label{b1g0}
\end{figure}

\subsubsection{Numerical results}

To obtain the best bounds, we shall of course use all available sum rules.
In Figure \ref{b1g0}, we plot the best causality bounds on the $|\bar{\beta}_1|^2$-$|\bar{\gamma}_0|^2$ ({\it left}) and $\breve{\beta}_1^2$-$\breve{\gamma}_0^2$ ({\it right}) planes for various angles $\varphi_{1,2}$.
The bounds in the $|\bar{\beta}_1|^2$-$|\bar{\gamma}_0|^2$ space are actually insensitive to the parity mixing angles $\varphi_{1,2}$. The angular $\varphi_{1,2}$ dependence of the causality bounds on the parity-conserving/violating coefficients is mainly due to the projection of the bounds from the complex coefficient plane to the $\beta_1^2$-$\gamma_0^2$ even-parity plane and $\breve{\beta}_1^2$-$\breve{\gamma}_0^2$odd-parity plane. The technical reason for this insensitivity will be explained momentarily.
Due to this projection, when $\varphi_{1,2}$ are small, the parameter space allowed by causality constraints on the sGB coefficients is large and the parameter space allowed for the dCS coefficients is small. As $\varphi_{1,2}$ increase, the causality bound region for the sGB coefficients shrinks, and that of the dCS coefficients expands. We note that this insensitivity with $\varphi_{1,2}$ is accidental, as some higher order coefficients do not share this property, as we will see later.

\begin{figure}[ht]
    \centering\!\!\!
    \begin{subfigure}{0.4\linewidth}
    \centering
    \includegraphics[width=0.99\linewidth]{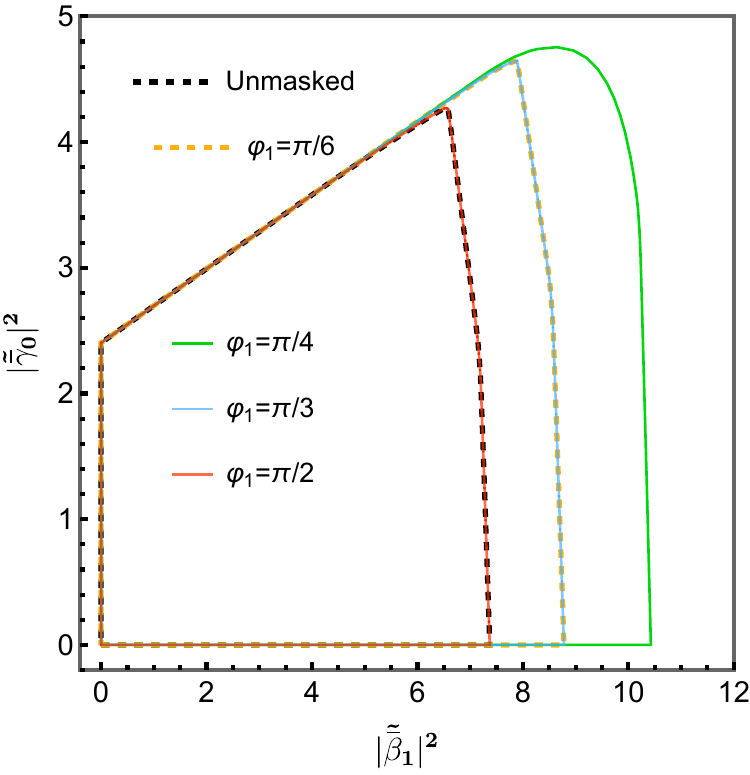}
    \end{subfigure}
\caption{Causality bounds on the $|\tilde{\bar{\beta}}_1|^2$-$|\tilde{\bar{\gamma}}_0|^2$ plane with the parity-violating $\beta_1 \breve{\beta}_1$ term masked.
 }
\label{fig:methcmp}
\end{figure}

\begin{figure}[ht]
    \centering\!\!\!
    \begin{subfigure}{0.4\linewidth}
    \centering
    \includegraphics[width=.99\linewidth]{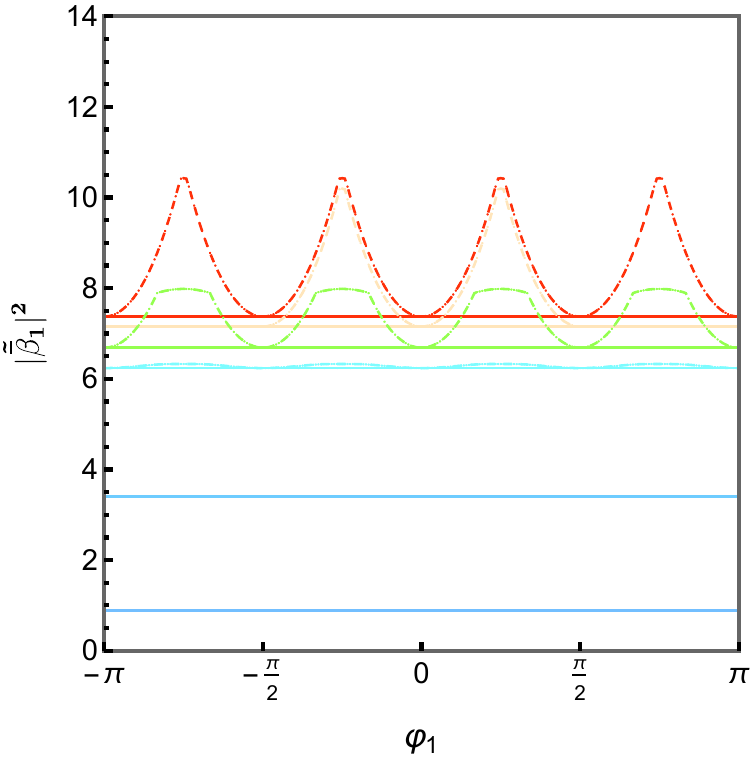}
    \end{subfigure}
        ~~~~
    \begin{subfigure}{0.4\linewidth}
    \centering~
    \includegraphics[width=0.97\linewidth]{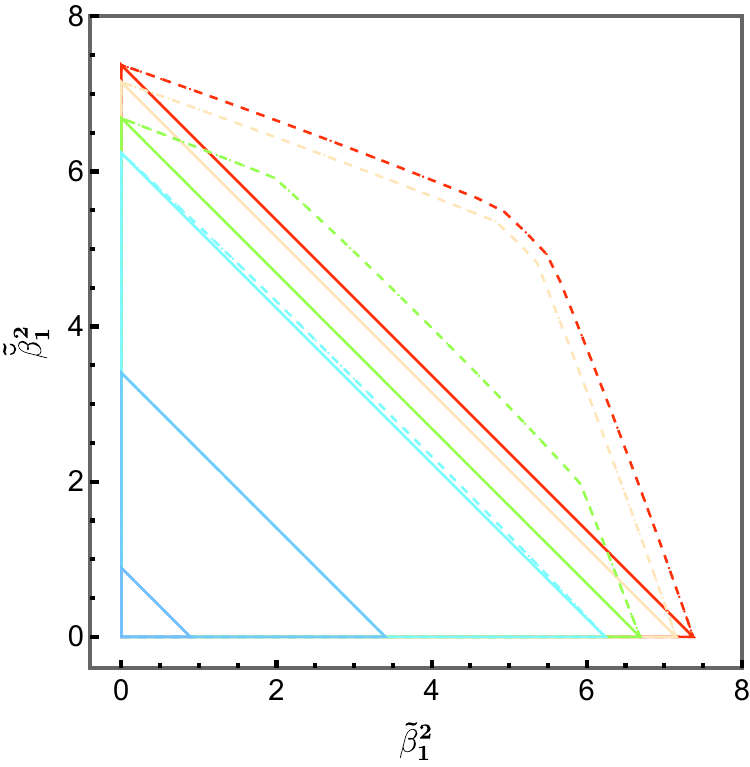}
    \end{subfigure}

    \centering
    \begin{subfigure}{0.8\linewidth}
    \centering
        \includegraphics[width=0.8\linewidth]{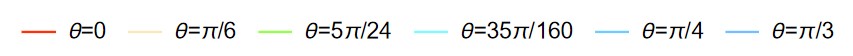}
    \end{subfigure}
\caption{({\it left}) Causality bounds on $|\tilde{\bar{\beta}}_1|^2$ for various angle $\varphi_1$ and $\theta$.
The dashed (solid) lines are the bounds with (without) the $\beta_1 \breve{\beta}_1$ term masked. ({\it right}) Same bounds as the left figure on the $\beta_1^2$-$\breve{\beta}_1^2$ plane.     }
\label{fb1phi}
\end{figure}

To probe how the parity-violating operators ({\it i.e.,} the imaginary parts of the tree-level amplitude) affect the causality bounds, we choose the weight functions to mask the $\beta_1 \breve{\beta}_1$ term in the dispersion relation for the amplitude $\mathcal{M}^{++++}$, that is, to impose extra constraints to eliminate it from the dispersion relation.
We also choose to be agnostic about the high order coefficient $\bar{g}_{1, 1}^{T_3}$. (Although the double three-point insertions $\bar{\gamma}_0^2$ contribute in this order, the four-point contact term is undetermined.) In Figure \ref{fig:methcmp}, we see that this masking significantly affects the bounds on the $|{\bar{\beta}}_1|^2$-$|{\bar{\gamma}}_0|^2$ plane, which would be almost invariant for different angles  $\varphi_{1,2}$ without this masking. Specifically, this masking enlarges the bounds along the $\bar{\beta}_1^2$ direction when $0\leq\varphi_1\leq\pi/4$ and reduces the bounds as $\varphi_1$ approaches $\pi/2$. The bounds are nearly unchanged when varying $\varphi_2$, which is due to the absence of $\bar{\gamma}_0^2$ term in $\mathcal{M}^{++++}$. Also, when varying $\varphi_1$ in the range $0\leq\varphi_1\leq\pi/2$, the bounds are nearly symmetric with respect to $\varphi_1=\pi/4$.
To see this clearly, in the left panel of Figure \ref{fb1phi}, the bounds on $|\bar{\beta}_1|^2$ for various  $\varphi_1$ and $\theta\equiv\arctan(|\bar{\gamma}_0|/|\bar{\beta}_1|)$ are shown.
Without masking the $\beta_1 \breve{\beta}_1$ term, we see that the bounds are nearly the same for different $\varphi_1$ at fixed $\theta$, with the differences being within $0.1\%$. After masking the $\beta_1 \breve{\beta}_1$ term, the bounds are enlarged and peak at $\varphi_1=\pi/4$.
The invariance of $|\bar{\beta}_1|^2$ with respect to $\varphi_1$ without masking, as well as the effects of the $\beta_1 \breve{\beta}_1$ masking, can also be clearly seen in the right panel of Figure \ref{fb1phi}.

\begin{figure}[ht]
    \centering\!\!\!
    \begin{subfigure}{0.42\linewidth}
    \centering
    \includegraphics[width=0.99\linewidth]{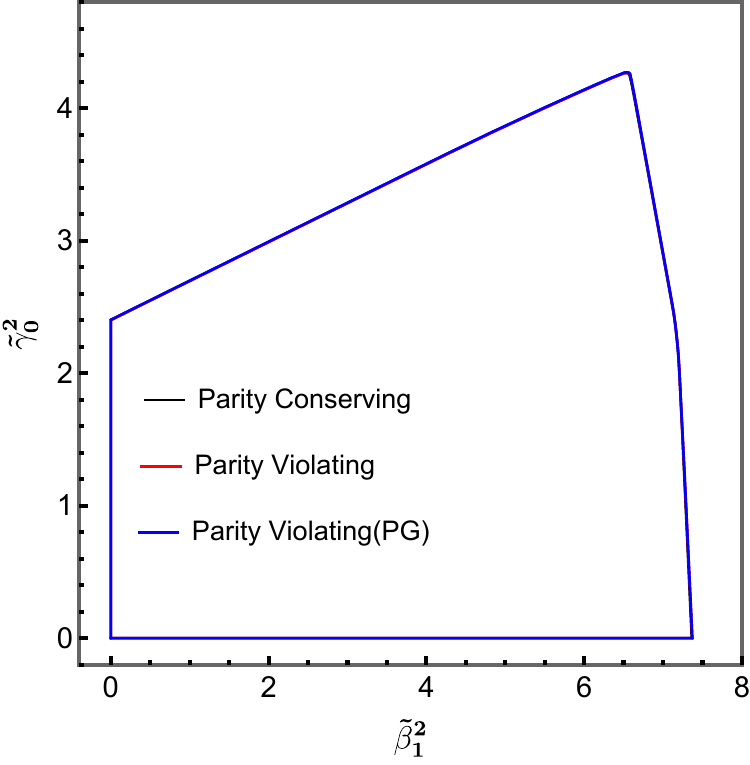}
    \end{subfigure}
\caption{Comparison of the causality bounds on $|\tilde{\beta}_1|^2$ and $|\tilde{\gamma}_0|^2$ for three cases: parity-conserving theory (black), parity-violating theory
({\it i.e.}, with parity-violating terms included; red) and parity-violating theory
using purely gravitational dispersion relations only (blue). The bounds for the three cases are almost the same.  }
\label{fb1fg0}
\end{figure}

In Figure \ref{fb1fg0}, we compare the causality bounds on $\beta_1^2$-$\gamma_0^2$ obtained in a parity-violating theory ({\it i.e.,} the theory with the parity-violating terms included) with those obtained in a parity-conserving theory \cite{Hong:2023zgm}. We see that parity-violating operators do not allow the parity-conserving coefficients to take larger values. Also, we see that the dispersion relations in the purely gravitational sector provide by far the dominant contribution to the bounds on ${\beta}_1^2$ and ${\gamma}_0^2$.
This is due to the fact that the sum rules with the scalar components in this case only contribute to the non-diagonal entries of the semi-definite matrices $\bar{\mathrm{B}}_{l,\mu}$, and the presence of these entries makes $\bar{\mathrm{B}}_{l,\mu}$ more difficult to satisfy the semi-positivity. This usually in turn makes the relevant decision variables highly suppressed in the optimization, and thus these sum rules often do not play a significant role in producing the final results.
In producing Figures \ref{b1g0} and \ref{fb1fg0}, 51 dispersion relations are used, including those with $F_{k\leq4,l}^{0000}$, $F_{k\leq4,l}^{+000}$, $F_{k\leq4,l}^{+++0}$, $F_{k\leq4,l}^{+++-}$, $F_{k\leq4,l}^{++++}$, $F_{k\leq4,l}^{++00}$, $F_{k\leq4,l}^{+0+0}$, $F_{k\leq4,l}^{+0-0}$, $F_{k\leq4,l}^{+-00}$, $F_{k\leq4,l}^{++0-}$, $F_{k\leq4,l}^{+0+-}$, $F_{k\leq6,l}^{+-+-}$ and $F_{k\leq6,l}^{++--}$, as we list in the Appendix \ref{secsr}.
As we mentioned above, many of these dispersion relations actually have little effect on the causality bounds on $|{\bar{\beta}}_1|^2$ and $|{\bar{\gamma}}_0|^2$, often less than $0.1\%$.
On the other hand, the evaluations of the bounds in parity-conserving cases are often computationally highly costly.
In the following, when computing the bounds on $|{\bar{\beta}}_1|^2$ and $|{\bar{\gamma}}_0|^2$,
we shall only use the purely gravitational dispersion relations in the numerical evaluations. This often presents a numerical speed-up factor as large as 50.

To intuitively understand the insensitivity of the $|\bar{\beta}_1|^2$-$|\bar{\gamma}_0|^2$ bounds with respective to $\varphi_{1,2}$ as well as the symmetry with respect to $\varphi_1=\pi/4$ for the $\beta_1 \breve{\beta}_1$ masked case. (The case with the $\beta_1 \breve{\beta}_1$ terms masked physically corresponds to a theory where the Gauss-Bonnet invariant and the Pontryagin invariant are coupled to different kinds of scalars.) Using only dispersion relations from the purely gravitational sector, let us consider a simple but representative bound:
\begin{equation}
        \int\!\! d p \left[f_1(p)|\langle F_{2,l}^{++--}\rangle|^2+f_2(p)|\langle F_{3,l}^{++--}\rangle|^2
    +f_3(p)|\langle F_{1,l}^{++++}\rangle|^2 +f_4(p)|\langle F_{2,l}^{++++}\rangle|^2\right]\geq0
\end{equation}
where the undetermined $f_i(p)$ functions are decision variables. Now, we replace one factor in each of the four squares $|\langle F_{k,l}^{\vmathbb{1234}}\rangle|^2$ with the left hand side of the dispersion relation. For example, we replace $\langle F_{2,l}^{++--}(\mu,t)\rangle$ with $-1/(M_P^2 t)$. Then neglecting the subleading linear terms in $\bar{\beta}_1$ and $\bar{\gamma}_0$ and higher order contributions, we get the following expression
\begin{align}\label{eq:helidecomp}
    &\Big\langle \(A-B|\bar{\beta}_1|^2-C|\bar{\gamma}_0|^2\)\((\mathrm{Re}c_{l,\mu}^{++})^2+(\mathrm{Im}c_{l,\mu}^{++})^2\) +\(A^\prime-B^\prime|\bar{\beta}_1|^2-C^\prime|\bar{\gamma}_0|^2\)\(c_{l,\mu}^{+-}\)^2 \notag\\
    &~~~ +D\((\beta_1^2-\breve{\beta}_1^2)((\mathrm{Re}c^{++}_{l,\mu})^2-(\mathrm{Im}c^{++}_{l,\mu})^2)+4\beta_1\breve{\beta}_1\mathrm{Re}c^{++}_{l,\mu}\mathrm{Im}c^{++}_{l,\mu}\)
    \Big\rangle\geq0
\end{align}
where $A,~B$, etc.~depend on $\mu$, $l$ and $f_i$. This positivity requires the following semi-definite condition for each $\mu$ and $l$:
\begin{equation}\label{fb1Bmat}
{\small
 \begin{pmatrix}
  A-B|\bar{\beta}_1|^2-C|\bar{\gamma}_0|^2+D(\beta_1^2-\breve{\beta}_1^2) & 0 & 2D\beta_1\breve{\beta}_1 \\
  0 & A^\prime-B^\prime|\bar{\beta}_1|^2-C^\prime|\bar{\gamma}_0|^2 & 0 \\
  2D\beta_1\breve{\beta}_1 & 0 & A-B|\bar{\beta}_1|^2-C|\bar{\gamma}_0|^2-D(\beta_1^2-\breve{\beta}_1^2)
\end{pmatrix}\succeq0,
}
\end{equation}
where the entries of the matrix are associated with $\text{Re}c_{l,\mu}^{++},~c_{l,\mu}^{+-}$ and $\text{Im}c_{l,\mu}^{++}$. The above semi-definite condition in turns is equivalent to the positivity of its eigenvalues:
{\small
\begin{equation}\label{fb1Bmatei}
 \(
A-B|\bar{\beta}_1|^2-C|\bar{\gamma}_0|^2+D|\bar{\beta}_1|^2, ~A^\prime-B^\prime|\bar{\beta}_1|^2-C^\prime|\bar{\gamma}_0|^2~, A-B|\bar{\beta}_1|^2-C|\bar{\gamma}_0|^2-D|\bar{\beta}_1|^2
 \).
\end{equation}
}
We see that the causality bounds inferred from this way are on $|\bar{\beta}_1|^2$ and $|\bar{\gamma}_0|^2$ only. On the other hand, if we mask the off-diagonal terms $2D\beta_1\breve{\beta}_1$, the eigenvalues will depend on $\beta_1^2-\breve{\beta}_1^2$. More specifically, the smallest eigenvalue in this case renders to be $A-r^2((B+|D\cos2\varphi_1|)\cos^2\theta+C\sin^2\theta)$, which leads to the bounds of the form $r^2\sim A/((B+|D\cos2\varphi_1|)\cos^2\theta+C\sin^2\theta)$, reaching the maxima at $\varphi_1=\pm\pi/4,~\pm 3\pi/4$.

\begin{figure}[ht]
    \centering\!\!\!
    \begin{subfigure}{0.42\linewidth}
    \centering
    \includegraphics[width=0.99\linewidth]{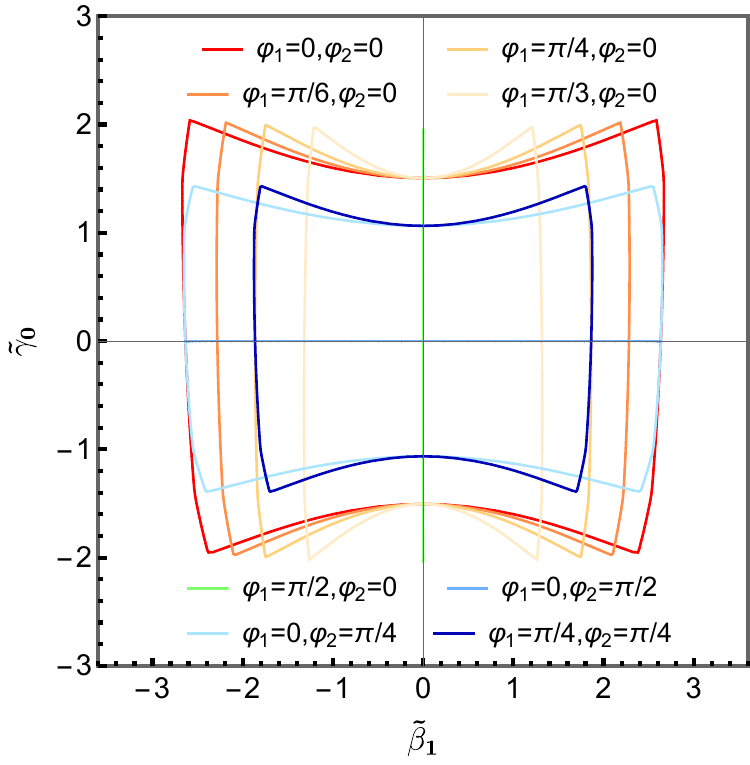}
    \end{subfigure}
    ~~~~
    \begin{subfigure}{0.42\linewidth}
    \centering
    \includegraphics[width=0.99\linewidth]{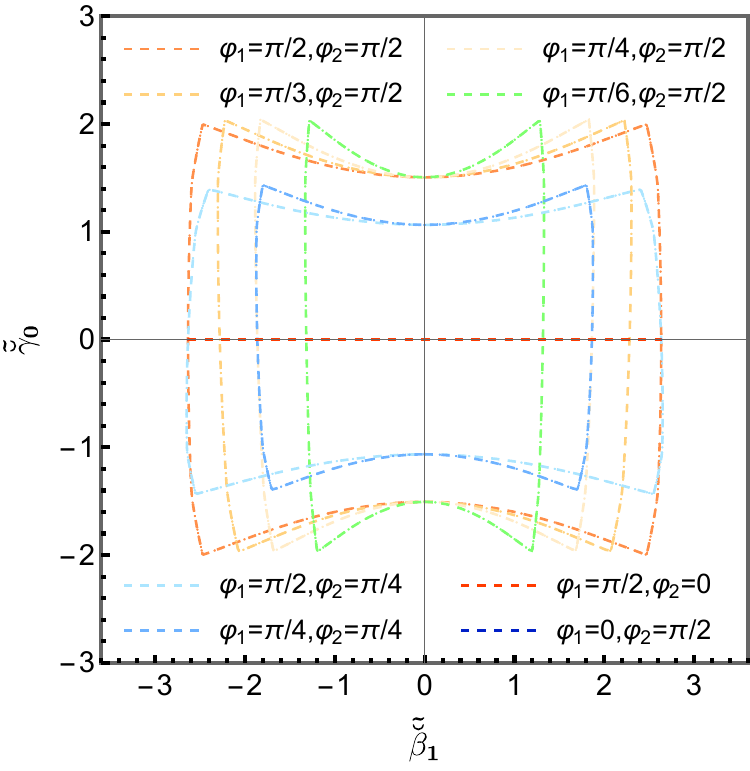}
    \end{subfigure}

\caption{ Causality bounds on $\tilde{\beta}_1$-$\tilde{\gamma}_0$ ({\it left}) and $\tilde{\breve{\beta}}_1$-$\tilde{\breve{\gamma}}_0$ ({\it right}) plane, improved by a linear iteration of \eqref{angli}.
We have set $\log(\Lambda/m_{\mathrm{IR}})=20$.    }
\label{litb1tg0}
\end{figure}

Previously, we have neglected the linear $r$ term in Eq.~\eqref{sdang}, assuming a relatively large $\log(\Lambda/m_{\mathrm{IR}})$.
In the case of a finite $\log(\Lambda/m_{\mathrm{IR}})$, we can perform a linear perturbation in $r$ on the previous results to obtain improved bounds. To this end, we take the previous result as an background $r_0$ and add a linear perturbation $\delta r$ to it:
\begin{equation}\label{angli}
  \frac{\Lambda^2|\bar{\beta}_1|}{M_P}=(r_0-\delta r)\cos\theta,\quad \frac{\Lambda^4|\bar{\gamma}_0|}{M_P^2}=(r_0-\delta r)\sin\theta,
\end{equation}
and similarly compute the bound on $\delta r$. Obviously, this can be done with more iterations to further improve the results if needed.
In Figure \ref{litb1tg0}, the first improved causality bounds on the ${\beta}_1$-${\gamma}_0$({\it left}) and ${\breve{\beta}}_1$-${\breve{\gamma}}_0$ plane are shown. For example, we may set $\log(\Lambda/m_{\mathrm{IR}})=20$ if we take $m_{\mathrm{IR}}$ to be the current Hubble scale. We see that, compared to the results of $\log(\Lambda/m_{\mathrm{IR}})=\infty$, the bounds differ only slightly. Although hardly visible, this breaks the symmetry with respect to the $\gamma_0=0$ axis, due to the presence of a linear $\bar{\gamma}_0$ term in the optimization.

\subsubsection{Anatomy of the bounds}\label{secana}

Before investigating the bounds on other coefficients, let us further analyze how the previous bounds depend on the choices of dispersion relations and SDP constraints.

\begin{figure}[ht]
    \centering
    \begin{subfigure}{0.45\linewidth}
    \centering
        \includegraphics[width=0.8\linewidth]{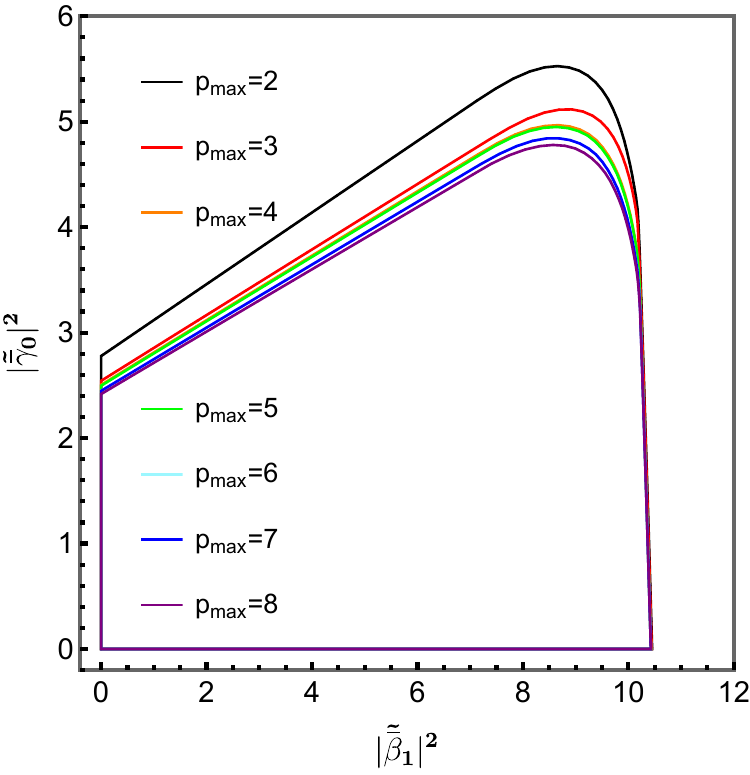}
    \end{subfigure}
     \centering
    \begin{subfigure}{0.45\linewidth}
    \centering
        \includegraphics[width=0.8\linewidth]{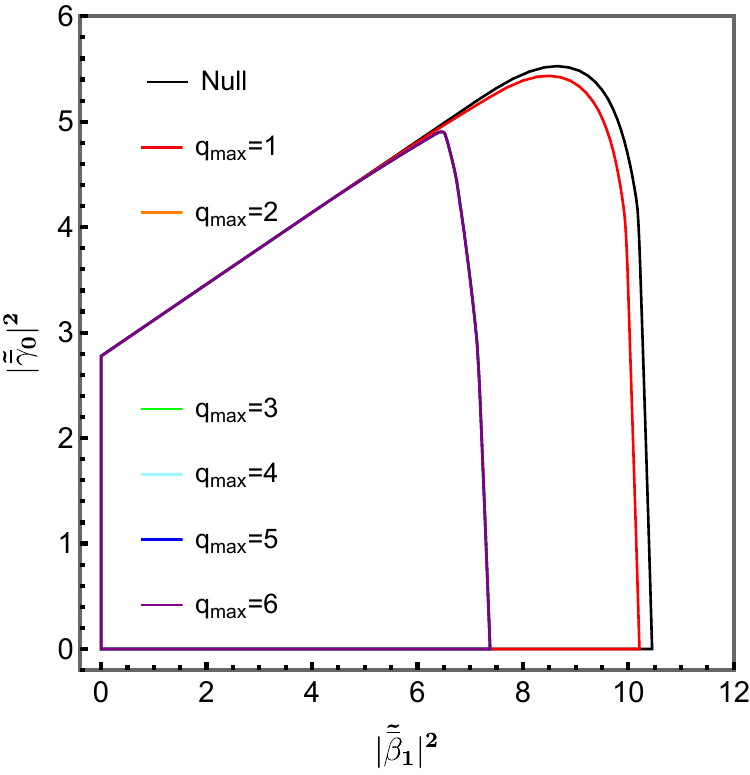}
    \end{subfigure}

     \centering
    \begin{subfigure}{0.45\linewidth}
    \centering
        \includegraphics[width=0.8\linewidth]{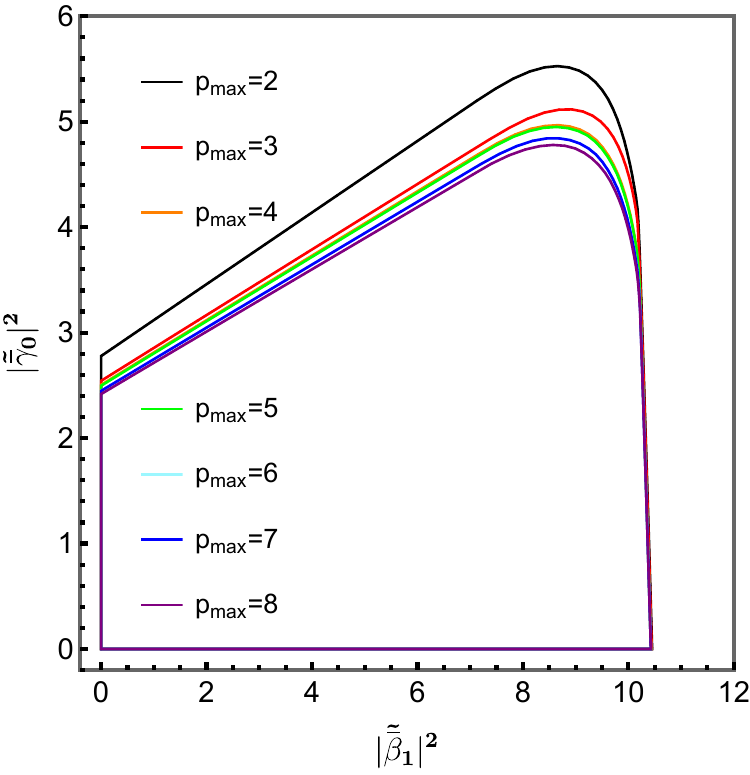}
    \end{subfigure}
     \centering
    \begin{subfigure}{0.45\linewidth}
    \centering
        \includegraphics[width=0.8\linewidth]{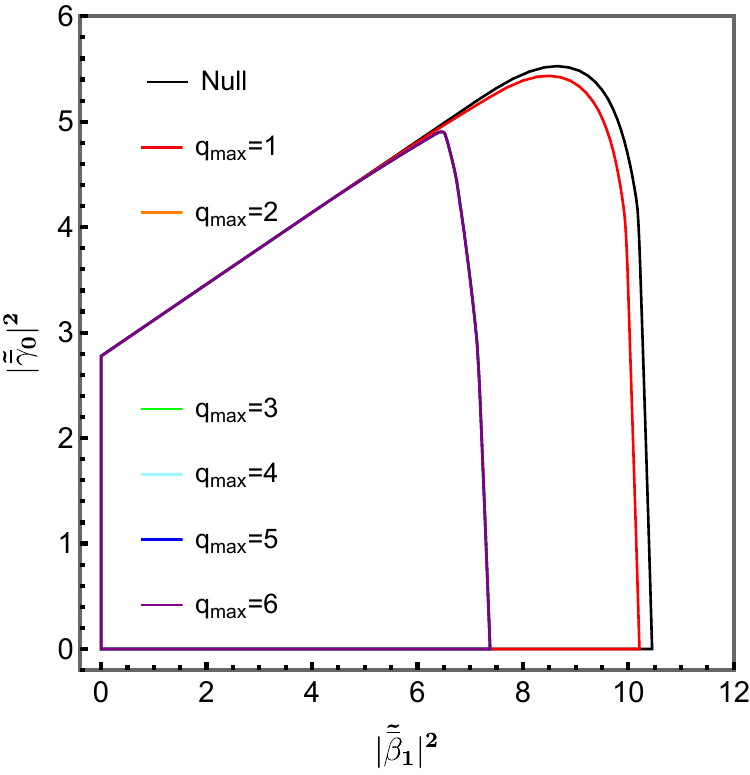}
    \end{subfigure}

    \caption{\label{figdis}Causality bounds on $|\tilde{\bar{\beta}}_1|^2$ and $|\tilde{\bar{\gamma}}_0|^2$ with different dispersion relations supplemented.
    The angles are $\varphi_1=\pi/2$, $\varphi_2=0$ for the top two figures, and $\varphi_{1}=\pi/4$, $\varphi_{2}=0$ for the bottom two. The left two panels show the optimal bounds obtained by adding the sum rules $F_{k\leq2,l}^{++--}$ and $F_{p\leq p_{\mathrm{max}},l}^{+-+-}$, and the right two panels corresponds the bounds obtained by supplementing the sum rules $F_{k\leq3,l}^{++--}$, $F_{p\leq 2,l}^{+-+-}$ and $F_{q\leq q_{\mathrm{max}},l}^{++++}$. }
\end{figure}

\noindent{\bf Choices of sum rules}

Let us first look at how the various dispersive sum rules affect the causality bounds.
In the optimization procedure, the decision variables are subjected to a series of semi-definite constraints $\bar{\mathbf{B}}_{l,\mu}\succeq0$ as in Eq.~(\ref{sdcond1}). In the purely gravitational sector, we shall adopt the following convention
\begin{equation}\label{Bconv}
  \bar{\mathbf{B}}^{i,j}_{l,\mu},~~~i,j=1(\mathrm{Re}c_{l,\mu}^{++}),\, 2(c_{l,\mu}^{+-}),\, 3(\mathrm{Im}c_{l,\mu}^{++}),
\end{equation}
to facilitate subsequent discussions. For these sum rules, the even $l$ ones contribute to the following $\bar{\mathbf{B}}_{l,\mu}$ components
\begin{eqnarray}
F_{k,l}^{+-+-}&\sim&
\left\{
\begin{aligned}\label{dispnpn}
  &(\cdots)|c_{l,\mu}^{++}|^2~&\Rightarrow& \bar{\mathbf{B}}_{l,\mu}^{11}\(=\bar{\mathbf{B}}_{l,\mu}^{33}\),\quad&  &l=2 \\
  &(\cdots)|c_{l,\mu}^{++}|^2+(\cdots)(c_{l,\mu}^{+-})^2~&\Rightarrow& \bar{\mathbf{B}}_{l,\mu}^{11}\(=\bar{\mathbf{B}}_{l,\mu}^{33}\),\bar{\mathbf{B}}_{l,\mu}^{22},\quad&  &l\geq4
\end{aligned}
\right.\\\label{disppnn}
F_{k,l}^{++--}&\sim&
\left\{
\begin{aligned}
  &(\cdots)|c_{l,\mu}^{++}|^2~&\Rightarrow& \bar{\mathbf{B}}_{l,\mu}^{11}\(=\bar{\mathbf{B}}_{l,\mu}^{33}\),\quad&  &l=0,2 \\
  &(\cdots)|c_{l,\mu}^{++}|^2+(\cdots)(c_{l,\mu}^{+-})^2~&\Rightarrow& \bar{\mathbf{B}}_{l,\mu}^{11}\(=\bar{\mathbf{B}}_{l,\mu}^{33}\),\bar{\mathbf{B}}_{l,\mu}^{22},\quad&  &l\geq4
\end{aligned}
\right.\\ \label{dispppm}
F_{k,l}^{+++-}&\sim&~
  (\cdots)c_{l,\mu}^{++}c_{l,\mu}^{+-}~\qquad\qquad\qquad~\;\Rightarrow \bar{\mathbf{B}}_{l,\mu}^{12},\bar{\mathbf{B}}_{l,\mu}^{23},\qquad  \qquad~~~~ l\geq4\\  \label{dispppp}
F_{k,l}^{++++}&\sim&~
  (\cdots)(c_{l,\mu}^{++})^2~\qquad\qquad\qquad~~~\Rightarrow \bar{\mathbf{B}}_{l,\mu}^{11}\(=-\bar{\mathbf{B}}_{l,\mu}^{33}\),\bar{\mathbf{B}}_{l,\mu}^{13},~~~  l\geq 0,
\end{eqnarray}
and the odd $l$ ones with $l\geq5$ contribute to
\begin{align}\label{frepo}
  F_{k,l}^{+-+-}&\sim
  (\cdots)\( c_{l,\mu}^{+-}\)^2\Rightarrow \bar{\mathbf{B}}_{l,\mu}^{22},\quad F_{k,l}^{+++-}=0  \\
  F_{k,l}^{++--}&\sim
  (\cdots)\( c_{l,\mu}^{+-}\)^2\Rightarrow \bar{\mathbf{B}}_{l,\mu}^{22},\quad F_{k,l}^{++++}=0 .
\end{align}
In Figure \ref{figdis}, we show how the causality bounds on the $|{\bar{\beta}}_1|^2$-$|{\bar{\gamma}}_0|^2$ plane vary with the choice of dispersive sum rules. In the top two panels, the bounds are obtained by using the sum rules with $F_{p\leq p_{\mathrm{min}},l}^{+-+-}$ and $F_{q\leq q_{\mathrm{min}},l}^{++++}$ with  $\varphi_{1}=\pi/2$, $\varphi_2=0$. In order to include necessary coefficients as well as null constraints in the optimization, the sum rules with $F_{k\leq 3,l}^{++--}$ and $F_{k\leq 2,l}^{+-+-}$ are also used.
The bottom two panels are similar except for the choice of $\varphi_{1}=\pi/4$, $\varphi_{2}=0$. Let us discuss how the addition of the various sum rules affect the bounds:

\begin{itemize}

\item The sum rules with $F_{p,l}^{+-+-}$ do not relate any low energy EFT coefficients to the UV S-matrix, so they are often referred to as null constraints. As we include more sum rules with $F_{p\geq3,l}^{+-+-}$, the bound on $|\tilde{\bar{\gamma_0}}|^2$ becomes smaller. This is because, as we see in Eq.~(\ref{dispnpn}), the sum rules with $F_{p,l}^{+-+-}$ contribute to $\bar{\mathrm{B}}_{l,\mu}^{11}$ and $\bar{\mathrm{B}}_{l,\mu}^{33}$ for $l\geq2$ and $\bar{\mathrm{B}}_{l,\mu}^{22}$ for $l\geq4$, making the positivity conditions easier to satisfy.
However, the sum rules with $F_{p,l}^{+-+-}$ only alter the bounds slightly. This is due to
the $F_{p,l}^{+-+-}$ sum rules starting with spin $l\geq2$, while the $l=0$ partial waves from other sum rules play the dominant role in this bound. The presence of the sum rules $F_{1,l}^{+-+-}$, $F_{2,l}^{+-+-}$ is essential for making the bounds of $|\tilde{\bar{\gamma}}_0|^2$ to be $\mathcal{O}(1)$, without which the relevant bounds are $\mathcal{O}(100)$. Note that the $F_{p,l}^{+-+-}$ sum rules start with spin $l\geq4$ for $p\geq3$.

\item The sum rules with $F_{q,l}^{++++}$  mainly influence the bounds in the $|\tilde{\bar{\beta}}_1|^2$ direction. The sum rules with $F_{1,l}^{++++}$ and $F_{2,l}^{++++}$ contribute to the normalization (\ref{kbcons}), leading to smaller bounds, while the sum rules with $F_{q,l}^{++++}$ with $q\geq3$ do not affect the optimal bounds significantly. The latter is because those sum rules do not contain the coefficients to optimize over and also contribute negagtively to $\bar{\mathrm{B}}_{l,\mu}^{11}$ or $\bar{\mathrm{B}}_{l,\mu}^{33}$, so the associated decision variables have to be suppressed to satisfy the semi-definite conditions.

\item The sum rules with $F_{k\geq4,l}^{++--}$, and $F_{k>2,l}^{+++-}$ have little effects on the bounds, as they do not contain the concerned coefficients and/or contribute negatively to the off-diagonal components of $\bar{\mathbf{B}}_{l,\mu}$ (see Eq.~(\ref{dispppm})).

\end{itemize}

\smallskip
\noindent{\bf The $\mu$ and $l$ constraint space}

The permissible configurations of the decision variables and hence the EFT coefficients are subject to the constraints from the positive semi-definite matrices $\bar{\mathbf{B}}_{l,\mu}$, which span different UV masses $\sqrt{\mu}$, spins $l$, with some of the regions partly sampled by the impact parameter $b= 2l/\sqrt{\mu}$ (see Appendix \ref{numde}). In the following we will investigate how these different constraint regions affect the causality bounds.

\begin{figure}[ht]
    \centering\!\!\!
    \begin{subfigure}{0.362\linewidth}
    \centering
    \includegraphics[width=0.99\linewidth]{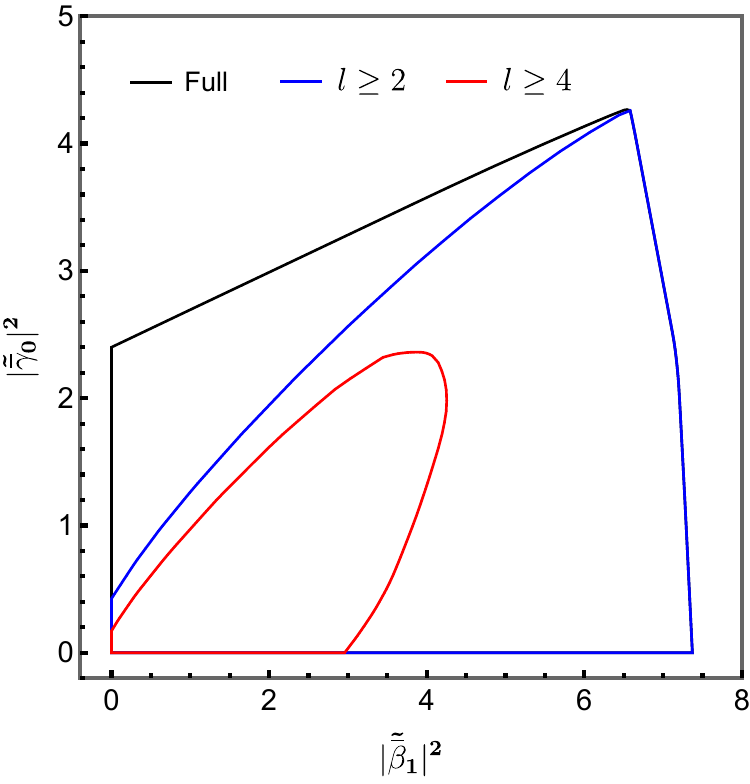}
    \end{subfigure}
    \begin{subfigure}{0.49\linewidth}
    \centering
    \includegraphics[width=0.99\linewidth]{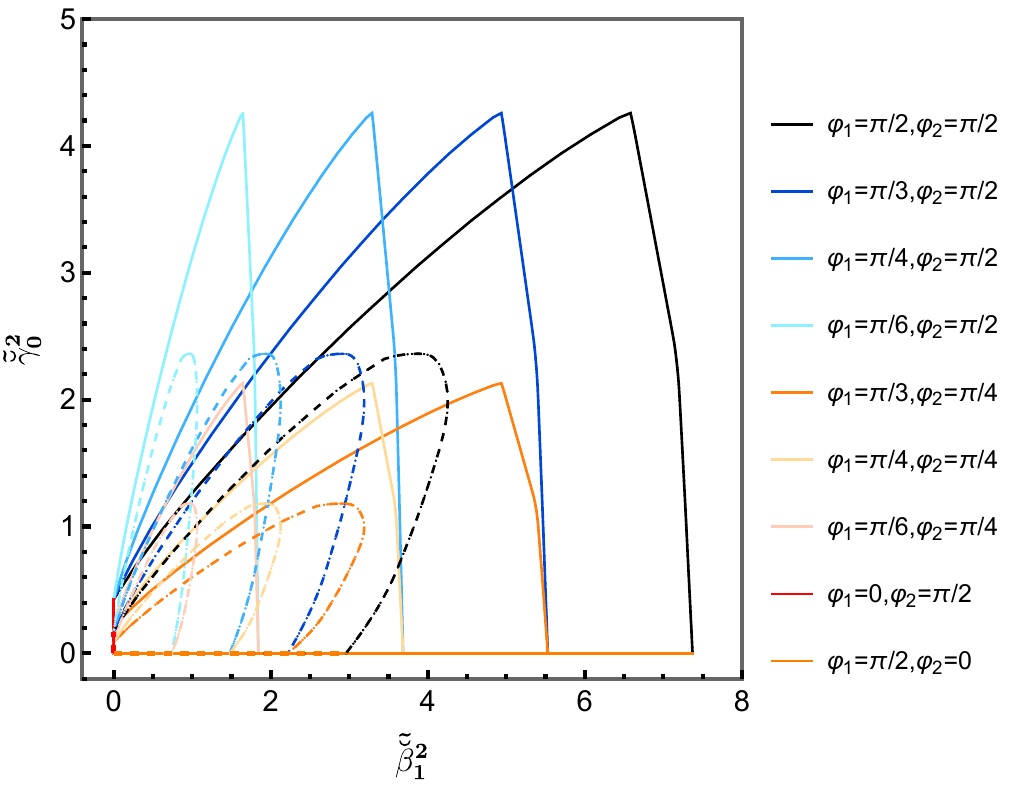}
    \end{subfigure}
\caption{({\it left}) Causality bounds on the $|\tilde{\bar{\beta}}_1|^2$-$|\tilde{\bar{\gamma}}_0|^2$ plane with certain UV low spin components suppressed.
The causality bounds without UV spin $l=0$ and $l\leq 2$ are shown. ({\it right}) The causality bounds without the UV $l=0$ (solid) and $l\leq2$ (dashed) components are plotted against the parity-violating coefficients. }
\label{diffl}
\end{figure}

In Figure \ref{diffl}, we plot the causality bounds on the $|{\bar{\beta}}_1|^2$-$|{\bar{\gamma}}_0|^2$ plane~(left panel) and the corresponding projections onto the ${\breve{\beta}}_1^2$-${\breve{\gamma}}_0^2$ plane~(right panel), with certain low spin UV states suppressed. Since a healthy scalar-tensor EFT requires support from both $l\leq4$ and $l\geq5$ UV spin states \cite{Hong:2023zgm} and the $l=1,~3$ spin states do not contribute to the bounds in a pure gravitational setup case (the $l=3$ spin component does contribute slightly when helicity-mixing sum rules are used),
we plot the bounds for UV spins $l\geq l_0$ with $l_0=2$ and $4$ respectively.  In the left pannel where UV scalars are suppressed (the $l\geq2$ line), the top right kink $(|\tilde{\bar{\beta}}_1|^2,|\tilde{\bar{\gamma}}_0|^2)=(6.58,4.26)$ remains, similar to the parity-conserving case.
However, this kink disappears when the UV spin-2 state  are also eliminated, leading to the suppression of the bounds in all directions, especially along the $|{\breve{\beta}}_1|^2$ direction. It is also observed that the large spins contribute only very slightly to the results.

\begin{figure}[ht]
    \centering\!\!\!
    \begin{subfigure}{0.43\linewidth}
    \centering
    \includegraphics[width=0.99\linewidth]{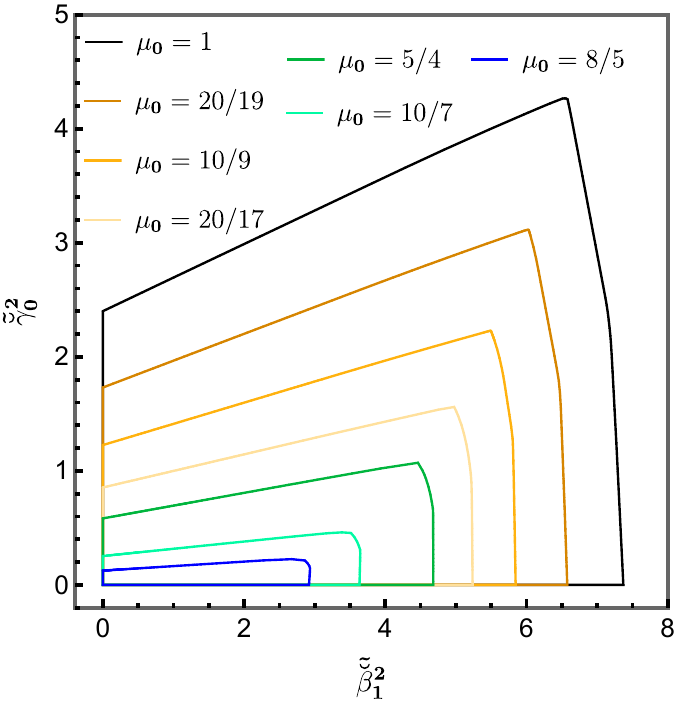}
    \end{subfigure}
    ~~~~
    \begin{subfigure}{0.45\linewidth}
    \centering
    \includegraphics[width=0.99\linewidth]{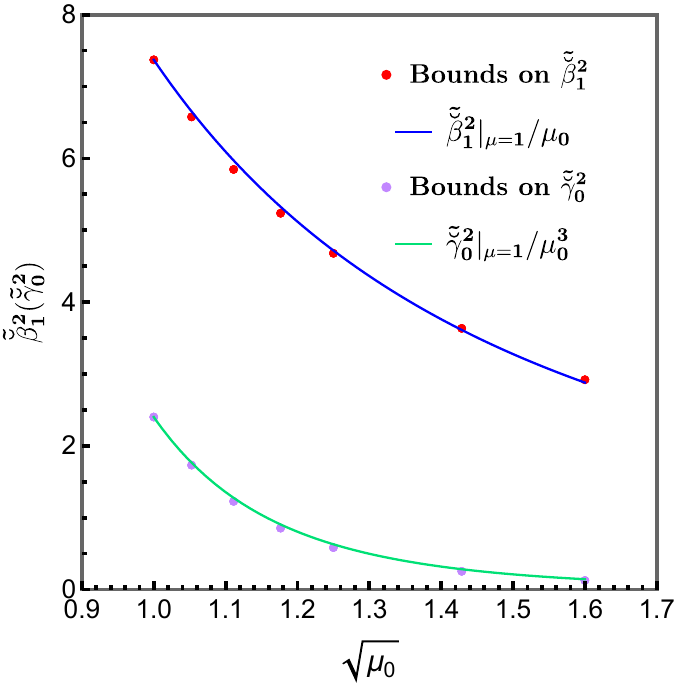}
    \end{subfigure}
\caption{({\it left}) Causality constraints on the $\tilde{\breve{\beta}}_1^2$-$\tilde{\breve{\gamma}}_0^2$ plane when the mass of the lowest-lying UV state is raised from $\Lambda$ to $\sqrt{\mu_0}$.
({\it right}) Scaling behaviors of  $\tilde{\breve{\beta}}_1^2$ (blue line) and $\tilde{\breve{\gamma}}_0^2$ (green line) with respect to $\mu_0$, matched to the numerical results (red and purple dots). }
\label{diffmu}
\end{figure}

To probe how the masses of the UV states, parameterized by $\sqrt{\mu}$, affect the bounds, in Figure \ref{diffmu}, we plot the bounds on the dCS coefficients plane when the part of the constraints closest to the cutoff $\Lambda$ is disregarded. That is, we drop the semi-definite conditions on $\bar{\mathbf{B}}_{l,\mu}$ within the range $\Lambda^2<\mu<\mu_0$ when performing the SDP optimization. We see that, as $\mu_0$ increases from $\Lambda^2$, the causality bounds are significantly reduced. On the other hand, if we only impose the semi-definite conditions within $\Lambda^2<\mu<\mu_{0}$, disregarding the $\bar{\mathbf{B}}_{l,\mu}$ constraints from $\mu>\mu_{0}$ in the finite $\mu$ and $l$ region, we find that the value of $\mu_{0}$ does not significantly affect the results for a relative large $\mu_{0}$. This highlights that the causality bounds are most sensitive to the part of the dispersive integral close to the EFT cutoff.

In fact, from the sum rules, one can easily infer the scaling behavior of the bounds when we drop the UV states with mass larger than $\mu_0$.
To see this, note that the leading scaling behavior of the relevant sum rules in the gravitational sector is given by $F_{k,l}^{++--}\sim\mathcal{O}\({1}/{\mu^{k+1}}\), ~F_{k,l}^{+-+-}\sim\mathcal{O}\({1}/{\mu^{k+4}}\),~  F_{1,l}^{++++},F_{2,l}^{++++}\sim\mathcal{O}\({1}/{\mu^{3}}\),~F_{3,l}^{++++},F_{4,l}^{++++}\sim\mathcal{O}\({1}/{\mu^{5}}\)$,  since the semi-definite constraints start with $\mu=\mu_0\geq\Lambda^2$.
If we want to compute the bounds on the coefficients $\breve{\beta}_1^2$, the associated semi-definite conditions $\bar{\mathbf{B}}_{l,\mu}\succeq0$ can be schematically written as
\begin{equation}
\label{ypoley}
y_{\text{pole}}\mathcal{O}\(\frac{1}{\mu^3}\)+y\mathcal{O}\(\frac{1}{\mu^4}\)\succeq0,
\end{equation}
where $y_{\text{pole}}$ and $y$ are the relevant decision variables associated with the $t$-channel pole and $\breve{\beta}_1^2$ respectively. The dual variable $y$ and $y_{\rm pole}$ have to be relatively large to balance the scaling hierarchy. Consequently, we have $y_{\text{pole}}\sim\mathcal{O}(\mu^3)$ and $y\sim\mathcal{O}(\mu^4)$. Acting $\langle\cdots\rangle$ on both sides of Eq.~(\ref{ypoley}) and making correspondence according to Eq.~(\ref{disf}), we can get positive conditions on $\breve{\beta}_1^2$: $y_{\text{pole}}\log(\Lambda/m_{\mathrm{IR}})+y \breve{\beta}_1^2\geq0$.
Combining these, we infer that schematically the bounds on ${\breve{\beta}}_1^2$ should scale as
\begin{equation}
{\breve{\beta}}_1^2\leq |y_{\text{pole}}/y|\sim\mathcal{O}\(\frac1{\mu}\) .
\end{equation}
A similar argument can also be applied to  ${\gamma}_0^2$, which gives
\begin{equation}
\tilde{\breve{\gamma}}_0^2\sim\mathcal{O}\(\frac{1}{\mu^3}\) .
\end{equation}
In the right pannel of Figure \ref{diffmu}, we see that these scaling behaviors match the rigorous numerical results very well.

\begin{figure}[ht]
    \centering\!\!\!
    \begin{subfigure}{0.49\linewidth}
    \centering
    \includegraphics[width=0.8\linewidth]{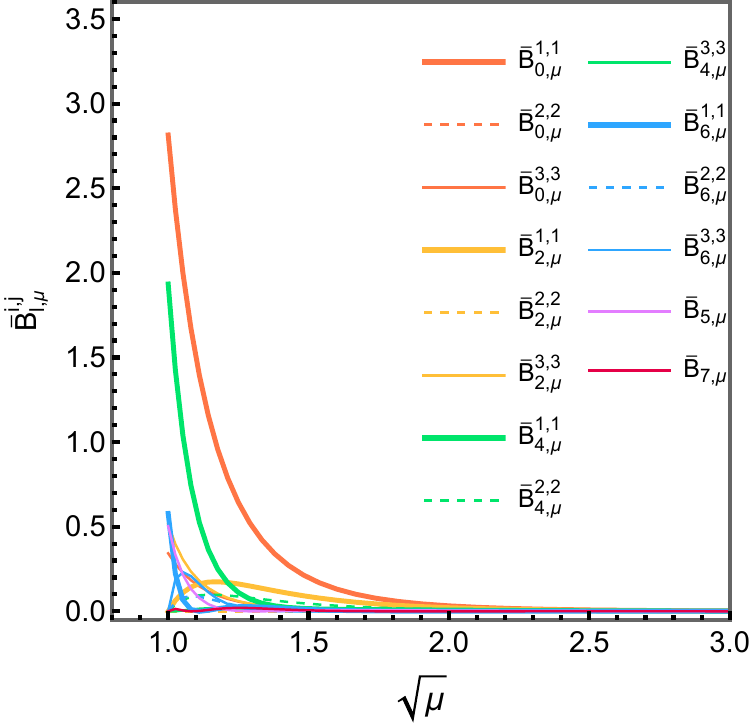}
    \end{subfigure}
    \begin{subfigure}{0.49\linewidth}
    \centering
    \includegraphics[width=0.8\linewidth]{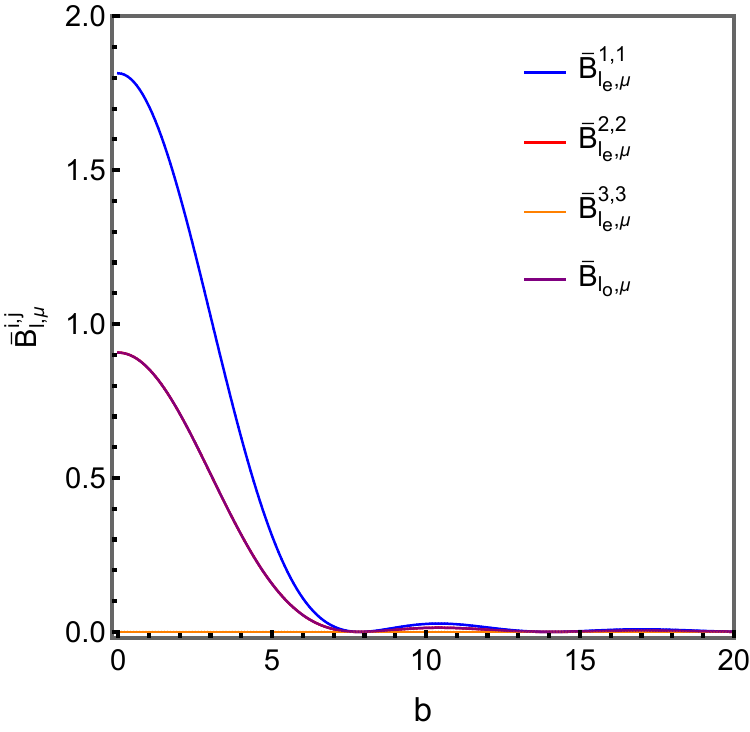}
    \end{subfigure}\\
    \begin{subfigure}{0.49\linewidth}
    \centering
    \includegraphics[width=0.8\linewidth]{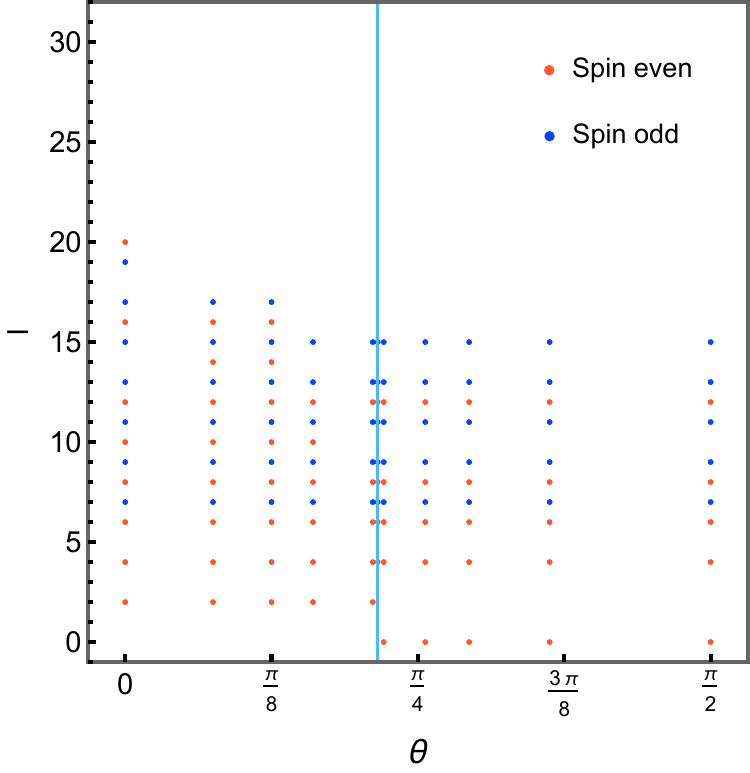}
    \end{subfigure}
    \begin{subfigure}{0.49\linewidth}
    \centering
    \includegraphics[width=0.8\linewidth]{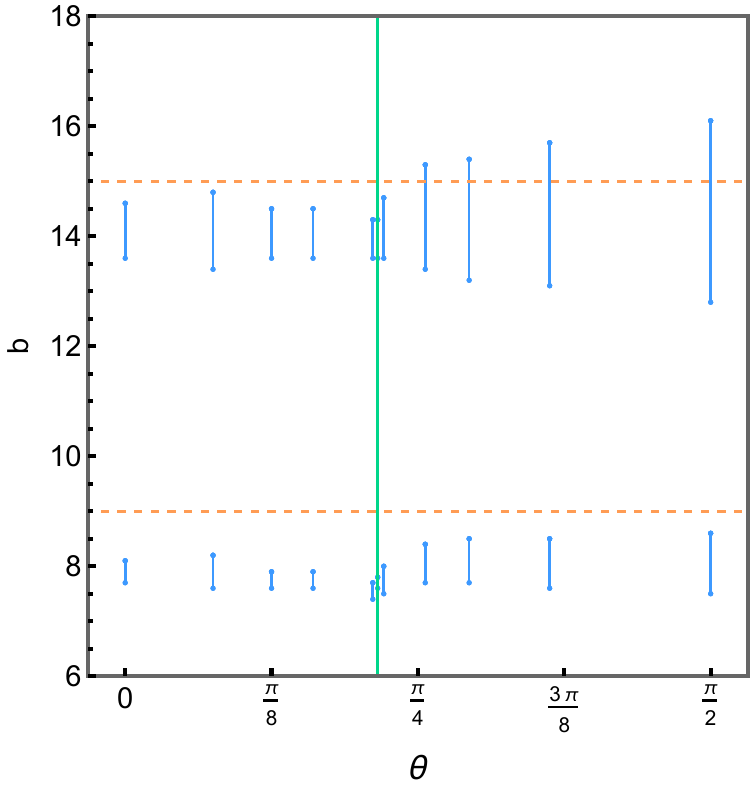}
    \end{subfigure}
\caption{Various intermediate quantities for computing the bounds of $\breve{\beta}_1^2$-$\breve{\gamma}_0^2$. For the top two panels we have set $\breve{\gamma}_0=0$. ({\it top left}) $\bar{\mathrm{B}}_{l,\mu}^{i\,j}$ (see Eq.~(\ref{Bconv})) against $\sqrt{\mu}$.
({\it top right}) $\bar{\mathrm{B}}_{l_*,\mu}^{i\,j}$ against the impact parameter $b= 2l/\sqrt{\mu}$, with $l_*=l_e,~l_o$ representing the oddness (even/odd) of the $\bar{\mathbf{B}}_{l_*,\mu}$. ({\it bottom left}) Dominating spins for various $\theta\equiv\arctan \tilde{\breve{\gamma}}_0/\tilde{\breve{\beta}}_1$. The kink $(\tilde{\beta}_1^2,\tilde{\gamma}_0^2)=(6.58,4.26)$ is marked as the vertical blue line. ({\it bottom right}) Dominating impact parameter $b$ range for various $\theta$. The two horizontal dashed lines indicate the necessary range of $b$ to make the SDP solvable. }
\label{Bmat}
\end{figure}

As specified in Appendix \ref{numde}, some of the constraint space is sampled by the impact parameter $b$. Particularly, the semi-definite constraints from the region with large $\mu$ and $l$ but a finite $b= 2l/\sqrt{\mu}$ are vital for the optimization. For example, generically  for all the computations in the pure gravity sector, we find that without the constraints from the region $9\leq b\leq15$, the numerical SDP is not solvable, while the exclusion of other regions of $b$ does not affect the bounds significantly.
More specifically, the bounds are most sensitive to the constraints in the neighbor of $b=14$, excluding which the bounds get significantly weakened.
This occurs because only one sum rule contains the $t$-channel pole in pure graviton scatterings, and these values of $b$ correspond to the semi-definite boundary of $\bar{\mathbf{B}}_{l,\mu}$, as shown in the top right panel of Figure \ref{Bmat}.

In the top left pannel of Figure \ref{Bmat}, we also show the values of the  $\bar{\mathbf{B}}_{l,\mu}$ components in the finite $\mu$,~$l$ region when the upper bound on $\breve{\beta}_1^2$~($\breve{\gamma}_0=0$) is reached. The relevant $\bar{\mathbf{B}}_{l,\mu}$ should reach its semi-definite boundary at the edge of the causality bounds, or at least one of the eigenvalues, which are simply the diagonal components in the pure dCS theory, equals to zero. The constraints from $l\geq2$ and $\mu=\Lambda^2$ play a dominant role here, as can be confirmed by the fact that $\bar{\mathrm{B}}_{2,\Lambda^2}^{1,1}$ vanishes, $\bar{\mathrm{B}}_{0,\mu}^{2,2}$ and $\bar{\mathrm{B}}_{2,\mu}^{2,2}$ are trivially zero,
and $\bar{\mathrm{B}}_{2,\mu\gtrsim(3\Lambda)^2}^{i,j}$ are polynomially ($\mathcal{O}(1/\mu^n))$ suppressed. On the other hand, $\bar{\mathrm{B}}_{0,\mu}^{1,1}$ and $\bar{\mathrm{B}}_{0,\mu}^{3,3}$ are positive for finite $\mu$, which indicates that spin-0 states have minor effects on the final bounds. As for higher spin states $4\leq l\leq 20$, for each $l$, there is at least one non-trivial eigenvalue reaching zero at $\mu=\Lambda^2$, indicating that they do affect the results, albeit slightly.
For the $l>20$ spin states, $\bar{\mathbf{B}}_{l,\mu}$ are mostly positive, making them negligible for the bounds.
On the other hand, the upper bound on $\breve{\gamma_0}^2$~($\breve{\beta}_1^2=0$) is sensitive to the $l=0$ state, as in this case $\bar{\mathrm{B}}_{0,\Lambda^2}^{11}=0$. A caveat is that although the (nontrivial) eigenvalues of $\bar{\mathbf{B}}_{2,\mu}$ are positive, the spin-2 states still affect the bounds slightly, due to the null constraint $0=\langle F_{k,l}^{+-+-}\rangle$, which contributes to the $l\geq2$ components. Although these null constraints do not contribute to $\bar{\mathbf{B}}_{2,\mu}$, they do affect the decision variables.

In the bottom left panel of Figure \ref{Bmat}, we plot the dominant contributing spins for the bounds along various directions on the $\breve{\beta}_1^2$-$\breve{\gamma}_0^2$ plane, where $\theta\equiv\arctan \tilde{\breve{\gamma}}_0/\tilde{\breve{\beta}}_1$. The kink $(\tilde{\beta}_1^2,\tilde{\gamma}_0^2)=(6.58,4.26)$ is marked as the vertical blue line. When $\theta < \theta_{\text{kink}}$, the lowest dominant contributing spin is $l = 2$. At those angles, the spin-2 states play a dominant role in determining the bounds.
On the other hand, when $\theta>\theta_{\text{kink}}$, the lowest contributing spin is $l=0$.
Notice that in this case the spin-2 states are negligible. However, when the spin-0 states are suppressed, the spin-2 states would be activated.
Also, we find that the suppression of the spin-2 states would allow higher spin states $14\leq l\leq 18$ to come into play, since less restrictions are imposed on the decision variables --- a larger decision variable space leads to a tighter bound.

To see how the constraints from the finite $b$ (and large $\mu$) region affect the bounds, in the bottom right panel of Figure \ref{Bmat}, we plot dominating ranges of $b$ that contribute to the bounds which are determined by the semi-definiteness of $\bar{\mathbf{B}}_{l,\mu}$.
In the practical computations, we find that the SDP cannot be solved without the constraints from $9\leq b\leq15$(marked by the two horizontal dashed lines), which is consistent with the fact that $\bar{\mathbf{B}}_{l,\mu}$ reaches its semi-positive definite boundary near $b=14$.

Note that we only keep leading $\mathcal{O}(1/\mu^3)$ contribution in this region, for which only the sum rule $-1/(M_P^2t)=\langle F_{2,l}^{++--}\rangle\sim\mathcal{O}(1/\mu^3)$ contributes. This sum rule arises from the graviton $t$-channel  exchange
and provides necessary restrictions on the associated decision variables such that the semi-definite optimization becomes solvable.

\vspace{0.3cm}

\subsection{Influence of higher order coefficients}\label{secho}

So far we have calculated the causality bounds on the low energy EFT couplings up to $\mathcal{O}(1/\Lambda^6)$, while being agnostic about the higher order coefficients. However, in some specific EFTs or experimental setups, one may have some prior knowledge about coefficients that are not of primal concern.
In this subsection, we shall investigate the effects of fixing higher order coefficients on the causality bounds of the lower order coefficients.

\begin{figure}[ht]
    \centering\!\!\!
    \begin{subfigure}{0.43\linewidth}
    \centering
    \includegraphics[width=0.99\linewidth]{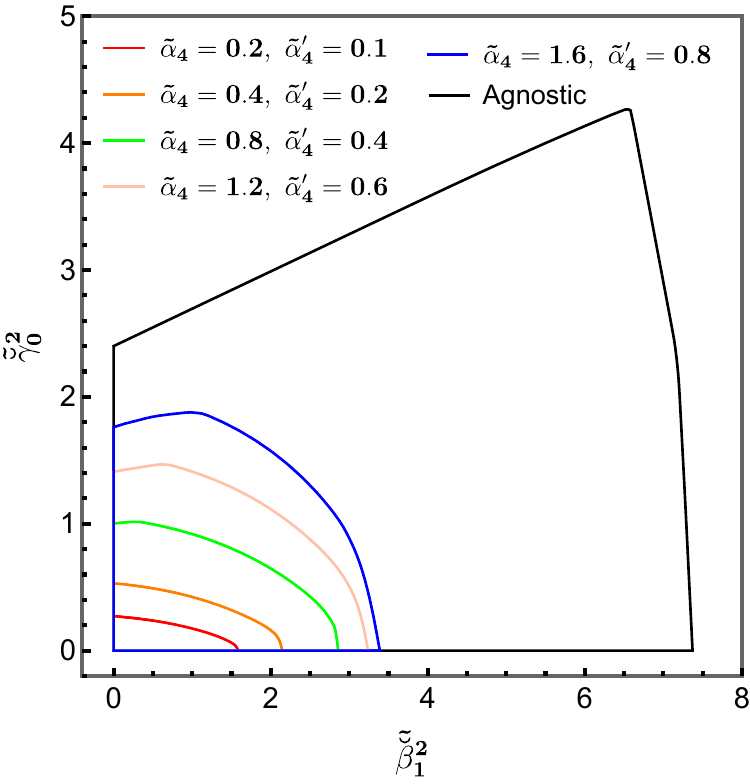}
    \end{subfigure}
    \centering
    ~~~~
    \begin{subfigure}{0.43\linewidth}
    \centering
    \includegraphics[width=0.99\linewidth]{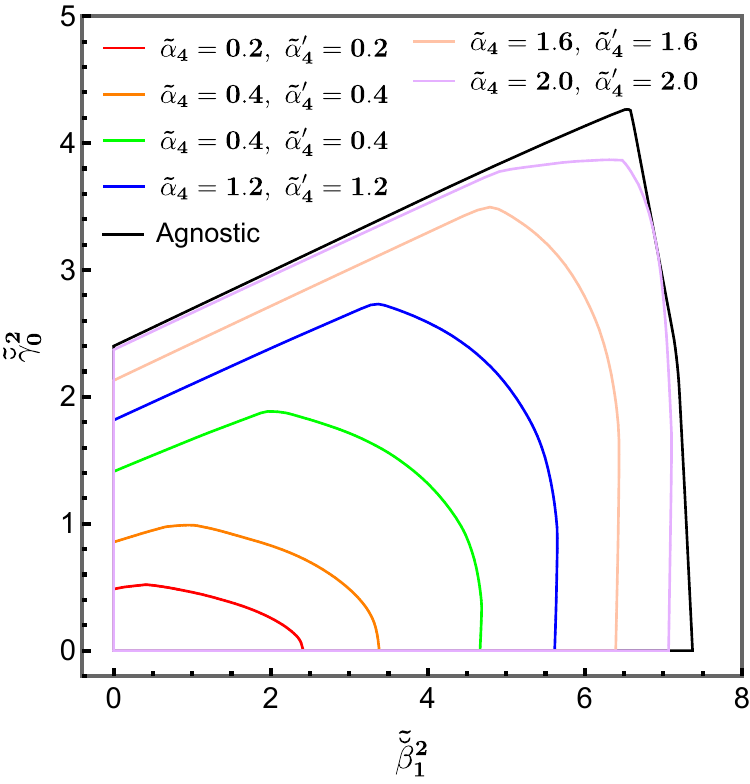}
    \end{subfigure}\\
    \begin{subfigure}{0.43\linewidth}
    \centering
    \includegraphics[width=0.99\linewidth]{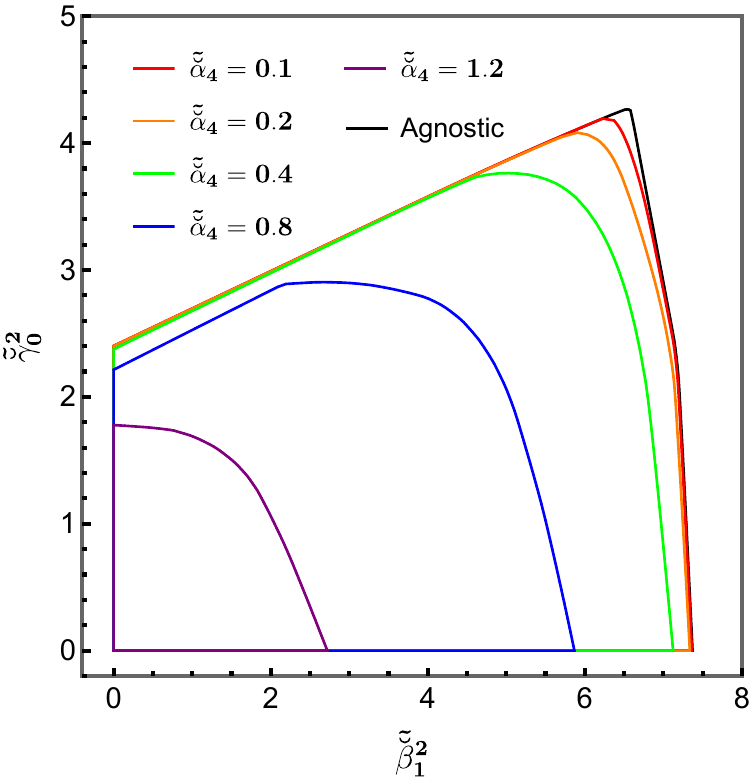}
    \end{subfigure}
    ~~~~
    \begin{subfigure}{0.43\linewidth}
    \centering
    \includegraphics[width=0.99\linewidth]{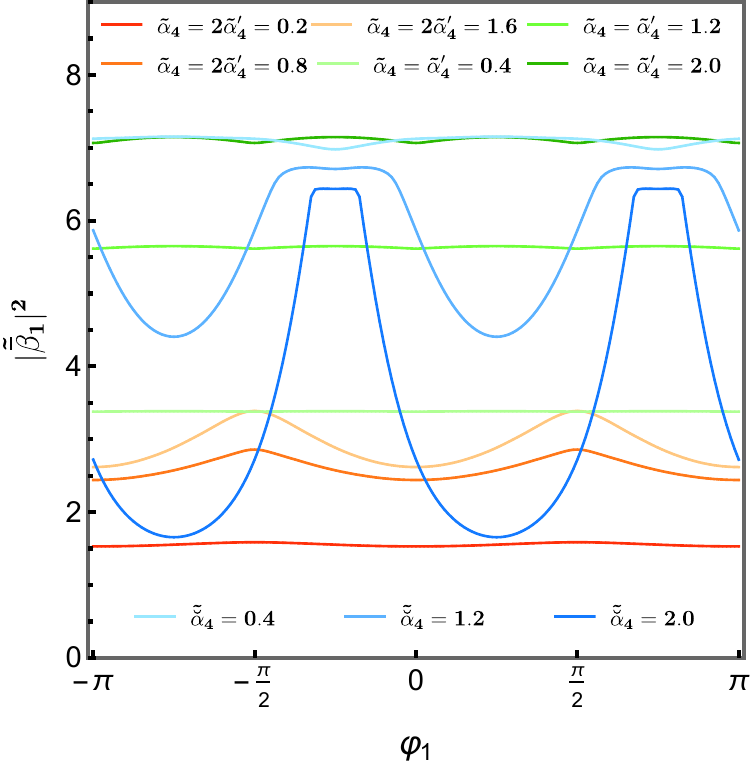}
    \end{subfigure}
\caption{({\it top left}) Causality constraints on $\tilde{\breve{\beta}}_1^2$ and $\tilde{\breve{\gamma}}_0^2$ for fixed quartic curvature couplings $\tilde{\alpha}_4$ and $\tilde{\alpha}^\prime_4$ along the direction of $\tilde{\alpha}_4=2\tilde{\alpha}^\prime_4$, where $\tilde{\alpha}_4\equiv \alpha_4 \Lambda^6/(M_P^2\log(\Lambda/m_{\mathrm{IR}}))$, $\tilde{\alpha}^\prime_4\equiv \alpha^\prime_4 \Lambda^6/(M_P^2\log(\Lambda/m_{\mathrm{IR}}))$.
({\it top right}) Same as the top left but along the $\tilde{\alpha}_4=\tilde{\alpha}^\prime_4$ direction. ({\it bottom left}) Similar to the top two panels but for fixed parity-violating quartic curvature couplings $\tilde{\breve{\alpha}}_4\equiv \breve{\alpha} \Lambda^6/(M_P^2\log(\Lambda/m_{\mathrm{IR}}))$. ({\it bottom right}) Causality constraints on the $\tilde{\bar{\beta}}_1^2$~(for $\bar{\gamma}_0=0$) for various $\varphi_1$ in the presence of fixed parity-conserving quartic curvature couplings $\tilde{\alpha}_4$, $\tilde{\alpha}^\prime_4$ and $\tilde{\breve{\alpha}}_4$. }
\label{quartic}
\end{figure}

Let us first look at the $\mathcal{O}((s,t,u)^4)$ terms in the pure gravitation scattering amplitudes: $g_{4,0}^{T_1} s^4 $ in $\mathcal{M}^{++--}$ and $(g_{2,0}^{T_3}+i\breve{g}_{2,0}^{T_3})(s^2+t^2+u^2)^2$ in $\mathcal{M}^{++++}$, which come from the quartic curvature operators $\alpha_4(\mathcal{R}^{(2)})^2/4$, $\alpha_4^\prime(\tilde{\mathcal{R}}^{(2)})^2/4$ and $\breve{\alpha}_4\mathcal{R}^{(2)}\tilde{\mathcal{R}}^{(2)}/4$, the first two being parity-conserving and the last one being parity-violating. Then relevant coefficients can be explicitly written as
\be
g_{4,0}^{T_1}=2(\alpha_4+\alpha_4^\prime)/M_P^4, ~~~~g_{2,0}^{T_3}=(\alpha_4-\alpha_4^\prime)/M_P^4,~~~~ \breve{g}_{2,0}^{T_3}=\breve{\alpha}_4/M_P^4 .
\ee
In Figure \ref{quartic} we show the causality bounds on the ${\breve{\beta}}_1^2$ and ${\breve{\gamma}}_0^2$ plane for various $\alpha_4$ and $\alpha_4^\prime$ along $\tilde{\alpha}_4=2\tilde{\alpha}_4^{\prime}$~(top left) and $\tilde{\alpha}_4=\tilde{\alpha}_4^{\prime}$~(top right) directions. We see that the bounds on ${\breve{\beta}}_1^2$ and
 ${\breve{\gamma}}_0^2$ are significantly reduced when $\alpha_4$ and $\alpha_4^\prime$ become smaller.
In the bottom left panel of Figure \ref{quartic}, we show how the bounds on ${\breve{\beta}}_1^2$ and ${\breve{\gamma}}_0^2$
are affected by fixing the parity-violating coupling $\breve{\alpha}_4$. We see that, different from the parity-conserving operators, small values of the parity-violating coupling $\breve{\alpha}_4$ do not alter the causality bounds significantly, while large values of the parity-violating coupling $\breve{\alpha}_4$ {\it do} significantly reduce the bounds. This originates from the fact the parity-violating operators give rise to imaginary parts of the amplitudes,
contributing to the off-diagonal entry $\bar{\mathbf{B}}_{l,\mu}^{13}$. When $\breve{\alpha}_4$ is small, little effects are leveraged against the semi-positive conditions.
When $\breve{\alpha}_4$ becomes relatively large, the off-diagonal components of $\bar{\mathbf{B}}_{l,\mu}^{13}$ become non-negligible, so the decision variables associated with the diagonal components have to be larger to make $\bar{\mathbf{B}}_{l,\mu}$ positive semi-definite, which gives rise to tighter bounds. This of course is consistent with the usual intuition that the parity-violating coefficients can be set to zero without affecting the parity-conserving coefficients.

\begin{figure}[ht]
    \centering\!\!\!
    \begin{subfigure}{0.43\linewidth}
    \centering
    \includegraphics[width=0.99\linewidth]{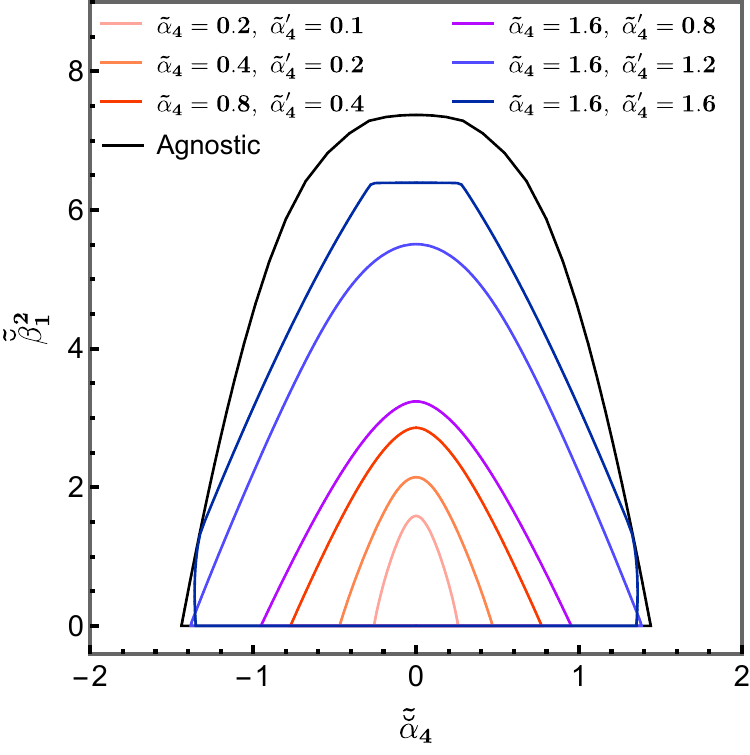}
    \end{subfigure}
\caption{Causality bounds on the $\tilde{\breve{\alpha}}_4$-$\breve{\beta}_1^2$ plane for various fixed $\tilde{\alpha}_4$ and $\tilde{\alpha}_4^\prime$. The parity-violating coefficients are upper bounded by the parity-conserving coefficients.
}
\label{fig:a4ta4}
\end{figure}

Fixing higher order couplings $\alpha_4$, $\alpha_4^\prime$ and $\breve{\alpha}_4$ will induce $\varphi_1$ dependence for the bounds on $|{\bar{\beta}}_1^2|$, as shown in the bottom right panel of Figure\ref{quartic}, which is absent when agnostic about these coefficients. To understand this, note that, in the presence of $\alpha_4$ and  $\alpha_4^\prime$, the smallest eigenvalue in Eq.~(\ref{fb1Bmatei}) for $\bar{\gamma}_0=0$ is modified to
\be
A-B|\bar{\beta}_1|^2-F(\alpha_4+\alpha_4^\prime)-\sqrt{(D|\bar{\beta}_1|^2-G(\alpha_4-\alpha_4^\prime))^2+4\beta_1^2G(\alpha_4-\alpha_4^\prime)}
\ee
When $\alpha_4=\alpha_4^\prime$, the bounds on $|\bar{\beta}_1|^2$ have no dependence on $\varphi_1$, in agreement with the numerical results. When $\alpha_4\neq\alpha_4^\prime$,
the causality bound on $|\bar{\beta}_1|^2$ reaches its minimum when $\varphi_1=0,\pm\pi$ and its maximum at $\varphi_1=\pm\pi/2$, again consistent with the numerical results. Similarly, the presence of a non-vanishing $\breve{\alpha}_4$ (but $\alpha_4=\alpha_4^\prime=0$) will alter the $\varphi_1$ dependence of $|\bar{\beta}_1|$ in a slightly different manner, where the smallest eigenvalue becomes $A-B|\bar{\beta}_1|^2-({D^2|\bar{\beta}_1|^2+F^2\breve{\alpha}_4^2+4DF\breve{\alpha}_4\beta_1\breve{\beta}_1})^{1/2}$.
Consequently, the bound on $|\bar{\beta}_1|^2$ reaches the maximum at $\varphi_1=-\pi/4,3\pi/4$, and the minimum at $\varphi_1=-3\pi/4, \pi/4$, which qualitatively matches the numerical results.

An interesting phenomenon is that the parity-violating couplings are often bounded by the parity-conserving couplings. For example, in Figure \ref{fig:a4ta4},  we show the causality bounds on the parity-violating coefficients $\breve{\alpha}_4$ and $\breve{\beta}_1^2$ for various fixed parity-conserving couplings ${\alpha}_4$ and ${\alpha}_4^{\prime}$. We see that as the parity-conserving couplings decrease, the bounds on the parity-violating couplings shrink. That is, in the absence of the ${\alpha}_4$ and ${\alpha}_4^{\prime}$ coefficients, the parity-violating coefficients $\breve{\alpha}_4$ and $\breve{\beta}_1^2$ are forced to be zero.

\begin{figure}[ht]
    \centering\!\!\!
    \begin{subfigure}{0.46\linewidth}
    \centering
    \includegraphics[width=0.99\linewidth]{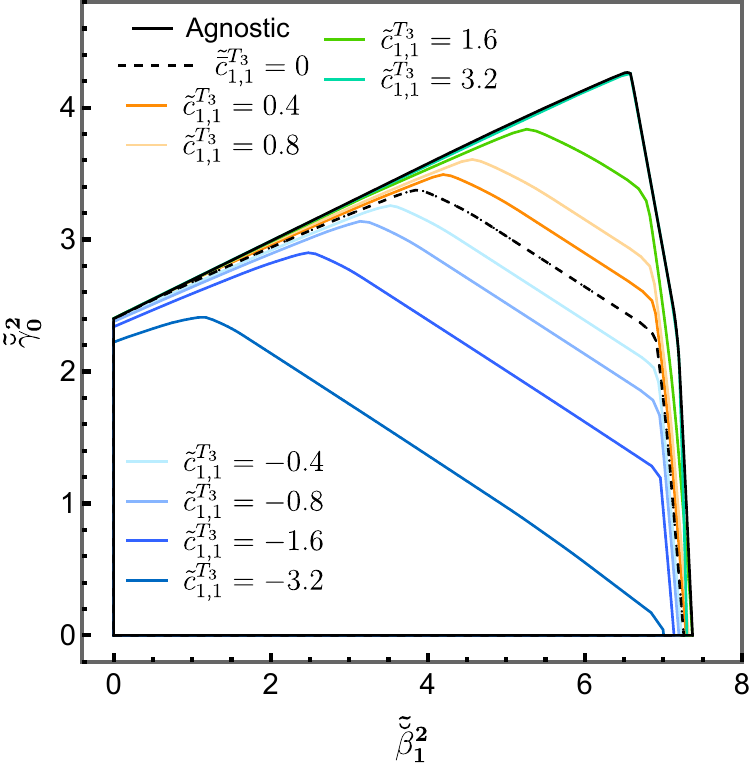}
    \end{subfigure}
    \begin{subfigure}{0.46\linewidth}
    \centering
    \includegraphics[width=0.99\linewidth]{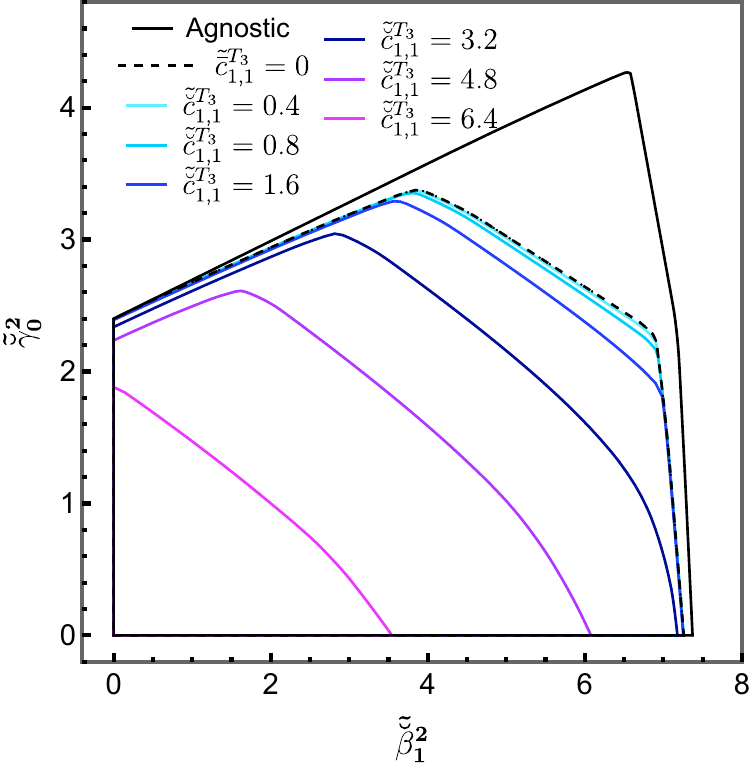}
    \end{subfigure}

\caption{Causality constraints on $\breve{\beta}_1^2\textbf{-}\breve{\gamma}_0^2$  for various non-vanishing dimension-10 coefficient $\tilde{c}^{T_3}_{1,1}$, where $\tilde{\bar{c}}^{T_3}_{1,1}\equiv  M_P^2\Lambda^8 \bar{c}^{T_3}_{1,1}/(\log(\Lambda/m_{\mathrm{IR}}))$. ({\it left}) Bounds for various $\tilde{c}^{T_3}_{1,1}$ with $\tilde{\breve{c}}^{T_3}_{1,1}=0$. ({\it right}) Bounds for various $\tilde{\breve{c}}^{T_3}_{1,1}$ with $\tilde{{c}}^{T_3}_{1,1}=0$.
}
\label{cxy}
\end{figure}

The effects of the even higher order coefficients on ${\breve{\beta}}_1^2$ and $\breve{\gamma}_0^2$ are similar. Let us look at, for example, the $\bar{g}_{1,1}^{T_3}stu(s^2+t^2+u^2)$ term contained in $\mathcal{M}^{++++}$. If we specify the four point interactions for this term, which are schematically $\nabla^2R^4$~(parity even) and $\nabla^2(R^2(R\tilde{R}))$~(parity odd), then this coefficient becomes
\be
\bar{g}_{1,1}^{T_3}=\bar{\gamma}_0^2/M_P^6+\bar{c}_{1,1}^{T_3}
\ee
where ${c}_{1,1}^{T_3}$ comes from the contributions of the four point contact terms. In Figure \ref{cxy}, we plot the causality bounds on $\breve{\beta}_1^2$, $\breve{\gamma}_0^2$ for some fixed $c_{1,1}^{T_3},\,\breve{c}_{1,1}^{T_3}$. We see that fixing $\bar{c}_{1,1}^{T_3}=0$ leads to the disappearance of the kink on the top right corner, which is only present for a relatively large $\bar{c}_{1,1}^{T_3}$. A positive $c_{1,1}^{T_3}$ enlarges the region of the causality bounds, while a negative $c_{1,1}^{T_3}$ further shrinks the region. As for the parity-violating counterpart $\breve{c}_{1,1}^{T_3}$, the region of the causality bounds always shrinks as $|\breve{c}_{1,1}^{T_3}|$ increases.

\subsection{Causality bounds on general gravitational EFT couplings}

Besides the linear sGB/dCS couplings $\beta_1\phi\mathcal{G}$ and  $\breve{\beta}_1\phi\mathcal{\tilde{R}}^{(2)}$, more general couplings of the form $f(\phi)\mathcal{G}$ and $\breve{f}(\phi)\mathcal{\tilde{R}}^{(2)}$ are of interest, particularly in the context of spontaneous scalarization \cite{Doneva:2017bvd, Silva:2017uqg}.
These general scalar coupling functions enable the black holes in these theories to develop scalar hair, depending on the curvatures of the systems. The essential ingredient for the spontaneous scalarization mechanism to work for the black hole case is the presence of the quadratic couplings $\phi^2\mathcal{G}$ or $\phi^2\mathcal{\tilde{R}}^{(2)}$, which can lead to an effective, tachyonic scalar mass in certain backgrounds. When hairy black holes are being formed, the tachyonic instability are quenched through the nonlinearity from higher order terms such that a stable scalar profile can be achieved. The causality bounds on the sGB coupling function $f(\phi)$ have been previously investigated\cite{Hong:2023zgm}, in this subsection we will focus on the case where parity-violating counterparts are also included.

We shall consider the following Lagrangian, up to quadratic order of $\phi$,
\begin{equation}\label{quadphi}
  \mathcal{L}\supset\sqrt{-g}\Big(\frac{\beta_1}{2!}\phi\mathcal{G}+\frac{\beta_2}{4}\phi^2\mathcal{G}+
  \frac{\breve{\beta}_1}{2!}\phi\tilde{\mathcal{R}}^{(2)}+\frac{\breve{\beta_2}}{4}\phi^2\tilde{\mathcal{R}}^{(2)} +\frac{\gamma_0}{3!}\mathcal{R}^{(3)}+\frac{\breve{\gamma}_0}{3!}\tilde{\mathcal{R}}^{(3)}+\frac{\alpha}{4}(\nabla\phi)^4
  \Big).
\end{equation}
In this subsection, we will use all available sum rules, including those involving the scalar. This is necessary because the relevant coefficients, particularly the higher-order ones, are often included in the mixed-helicity sum rules and cannot be bounded using only the pure graviton sum rules.

\begin{figure}[ht]
    \centering\!\!\!
    \begin{subfigure}{0.43\linewidth}
    \centering
    \includegraphics[width=0.9\linewidth]{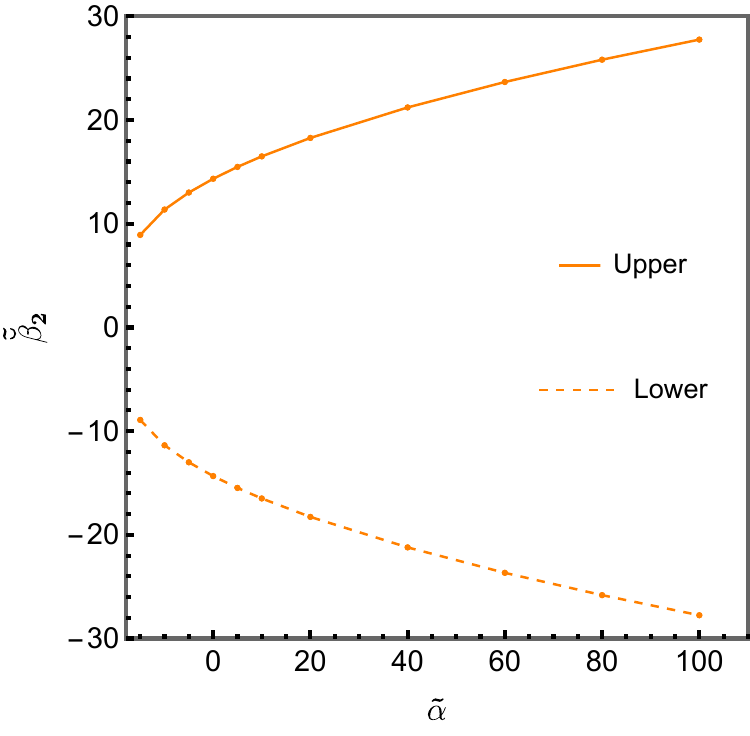}
    \end{subfigure}
    ~~~
    \begin{subfigure}{0.43\linewidth}
    \centering
    \includegraphics[width=0.9\linewidth]{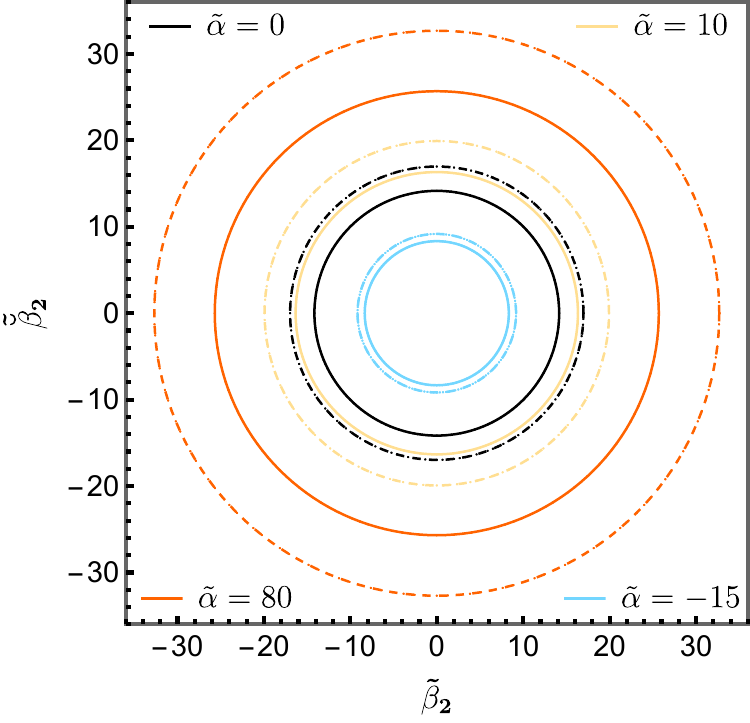}
    \end{subfigure}\\
    \begin{subfigure}{0.43\linewidth}
    \centering
    \includegraphics[width=0.9\linewidth]{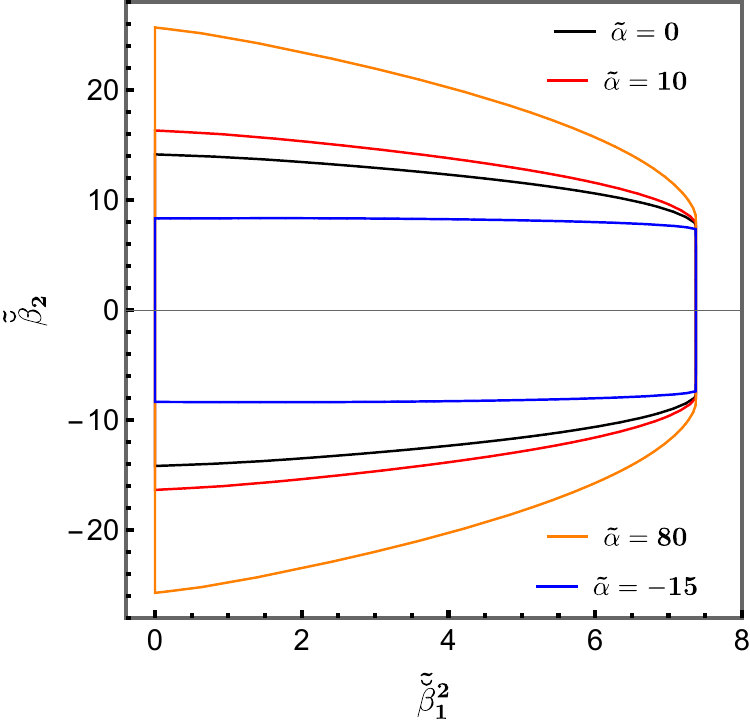}
    \end{subfigure}
    ~~~
    \begin{subfigure}{0.43\linewidth}
    \centering
    \includegraphics[width=0.9\linewidth]{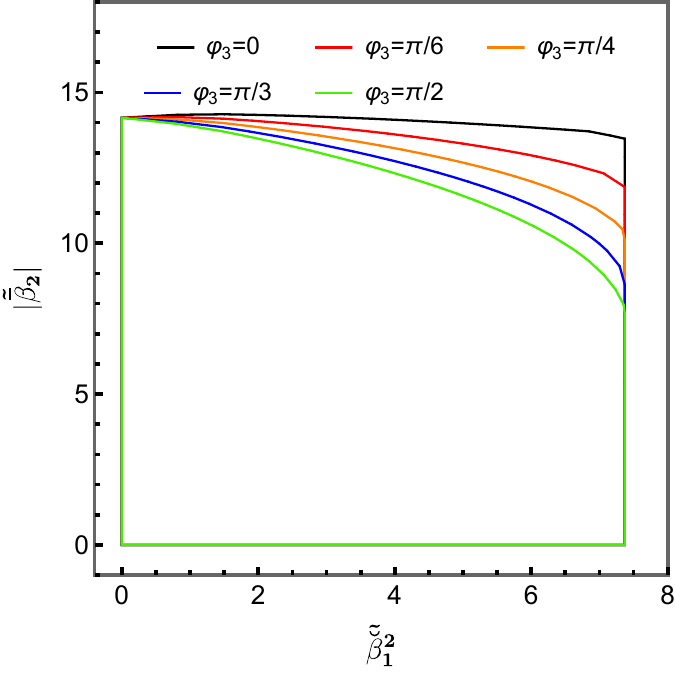}
    \end{subfigure}
\caption{ Causality bounds on the coefficient $\tilde{\breve{\beta}}_2$. We have defined $\tilde{\alpha}\equiv M_P^2\Lambda^2 \alpha/(\log(\Lambda/m_{\mathrm{IR}}))$,  $\bar{\beta}_2\equiv|\bar{\beta}_2|e^{i\varphi_3}=\beta_2+i\breve{\beta}_2$, $\tilde{\bar{\beta}}_{2}\equiv \bar{\beta}_{2}\Lambda^4M_P/(\log(\Lambda/m_{\mathrm{IR}}))$.
({\it top left}) Bounds on the $\tilde{\bar{\beta}}_2$ for various $\tilde{\alpha}$ when $\bar{\beta}_1=\bar{\gamma}_0=0$. ({\it top right}) Causality bounds on the $\tilde{{\beta}}_2$-$\tilde{\breve{\beta}}_2$ plane, the solid lines for $\bar{\beta}_1=0$ and the dashed lines for being agnostic about $\bar{\beta}_1$.
({\it bottom left}) Causality bounds on the parity-violating couplings $\tilde{\breve{\beta}}_1^2$ and $\tilde{\breve{\beta}}_2$ for various $\tilde{\alpha}$. ({\it bottom right}) Causality constraints on the coefficients $\tilde{\breve{\beta}}_1^2$ and $|\tilde{\bar{\beta}}_2|$ for various angle $\varphi_3\equiv \arg \bar{\beta}_2$ when  $\alpha=0$.
}
\label{tb1tb2}
\end{figure}

In Figure \ref{tb1tb2}, we show the causality bounds on the parity-violating couplings $\breve{\beta}_1^2$, $\breve{\beta}_2$. The $(\nabla\phi)^4$ coefficient has to be specified, otherwise $\breve{\beta}_2$ can not be bounded.
This can be easily seen by following an argument similar to that around Eq.~\eqref{eq:helidecomp}. To that end, we can construct an inequality involving the sum rules with $\langle F_{1,l}^{0000}\rangle$, $\langle F_{2,l}^{0000}\rangle$, $\langle F_{2,l}^{+0-0}\rangle$, $\langle F_{3,l}^{+0-0}\rangle$, $\langle F_{2,l}^{++--}\rangle$ and $\langle F_{2,l}^{++00}\rangle$. Dropping the higher order contributions and setting $\bar{\beta}_1=0$, we get the semi-definite condition~(with fixed $\mu$ and $l$):
\begin{equation}\label{b2sdmat}
  \begin{pmatrix}
    A&-E\beta_2&-E\breve{\beta}_2&0&0\\
    -E\beta_2&B&0&0&0\\
    -E\breve{\beta}_2&0&B&0&0\\
    0&0&0&C-D\beta_2&-D\breve{\beta}_2\\
    0&0&0&-D\breve{\beta}_2&C+D\beta_2
  \end{pmatrix}\succeq0
\end{equation}
where the ordering of the matrix components are with respect to  $(c^{00}_{l,\mu},\mathrm{Re}c^{++}_{l,\mu}, \mathrm{Im}c^{++}_{l,\mu},\mathrm{Re}c^{+0}_{l,\mu},\mathrm{Im}c^{+0}_{l,\mu})$ and $A,B,C,D,E$ are a new set of quantities.
The relevant eigenvalues of the matrix in Eq.~\eqref{b2sdmat} are given by
{\small
\begin{equation}\label{fb2eig}
  \(\frac{A+B+\sqrt{((A-B)^2+4E^2|\bar{\beta}_2|^2)}}{2},B,\frac{A+B-\sqrt{((A-B)^2+4E^2|\bar{\beta}_2|^2)}}{2},C+|D\bar{\beta}_2|,C-|D\bar{\beta}_2|\)
\end{equation}
}
Note that the $(\nabla\phi)^4$ coupling $\alpha$ is contained in $A$. So if we are agnostic about $\alpha$,
$A$ needs to be suppressed to zero in the optimization.
Then, in the presence of a nonzero $\bar{\beta}_2$ and $A$ is negligible, the eigenvalue becomes $B-\sqrt{B+4E^2|\bar{\beta}_2|^2}$, which is negative, so the semi-definiteness of the matrix cannot be achieved. Therefore, to get the bounds on $|\bar{\beta}_2|$, we need to specify the value of $\alpha$.

The causality bounds on $\bar{\beta}_2\equiv|\bar{\beta}_2|e^{i\varphi_3}=\beta_2+i\breve{\beta}_2$ are shown in Figure \ref{tb1tb2}.
In the top left pannel of Figure \ref{tb1tb2}, we see that the bounds on $|\breve{\beta}_2|$ are enlarged as $\alpha$ increases. In the top right panel of Figure \ref{tb1tb2}, it can be seen that the bounds on $\bar{\beta}_2$ are almost invariant with respect to $\arg\bar{\beta}_2$ when $\bar{\beta}_1=0$, that is, the bounds can be formulated as constraints on $|\bar{\beta}_2|$.
When agnostic about $\bar{\beta}_1$, the bounds on $\bar{\beta}_1$ enlarge but are still insensitive to $\arg\bar{\beta}_2$. However, as we will see in the bottom right panel of  Figure \ref{tb1tb2}, in the presence of fixed values of $\breve{\beta}_1^2$, the bounds can no longer be cast as constraints on $|\bar{\beta}_2|$.
In the bottom left panel of Figure \ref{tb1tb2}, we plot the bounds on the $\breve{\beta}_2$-$\breve{\beta}_1^2$ plane for a few $\alpha$. As the lower order parity-violating coupling $\breve{\beta}_1^2$ increases, the  bounds on $\breve{\beta}_2$ shrink. Generally, as $\alpha$ increases, the bounds on $\breve{\beta}_2$ enlarge. But when $\breve{\beta}_1^2$ reaches its upper bounds, the bounds on $\breve{\beta}_2$ are fixed for all $\alpha$, indicating that $\breve{\beta}_1^2$ is insensitive to the constraints from pure helicity-0 sum rules. As $\alpha$ reaches its lower bound, the bounds on $\breve{\beta}_2$, $\breve{\beta}_1^2$ become nearly independent of each other. Similar behaviors have also been observed in the parity-conserving counterparts.
We also plot how $\varphi_3=\arg\bar{\beta}_2$ affects the bounds on the $|\bar{\beta}_2|$-$\breve{\beta}_1^2$ plane. Visible differences occur at relatively large $\breve{\beta}_1^2$, where the bounds on $|\bar{\beta}_2|$ become smaller as $\varphi_3$ varying from $0$ to $\pi/2$.

\begin{figure}[ht]
    \centering\!\!\!
    \begin{subfigure}{0.4\linewidth}
    \centering
    \includegraphics[width=0.99\linewidth]{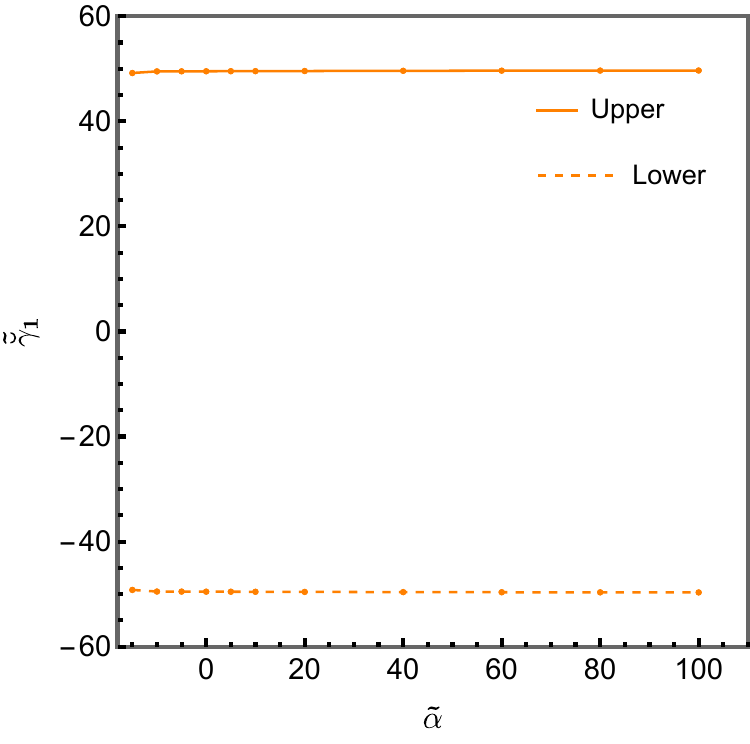}
    \end{subfigure}
    \centering
    \begin{subfigure}{0.4\linewidth}
    \centering
    \includegraphics[width=0.99\linewidth]{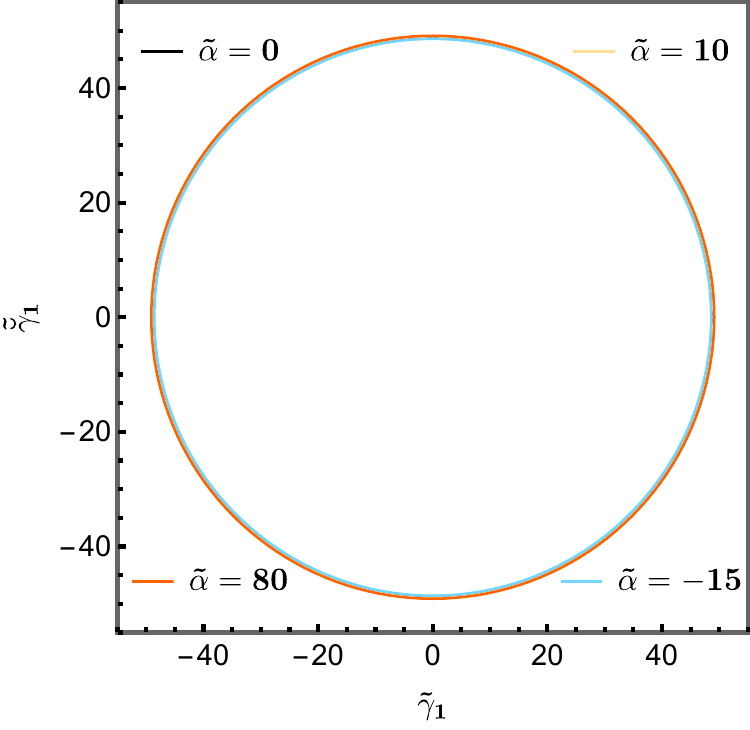}
    \end{subfigure}\\

    \centering
    \begin{subfigure}{0.4\linewidth}
    \centering
    \includegraphics[width=0.99\linewidth]{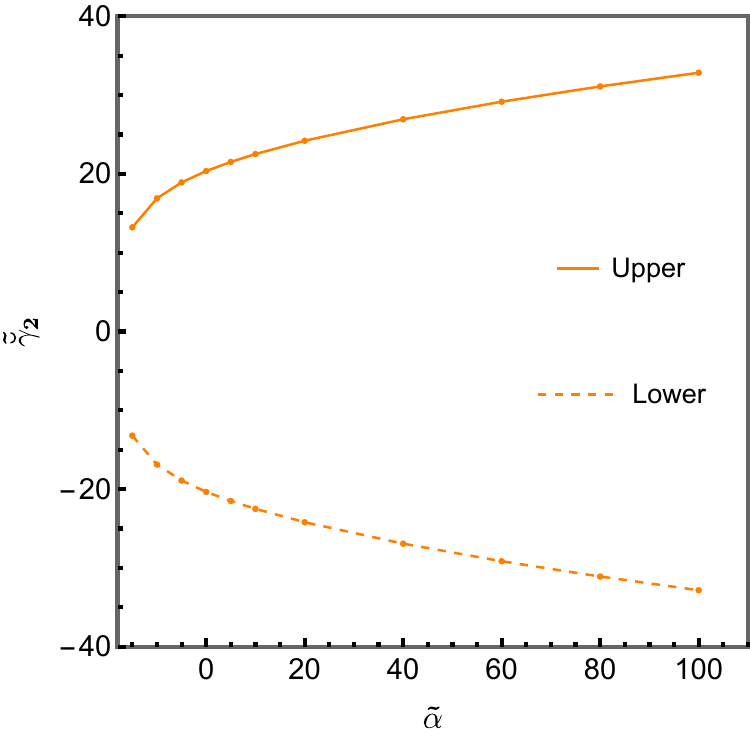}
    \end{subfigure}
    \centering
    \begin{subfigure}{0.4\linewidth}
    \centering
    \includegraphics[width=0.99\linewidth]{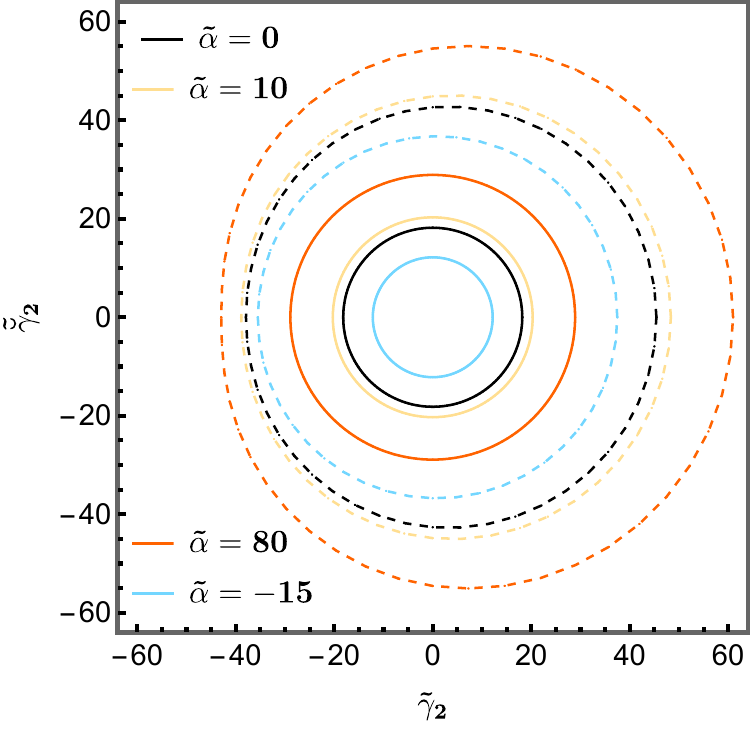}
    \end{subfigure}\\

    \centering
    \begin{subfigure}{0.4\linewidth}
    \centering
    \includegraphics[width=0.99\linewidth]{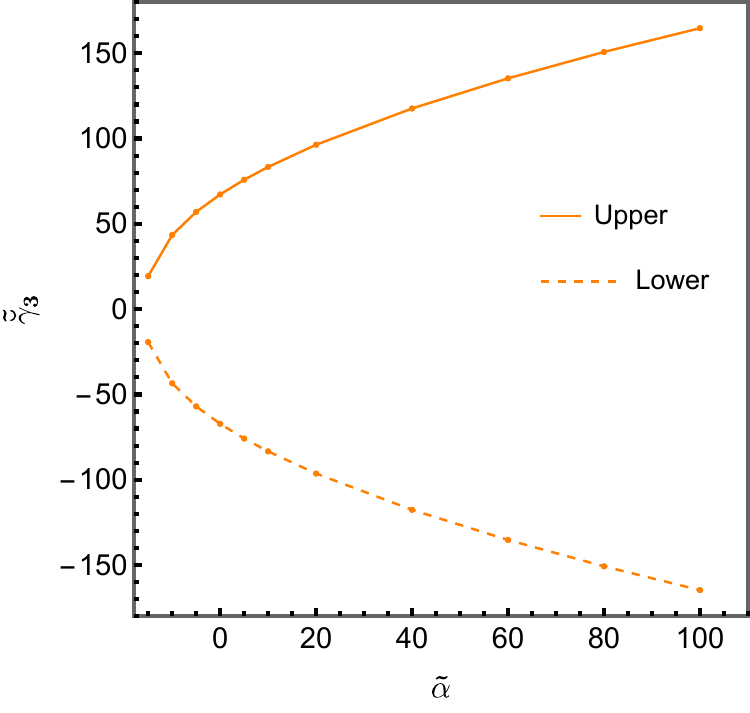}
    \end{subfigure}
    \centering
    \begin{subfigure}{0.4\linewidth}
    \centering
    \includegraphics[width=0.99\linewidth]{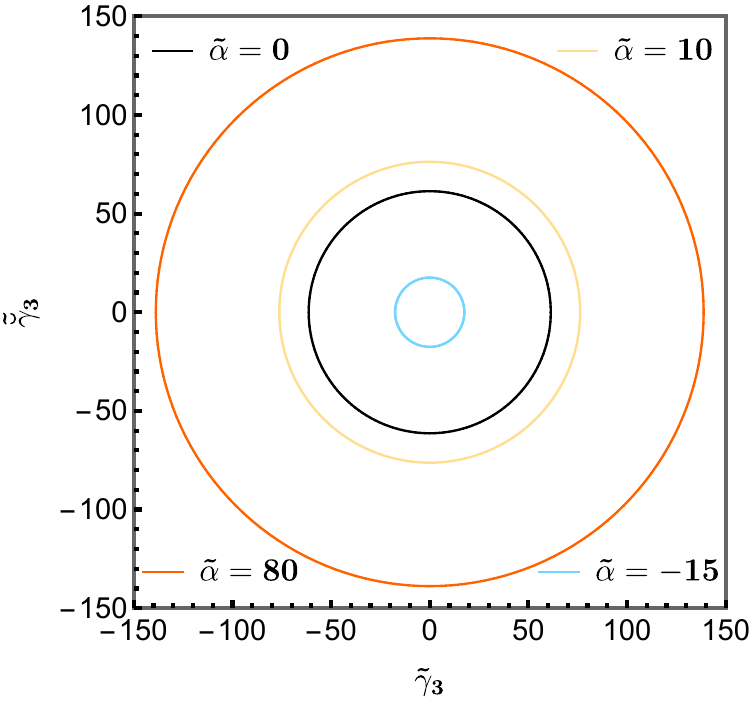}
    \end{subfigure}
\caption{ The left three panels are the causality bounds on parity-violating couplings $\tilde{\breve{\gamma}}_{i}\equiv \breve{\gamma}_{i}\Lambda^4M_P^{i-2}/(\log(\Lambda/m_{\mathrm{IR}}))$, $i=1,2,3$ for various $\tilde{\alpha}$ when $\bar{\beta}_1=\bar{\gamma}_0=0$. The right three panels correspond to causality bounds on the $\tilde{\gamma}_i$-$\tilde{\breve{\gamma}}_i$ for various $\alpha$. Note that the dashed lines on the $\tilde{\gamma}_2$-$\tilde{\breve{\gamma}}_2$ plane correspond to being agnostic about $\bar{\beta}_1$ and $\bar{\gamma}_0$.
When agnostic about $\bar{\beta}_1$, the bounds on the other two coefficients $\bar{\gamma}_1$ and $\bar{\gamma}_3$ do not alter significantly.  }
\label{tgn}
\end{figure}

Next, we also compute the bounds on the following higher order operators, which contain more derivatives on the scalar,
\begin{equation}\label{Lfgn}
\begin{aligned}
\mathcal{L}\supset\sqrt{-g}\Big(&\frac{\gamma_1}{3!}\phi\mathcal{R}^{(3)} +\frac{\gamma_2}{2}\nabla_\mu\phi\nabla^\mu\phi\mathcal{R}^{(2)}
  +\frac{4\gamma_3}{3}\phi\nabla_\mu\nabla_\rho\phi\nabla_\nu\phi\nabla_\sigma\phi R^{\mu\nu\rho\sigma} \\ &+\frac{\breve{\gamma}_1}{3!}\phi\tilde{\mathcal{R}}^{(3)} +\frac{\breve{\gamma}_2}{2}\nabla_\mu\phi\nabla^\mu\phi\tilde{\mathcal{R}}^{(2)}
   +\frac{4\breve{\gamma}_3}{3}\phi\nabla_\mu\nabla_\rho\phi\nabla_\nu\phi\nabla_\sigma\phi \tilde{R}^{\mu\nu\rho\sigma}\Big)
\end{aligned}
\end{equation}
In Figure \ref{tgn}, we show the causality constraints on these EFT couplings for various $\alpha$. Note that the coefficient $|\bar{\gamma}_1|$ is insensitive to the dimension-8 scalar coupling $\alpha$, which is due to the fact $\bar{\gamma}_1$ is contained in the following sum rules
\begin{equation}\label{tg1sr}
   -\frac{\bar{\gamma}_1t^2}{M_P^3}=\left\langle F_{1,l}^{+++0}(\mu,t) \right\rangle,\quad
   -\frac{\bar{\gamma}_1t}{M_P^3}=\left\langle F_{2,l}^{+++0}(\mu,t) \right\rangle,
\end{equation}
To understand this intuitively, one can follow an argument similar to that around Eq.~\eqref{eq:helidecomp}.
For fixed $\mu$ and $l$, one can schematically get a semi-positive condition of the form
\begin{equation}\label{tb1mat}
\begin{pmatrix}
  A & -C\gamma_1 & 0 & -C\breve{\gamma}_1 \\
 -C\gamma_1 & B & -C\breve{\gamma}_1 & 0 \\
  0 & -C\breve{\gamma}_1 & A & C\gamma_1 \\
  -C\breve{\gamma}_1 & 0 & C\gamma_1 & B
\end{pmatrix}\succeq0
\end{equation}
where the entries are ordered as $\(\mathrm{Re}c_{l,\mu}^{+0},\mathrm{Re}c_{l,\mu}^{++},\mathrm{Im}c_{l,\mu}^{+0},\mathrm{Im}c_{l,\mu}^{++},\)$.
{\small
\begin{equation}\label{tb1eig}
\begin{aligned}
  \Big(&\frac{A+B+\sqrt{(A-B)^2+4C^2|\gamma_1|^2}}{2},\frac{A+B+\sqrt{(A-B)^2+4C^2|\gamma_1|^2}}{2}, \\
  &\frac{A+B-\sqrt{(A-B)^2+4C^2|\gamma_1|^2}}{2},\frac{A+B-\sqrt{(A-B)^2+4C^2|\gamma_1|^2}}{2},\Big).
\end{aligned}
\end{equation}
}
Setting the lowest eigenvalue to zero, one gets $|\bar{\gamma}_1|^2\sim AB/C^2$, which is insensitive $\alpha$ because $A$, $B$ and $C$ do not contain $\alpha$.
This also implies that the bounds on $\bar{\gamma}_1$ are insenstive to $\arg \bar{\gamma}_1$, as shown in the right panel of Figure \ref{tgn}.

\begin{figure}[ht]
    \centering\!\!\!
    \begin{subfigure}{0.46\linewidth}
    \centering
    \includegraphics[width=0.99\linewidth]{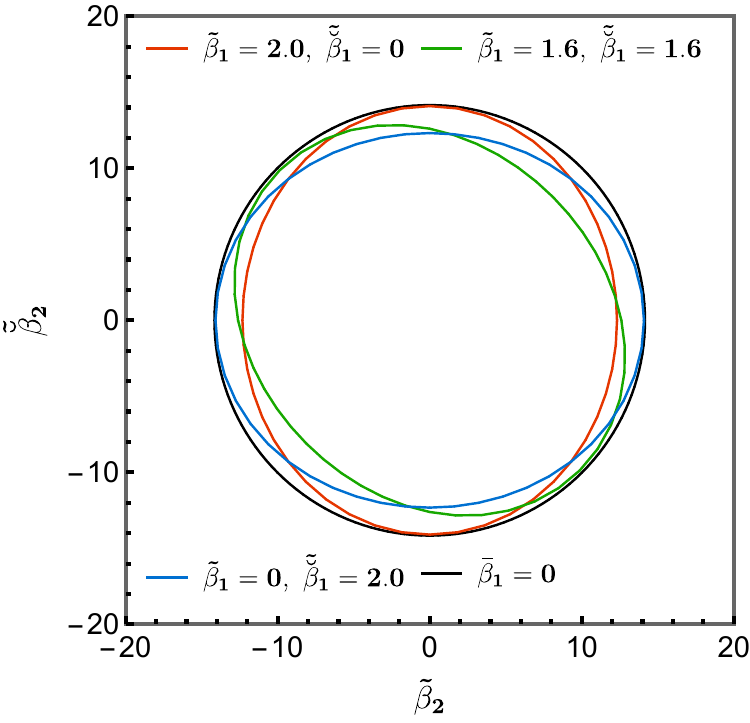}
    \end{subfigure}
    \begin{subfigure}{0.46\linewidth}
    \centering
    \includegraphics[width=0.99\linewidth]{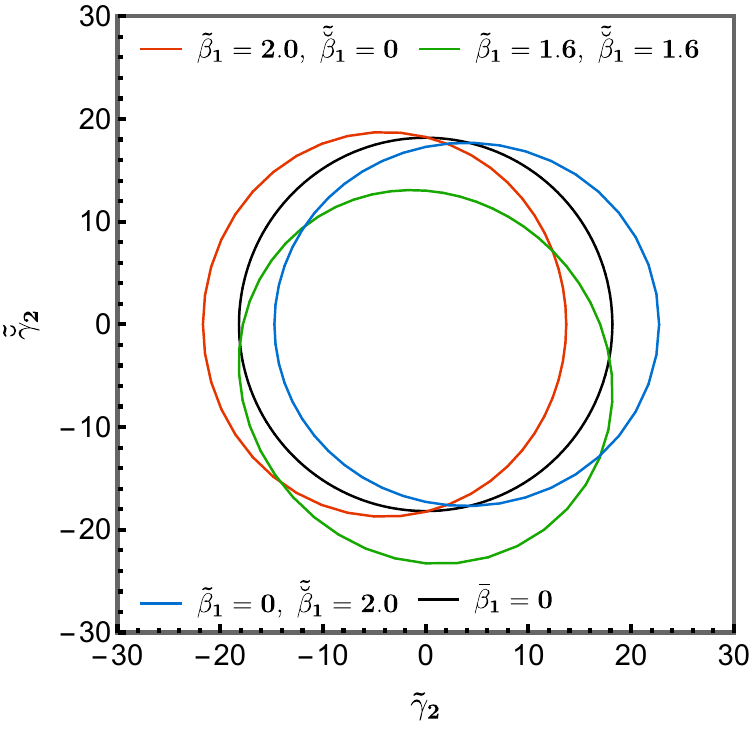}
    \end{subfigure}

\caption{Causality constraints on the ${\beta}_2\textbf{-}\breve{\beta}_2$ ({\it left}) and ${\gamma}_2\textbf{-}\breve{\gamma}_2$ ({\it right}) planes for some fixed values of $\tilde{\bar{\beta}}_1$, where we have set $\alpha=0$ and $\bar{\gamma}_0=0$. The bounds on $\bar{\beta}_2$ ($\bar{\gamma}_2$) are no longer in terms of $|\bar{\beta}_2|$ ($|\bar{\gamma}_2|$) for a specific choice of $\bar{\beta}_1$.
}
\label{fixb1}
\end{figure}

As for the remaining two coefficients, the associated sum rules are given by
\begin{align}\label{tg2tg3sr}
\left\langle F_{3,l}^{++00}(\mu,t) \right\rangle &=\frac{M_P^2\bar{\gamma}_2+\bar{\beta}_1^2}{M_P^4}-\frac{\bar{\gamma}_0}{M_P^4}t-\bar{g}^{M_3}_{2,1}t-\bar{g}^{M_3}_{3,1}t^2\\
 \left\langle F_{1,l}^{+000}(\mu,t) \right\rangle&= \frac{\bar{\beta}_1t}{M_P^3}+\frac{\bar{\gamma}_3t^2}{M_P},\quad \quad \left\langle F_{2,l}^{+000}(\mu,t) \right\rangle= \frac{\bar{\beta}_1}{M_P^3}+\frac{\bar{\gamma}_3t}{M_P},
\end{align}
The coupling $\bar{\gamma}_2$ is contained in the sum rule with $F_{3,l}^{++00}$, which has a similar partial wave structure as $F_{2,l}^{++00}$ that contains $\bar{\beta}_2$. Consequently, the bounds on $|\bar{\gamma}_2|$ are similar to that of $|\bar{\beta}_2|$, as seen in Figure \ref{tgn}. The causality bounds on $|\bar{\gamma}_2|$ are sensitive to $\arg \bar{\gamma}_2$ when $\bar{\beta}_1$ play roles, and are enlarged as the $(\nabla\phi)^4$ coupling $\alpha$ increases. The main feature of the $|\bar{\gamma}_2|$ bound is that it is susceptible to the $\bar{\beta}_1^2$ coupling, as see in Figure \ref{tgn}.
This originates from the fact that $\bar{\beta}_1^2$ co-exists with $\bar{\gamma}_2$ in the sum rule $F_{3,l}^{++00}(\mu,t)$. To see this more clearly, in Figure \ref{fixb1}, we plot the causality bounds on $\bar{\beta}_2$ and $\bar{\gamma}_2$ with some specific choices of $\bar{\beta}_1$, and we see that the presence of fixed nonzero $\bar{\beta}_1$ breaks the phase independence, as the eigenvalues in the related matrices can no longer be cast in terms of absolute values of the coefficients, similar to the previous discussions.

As for the $\bar{\gamma}_3$ coupling,
the bounds on $|\bar{\gamma}_3|$ are also insensitive to the angle $\arg \bar{\gamma}_3$, but are positively correlated with $\alpha$, as shown in the bottom right panel of Figure \ref{tgn}. This again can be understood along the lines of the argument around Eq.~\eqref{eq:helidecomp}.

\subsection{Implications for observational constraints}

While a full-scale phenomenological exploration of these causality bounds is beyond the scope of this paper, we shall briefly confront the bounds with existing astronomical observations. Particularly, we will be concerned with the scalar-Gauss-Bonnet coefficient $\beta_1$ and dynamical-Chern-Simons coefficient $\breve{\beta}_1$, which have generated a lot of interests recently in relativisitc astrophysics.
Our goal here is very modest. In the absence of more constraining data, we shall utilize the previously obtained sharp causality bounds to convert the experimental bounds into the lower bounds on the cutoff of some {\it specific} EFTs.

\begin{table}[]
\centering
\renewcommand\arraystretch{1.5}
\scalebox{1}{
\begin{tabular}{|c||c|c|}
\hline
$\Lambda$($10^{-11}\mathrm{eV}$)& EM+GW\cite{Silva:2020acr} & Ringdown\cite{Silva:2022srr} \\ \hline\hline
\text{Conservative}& 6.5 &1.4 \\ \hline
\text{Kink}& 6.3 &1.4\\ \hline
\text{Fixed}&4.4& 0.96\\ \hline
\end{tabular}
}
\caption{Lower bounds on the EFT cutoff $\Lambda$ (in units of $10^{-11}$eV) from the observational constraints for some specific theories. "Conservative", "Kink" and "Fixed" refer to theories located at three different places within the causality bounds on $\tilde{{\beta}_1}^2$ and $\tilde{\breve{\beta}}_1^2$: the global upper bound, the top right kink and the case with fixed $\tilde{\alpha}_4=0.2$ and $\tilde{\alpha}^\prime_4=0.1$. ``EM+GW'' denotes the observational bound obtained by combing the pulsar X-ray data from PSR J0030+0451 with the GW events from the binary neutron star GW170817. ``Ringdown'' indicates the gravitational wave constraints from the ringdown stage using the events GW150914 and GW200129.}\label{tabexp}
\end{table}

To this end, we first express the coefficients $\beta_1$ and $\breve{\beta}_1$ in terms of the corresponding dimensionless ones and require their values to lie within the observational constraints
\begin{equation}\label{fb1obs}
  \frac{\tilde{\beta}_1M_P\sqrt{\log(\Lambda/m_{\mathrm{IR}})}}{\Lambda^2}=\beta_1\leq \beta_1^{\text{obs}}, \quad \frac{\tilde{\breve{\beta}}_1M_P\sqrt{\log(\Lambda/m_{\mathrm{IR}})}}{\Lambda^2}=\breve{\beta}_1\leq\breve{\beta}_1^{\text{obs}},
\end{equation}
where $m_{\rm IR}$ is conservatively chosen to be the current Hubble scale.
In Table \ref{tabexp} we list the observational constraints on the EFT cutoff $\Lambda$ for three representative theories that satisfying the causality bounds. The observational constraints are extracted from the pulsar X-ray data of PSR J0030+0451, the GW events from the binary neutron star GW170817 \cite{Silva:2020acr} and the ringdown stage of GW150914 and GW200129 \cite{Silva:2022srr}.

\begin{figure}[ht]
    \centering\!\!\!
    \begin{subfigure}{0.42\linewidth}
    \centering
    \includegraphics[width=0.99\linewidth]{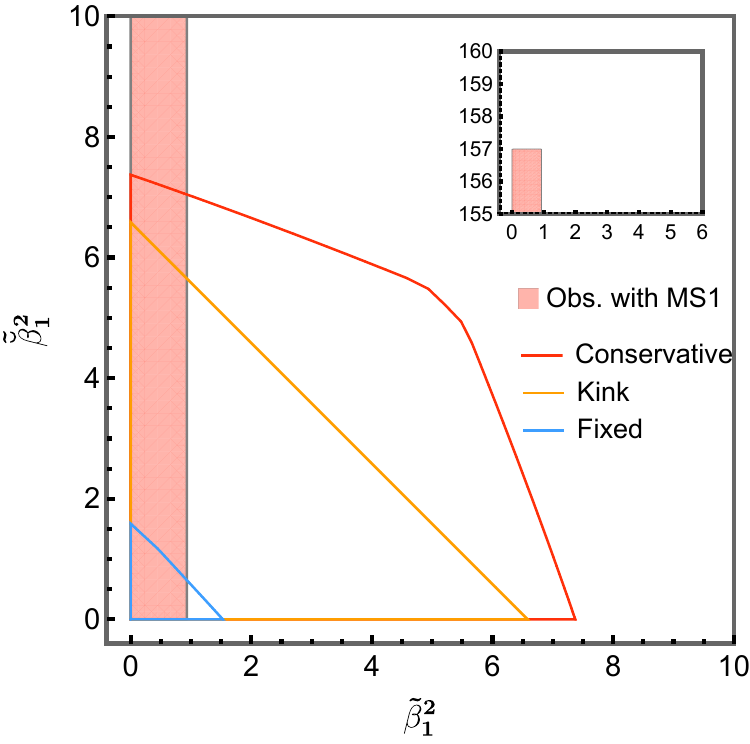}
    \end{subfigure}
    \centering~~~~~~~~
    \begin{subfigure}{0.42\linewidth}
    \centering
    \includegraphics[width=0.99\linewidth]{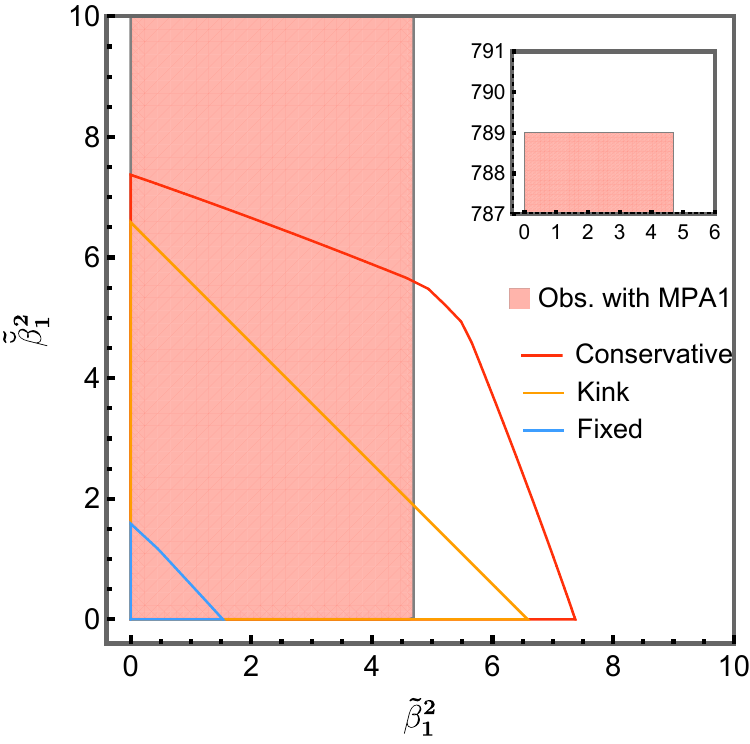}
    \end{subfigure}
\caption{Comparison of the causality bounds and observational constraints on the $\tilde{\beta}_1^2$-$\tilde{\breve{\beta}}_1^2$ plane, when the observational constraints on the quadratic sGB coupling $\beta_2$ saturate the causality bound.
"Obs. with MS1/MPA1" denotes the case where the MS1/MPA1 equations of state are utilized to obtain the observational constraints on the EFT cutoff \cite{Danchev:2021tew}. The causality bounds are those with $\alpha=0$ and $\bar{\gamma}_0=0$. The observation constraints for $\beta_1/\breve{\beta}_1$ come from the most stringent GW constraints in Ref.~\cite{Perkins:2021mhb, Silva:2020acr} respectively.
The boundaries of the triangles marked by "Conservative", "Kink" and "Fixed" correspond to the three models mentioned in Table \ref{tabexp} with the crossing term $\beta_1\breve{\beta}_1$ masked. }
\label{obscmp}
\end{figure}

The $\beta_2$ and $\breve{\beta}_2$ terms can induce spontaneous scalarization in black holes. The quadratic sGB coupling $\beta_2$ has been constrained using the observational data from binary pulsars. Ref.~\cite{Danchev:2021tew} has set several bounds on $|\beta_2|$ using three neutron star-white dwarf binaries with both $\beta_2<0$ and $\beta_2>0$ cases considered. For hairy black hole solutions, $\beta_2$ has to be strictly positive such that a tachyonic instability can be induced for the spontaneous scalarization to happen. If we require the observational constraints on $\beta_2$ to saturate the causality bounds, which fixes the EFT cutoff $\Lambda$, we can put observational constraints on the coefficients $\tilde{\beta}_1$ and $\tilde{\breve{\beta}}_1$, as shown in Figure \ref{obscmp}. We see that the current observational constraints provide relatively looser restrictions on the dCS coupling $\breve{\beta}_1^2$, in comparison with its parity-conserving counterpart $\beta_1^2$. Because of this, if the constraining power of the causality bounds is comparable to that of the observation constraints for the parity-conserving coefficients, the causality bounds are much more effective in eliminating the parameter space for the parity-violating coefficients.

\acknowledgments

We would like to thank Shi-Lin Wan for helpful discussions. SYZ acknowledges support from the National Key R\&D Program of China under grant No.~2022YFC2204603 and from the National Natural Science Foundation of China under grant No.~12075233 and No.~12247103. \\
~\\

\appendix

\noindent{\bf \Large Appendices}

\section{Numerical details}\label{numde}

In this appendix we shall briefly discuss how to formulate the semi-definite program for matrix $\mathbf{B}_{l,\mu}$.
We will use the SDPB package to perform the numerical SDPs. As this package can only deal with SDPs whose entries are polynomials with a single continuous decision variable, following \cite{Caron-Huot:2021rmr,Caron-Huot:2022ugt}, we divide the $\mu$-$l$ plane into five regions and apply a different approximation scheme in each region. For simplicity, we set $\Lambda=1$ in most of the following discussions, but will restore it in the final results.

\begin{figure}[ht]
    \centering\!\!\!
    \begin{subfigure}{0.46\linewidth}
    \centering
    \includegraphics[width=0.99\linewidth]{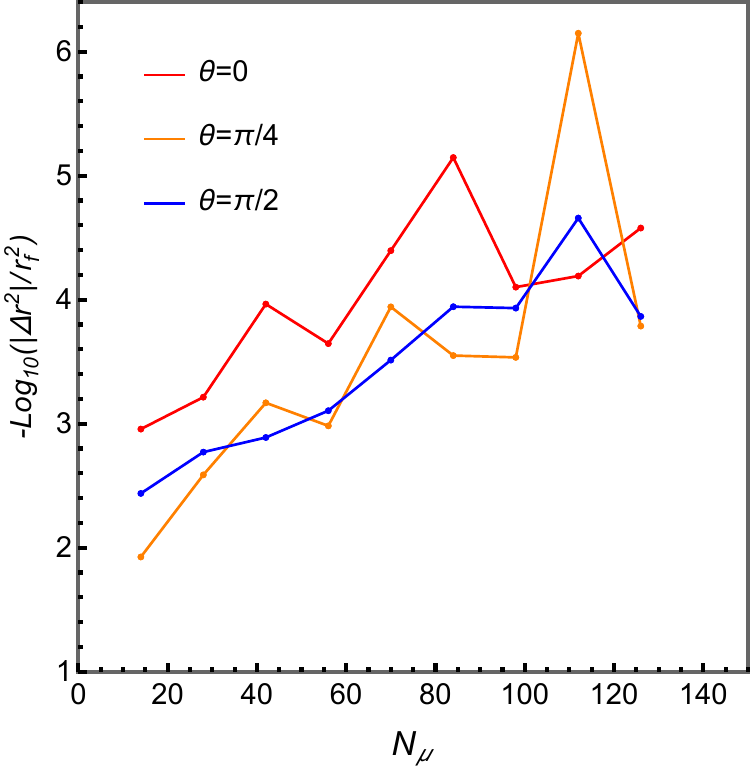}
    \end{subfigure}
    \begin{subfigure}{0.46\linewidth}
    \centering
    \includegraphics[width=0.99\linewidth]{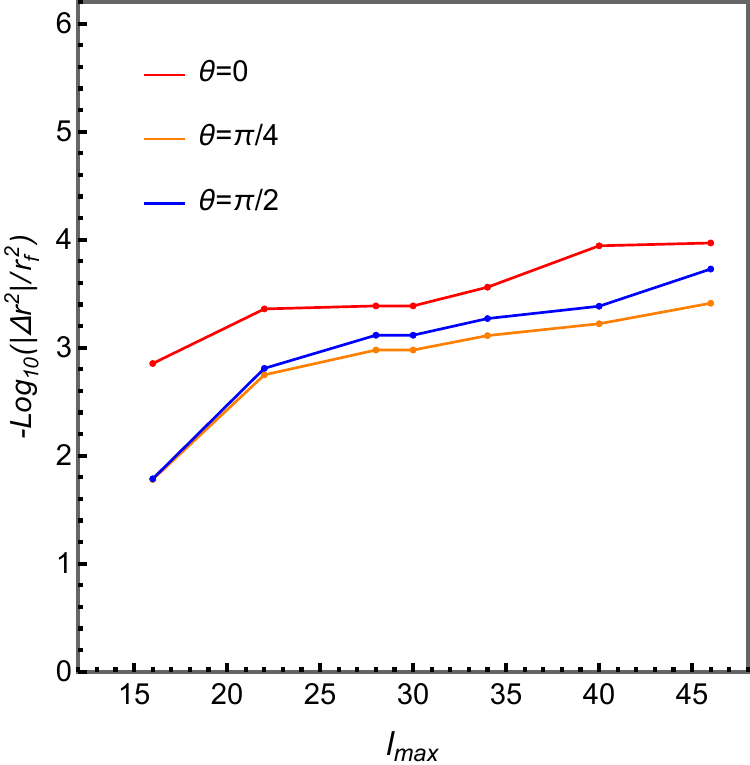}
    \end{subfigure}
\caption{Convergence tests for the causality bounds on $|\tilde{\bar{\beta}}_1|^2$ and $|\tilde{\bar{\gamma}}_0|^2$($\varphi_1=\pi/2,\varphi_2=\pi/2$) for various $N_\mu$ (number of discrete $\mu$), maximal spin $l_{\mathrm{max}}$({\it right}) and $\theta\equiv \arctan(|\tilde{\bar{\gamma}}_0|/|\tilde{\bar{\beta}}_1|)$. $\Delta r^2\equiv r^2-r_f^2$ and $r_f^2$ correspond to the causality bounds in the angular optimization (see Eq.~(\ref{angular})).
}
\label{conver}
\end{figure}

The convergence tests for our numerical setup are shown in Figure \ref{conver}.

\noindent\textbf{Finite $\mu$ and $l$}:

\noindent We simply take discrete values of $\sqrt{\mu}$ starting from $\Lambda$ to several times of the EFT cutoff. The partial wave number $l$ is a positive integer, taken up to a maximal value $l_{\mathrm{max}}$, which is chosen to be $30$. Also the weight functions $f_k^{\vmathbb{1234}}(p)$ are finite dimension polynomials with the base functions of the form $(1-p)^2p^i$, where $i$ ranges from $i_{\mathrm{min}}$ to a truncated order $i_{\mathrm{max}}$. The discrete values of $\mu$ are taken to be $\{\sqrt{\mu}=1/(1-2k/80)|0\leq k\leq 36,k\in \mathbb{Z}\}\cup\{\sqrt{\mu}=1/(1-(2k+1)/320)|144\leq k\leq 159,k\in \mathbb{Z}\}$.
\vspace{0.5cm}

\noindent\textbf{Large $\mu$ and finite $l$}:

\noindent In this case $\mu$ is relatively large as compared to the UV scale $\Lambda$, so $1/\mu$ is small and can be taken as a perturbative variable. We approximate the matrix $\mathbf{B}_{l,\mu}$ by taking a few leading terms in the $1/\mu$ expansion. Then we multiply truncated $\mathbf{B}_{l,\mu}$ with appropriate power of $\mu$ to make all the components of $\mathbf{B}_{l,\mu}$ polynomials of $\mu$, which can then be handled by SDPB.
We then construct $\mathbf{B}_{l,\mu}$ with $\mu$ entries for each partial wave spin number $l\leq l_{\mathrm{max}}$ in this region, with $l_{\mathrm{max}}=30$.
\vspace{0.5cm}

\noindent\textbf{Finite $\mu$ and large $l$}:

\noindent The asymptotic behavior of the Wigner d-functions in the large $l$ region has an oscillatory pattern in terms of $p^2/\mu$. Integrating over $p$, the oscillations tend to cancel out. Because of this, it is useful to include the supplementary forward limit sum rules in this region.
When $l$ is large, the spin number $l$ can be taken as a continuous parameter, to a good approximation. Note that spin number $l$ may appear in the form $\sqrt{(l+c_1)(l+c_2)\cdots(l+c_n)}$, which are not polynomials. In this case, we need to take a Laurent expansion at $l\rightarrow\infty$ and take several leading terms as an approximation, schematically of the form $l^{n/2}(1+(c_1+c_2+\cdots+c_n)l^{-1}/2+\mathcal{O}(l^{-2}))$. Further making the substitution $l\rightarrow(y+\sqrt{l_{\mathrm{max}}})^2$, we can convert it to a polynomial of $y$ with $y>0$. Then the entries of the matrices $\mathbf{B}_{l,\mu}$ become the polynomials of the continuous variable $y$ for each discretized $\mu$, and thus the semi-definite conditions for $\mathbf{B}_{l,\mu}$ are admissible by SDPB. The discrete set of $\mu$ is chosen to be $\{\sqrt{\mu}=1/400+1/\sqrt{1-k/100}|0\leq k\leq99,k\in\mathbb{Z}\}$.

\vspace{0.5cm}

\noindent\textbf{Large $\mu$ and $l$ with finite $b$}:

\noindent In order to effectively sample the $\mathbf{B}_{l,\mu}$ constraints at large $\mu$ and $l$, one can introduce the impact parameter $b\equiv 2\mu/\sqrt{l}$ to parameterize this region \cite{Caron-Huot:2021rmr}. For a finite $b$, the asymptotic behavior of the hypergeometric function contained in the Wigner d-matrices has the following form
\begin{equation}\label{asymplb}
  \lim _{\mu, \ell \rightarrow \infty,b<\infty}{ }_2 F_1\left(h_1-\ell, \ell+h_1+1 ; h_1-h_2+1 ; p^2 / \mu\right)=\frac{\Gamma\left(h_1-h_2\right)}{(b p / 2)^{h_1-h_2}} J_{h_1-h_2}(b p),
\end{equation}
where $J_h(x)$ is the Bessel function of the first kind. We take finite discrete values of $b=2\ell/\sqrt{\mu}$ in this case,
and keep the leading order contribution in the Taylor expansion of the matrices $\mathbf{B}_{l,\mu}$ in the limit $\mu\rightarrow\infty$, that is, the $\mathcal{O}(1/\mu^3)$ contribution.
The semi-definite constraints become
\begin{equation}\label{sdfb}
  \tilde{\mathbf{B}}_{\tilde{l}}(b)\succeq 0, \,\text{for all $b>0$, $\tilde{l}$=even/odd},
\end{equation}
where $\tilde{\mathbf{B}}_{\tilde{l}}(b)\equiv\mu^3\mathbf{B}_{\tilde{l},\mu}$ and $\tilde{l}$ denotes the matrices only depend on whether $l$ is odd or even. The discrete values of $b$ are taken to be $b=\epsilon_b+k\delta_b$, with $\epsilon_b$ being a small number and $k$ being integers, up to a maximal $b_{\mathrm{max}}$. In this paper we set $\epsilon_b=1/250$, $\delta_b=1/10$ and $b_{\mathrm{max}}=40$.
\vspace{0.5cm}

\noindent\textbf{Large $\mu$ and $l$ with large $b$}:

\noindent For a very large $b$, we can also make use of the asymptotic behavior of the hypergeometric function to simplify the expression of $\tilde{\mathbf{B}}_{\tilde{l}}(b)$ into the form
\begin{equation}\label{lbreg}
  \tilde{\mathbf{B}}_{\tilde{l}}(b)=F(b)+G(b)\cos(b)+H(b)\sin(b)
\end{equation}
where $F(b)$, $G(b)$ and $H(b)$ are matrices whose entries are the polynomials of $1/b$, truncated up to $\mathcal{O}(1/b^{M_b})$. The oscillatory behavior of $b$ with $\cos(b)$ or $\sin(b)$ can not dominate for a large $b$; otherwise the positive conditions in Eq.~(\ref{sdcond1}) cannot be satisfied. The oscillatory terms with $\cos(b)$ or $\sin(b)$ are suppressed if we choose the functions basis of $f_{k}^{\vmathbb{1234}}(p)$ appropriately.
Since the $\cos(b)$ and $\sin(b)$ terms now have minor effects in the optimization, we can further approximate $(\sin(b),\cos(b))$ with $(\pm\sqrt{2},0)$, which extracts the most important contribution from the oscillatory terms.  Upon multiplying $\tilde{\mathbf{B}}_{\tilde{l}}(b)$ with a common factor $b^{M_b}$, the constraints become directly admissible by SDPB. In this paper we have chosen $M_b=11$.

\section{Explicit sum rules}\label{secsr}

Here we explicitly list the dispersion relations/sum rules with the $s$-$t$ crossing symmetry implemented (see Eq.~\eqref{disf}). Apart from computing the final causality bounds, these dispersion relations can be used to perform dimension analysis on the Wilson coefficients (see Section \ref{powercounting}). Some of these dispersion relations arise from amplitudes with triple crossing symmetry, while the rest come from amplitudes with only one crossing symmetry. The sum rules originating from full $stu$ crossing symmetry are
{\small
\begin{multicols}{2}
\noindent
\begin{align}\label{srfull}
  -\frac{1}{M_P^2}+2\alpha t-\gamma_4 t^2&=\left\langle F_{1,l}^{0000}(\mu,t) \right\rangle \\
  \begin{split}
     -\frac{1}{M_P^2 t}+2\alpha-\gamma_4 t+&12 g^S_{0,2} t^2\\ &=\left\langle F_{2,l}^{0000}(\mu,t) \right\rangle
  \end{split}\\
  8g^S_{0,2} t-4g^S_{1,1}t^2&=\left\langle F_{3,l}^{0000}(\mu,t) \right\rangle \\
  \begin{split}
  4g^S_{0,2}-2g^S_{1,1}t+(48g^S_{0,3}+&g^S_{2,0}) t^2 \\ &=\left\langle F_{4,l}^{0000}(\mu,t) \right\rangle
  \end{split}\\
  \frac{\bar{\beta}_1t}{M_P^3}-\frac{\bar{\gamma}_3t^2}{M_P}&=\left\langle F_{1,l}^{+000}(\mu,t) \right\rangle\\
  \frac{\bar{\beta}_1}{M_P^3}-\frac{\bar{\gamma}_3t}{M_P}&=\left\langle F_{2,l}^{+000}(\mu,t) \right\rangle\\
  -4\bar{g}^{M_5}_{1,1}t^2&=\left\langle F_{3,l}^{+000}(\mu,t) \right\rangle\\
  -2\bar{g}^{M_5}_{1,1}t+\bar{g}^{M_5}_{2,0}t^2&=\left\langle F_{4,l}^{+000}(\mu,t) \right\rangle\\
  -\frac{\bar{\gamma}_1t^2}{M_P^3}&=\left\langle F_{1,l}^{+++0}(\mu,t) \right\rangle\\
   -\frac{\bar{\gamma}_1t}{M_P^3}&=\left\langle F_{2,l}^{+++0}(\mu,t) \right\rangle\\
    -4\bar{g}^{M_1}_{1,1}t^2&=\left\langle F_{3,l}^{+++0}(\mu,t) \right\rangle\\
    -2\bar{g}^{M_1}_{1,1}t+\bar{g}^{M_1}_{2,0}t^2&=\left\langle F_{4,l}^{+++0}(\mu,t) \right\rangle\\
    -\frac{\bar{\gamma}_0t^2}{M_P^4}&=\left\langle F_{1,l}^{+++-}(\mu,t) \right\rangle\\
    -\frac{\bar{\gamma}_0t}{M_P^4}&=\left\langle F_{2,l}^{+++-}(\mu,t) \right\rangle\\
   0&=\left\langle F_{3,l}^{+++-}(\mu,t) \right\rangle\\
   \bar{g}^{T_2}_{2,0}t^2&=\left\langle F_{4,l}^{+++-}(\mu,t) \right\rangle\\
   \frac{3\bar{\beta}_1^2-10\bar{\gamma}_0}{M_P^4}t^2&=\left\langle F_{1,l}^{++++}(\mu,t) \right\rangle\label{pppp1}\\
   \begin{split}
         \frac{12(\alpha_4-\alpha_4^\prime+i\, \breve{\alpha}_4)}{M_P^4}t^2&+\frac{3\bar{\beta}_1^2-10\bar{\gamma}_0}{M_P^4}t\\&=\left\langle F_{2,l}^{++++}(\mu,t) \right\rangle
   \end{split}\label{pppp2}\\
   \begin{split}
         \frac{8(\alpha_4-\alpha_4^\prime+i\, \breve{\alpha}_4)}{M_P^4}t-&4\(\frac{\bar{\gamma}_0^2}{M_P^6}+\bar{c}_{1,1}^{T_3}\)t^2\\&=\left\langle F_{3,l}^{++++}(\mu,t) \right\rangle
   \end{split}\label{pppp3}\\
   \begin{split}
         \frac{4(\alpha_4-\alpha_4^\prime+i \, \breve{\alpha}_4)}{M_P^4}-&2\(\frac{\bar{\gamma}_0^2}{M_P^6}+\bar{c}_{1,1}^{T_3}\)t\\+\(48\bar{g}^{T_3}_{0,3}+\bar{g}^{T_3}_{2,0}\)t^2&=\left\langle F_{4,l}^{++++}(\mu,t) \right\rangle
   \end{split}
\end{align}\label{pppp4}
\end{multicols}
}
The definition of $F_{k,l}^{\vmathbb{1234}}$ is shown in Eq.~(\ref{Fdef}). The sum rules coming from the amplitudes with only one crossing symmetry are
{\small
\begin{multicols}{2}
\noindent
\begin{align}\label{srone}
\frac{\bar{\beta}_2}{M_P^2}-\frac{\bar{\gamma}_0}{M_P^4}t-\bar{g}^{M_3}_{2,1}t^2&=\left\langle F_{2,l}^{++00}(\mu,t) \right\rangle\\
   \frac{M_P^2\bar{\gamma}_2+\bar{\beta}_1^2}{M_P^4}-\frac{\bar{\gamma}_0}{M_P^4}t & -\bar{g}^{M_3}_{2,1}t-\bar{g}^{M_3}_{3,1}t^2\nonumber\\&=\left\langle F_{3,l}^{++00}(\mu,t) \right\rangle\\
   \bar{g}^{M_3}_{4,0}-\bar{g}^{M_3}_{3,1}t+&\(\bar{g}^{M_3}_{2,2}-\bar{g}^{M_3}_{4,1}\)t^2\nonumber\\&=\left\langle F_{4,l}^{++00}(\mu,t) \right\rangle\\
-\frac{\bar{\gamma}_0 t^2}{MP^4}&=\left\langle F_{1,l}^{+0+0}(\mu,t) \right\rangle\\
-\frac{\bar{\gamma}_0 }{MP^4}t-\bar{g}^{M_3}_{2,1}t^2&=\left\langle F_{2,l}^{+0+0}(\mu,t) \right\rangle\\
0&=\left\langle F_{3,l}^{+0+0}(\mu,t) \right\rangle\\
\bar{g}^{M_3}_{2,2}t^2&=\left\langle F_{1,l}^{+0+0}(\mu,t) \right\rangle\\
-\frac{|\bar{\beta}_1|^2t}{M_P^4}+g^{M_4}_{0,2}t^2&=\left\langle F_{2,l}^{+-00}(\mu,t) \right\rangle\\
g^{M_4}_{1,2}t^2&=\left\langle F_{3,l}^{+-00}(\mu,t) \right\rangle\\
g^{M_4}_{2,2}t^2&=\left\langle F_{4,l}^{+-00}(\mu,t) \right\rangle\\
-\frac{1}{M_P^2}+\frac{|\bar{\beta}_1|^2}{M_P^4}t^2&=\left\langle F_{1,l}^{+0-0}(\mu,t) \right\rangle\\
-\frac{1}{M_P^2t}-\frac{|\bar{\beta}_1|^2}{M_P^4}t+g^{M_4}_{0,2}t^2&=\left\langle F_{2,l}^{+0-0}(\mu,t) \right\rangle\\
2g^{M_4}_{0,2}t+2g^{M_4}_{1,2}t^2&=\left\langle F_{3,l}^{+0-0}(\mu,t) \right\rangle\\
\begin{split}
   g^{M_4}_{0,2}+g^{M_4}_{1,2}t+&(-3g^{M_4}_{0,3}+g^{M_4}_{2,2})t^2\\&=\left\langle F_{4,l}^{+0-0}(\mu,t) \right\rangle
\end{split}\\
\frac{\bar{\beta}_1}{M_P^3}+\frac{\bar{\beta}_1^*\bar{\gamma}_0}{M_P^5}t^2&=\left\langle F_{2,l}^{++0-}(\mu,t) \right\rangle\\
\frac{\bar{\beta}_1^*\bar{\gamma}_0}{M_P^5}t-\bar{g}^{M_2}_{3,1}t^2&=\left\langle F_{3,l}^{++0-}(\mu,t) \right\rangle\\
-\bar{g}^{M_2}_{3,1}t-\bar{g}^{M_2}_{4,1}t^2&=\left\langle F_{4,l}^{++0-}(\mu,t) \right\rangle\\
\frac{\bar{\beta}_1^*\bar{\gamma}_0}{M_P^5}t^2&=\left\langle F_{2,l}^{+0+-}(\mu,t) \right\rangle\\
0&=\left\langle F_{3,l}^{+0+-}(\mu,t) \right\rangle\\
0&=\left\langle F_{4,l}^{+0+-}(\mu,t) \right\rangle\\
0&=\left\langle F_{1,l}^{+-+-}(\mu,t) \right\rangle\label{pmpm1}\\
0&=\left\langle F_{2,l}^{+-+-}(\mu,t) \right\rangle\label{pmpm2}\\
0&=\left\langle F_{3,l}^{+-+-}(\mu,t) \right\rangle\\
0&=\left\langle F_{4,l}^{+-+-}(\mu,t) \right\rangle\\
0&=\left\langle F_{5,l}^{+-+-}(\mu,t) \right\rangle\\
0&=\left\langle F_{6,l}^{+-+-}(\mu,t) \right\rangle\\
-\frac{1}{M_P^2t}&=\left\langle F_{2,l}^{++--}(\mu,t) \right\rangle\label{ppmm2}\\
-\frac{|\bar{\beta}_1|^2}{M_P^4}-\frac{|\bar{\gamma}_0|^2}{M_P^6}t^2&=\left\langle F_{3,l}^{++--}(\mu,t) \right\rangle\label{ppmm3}\\
\begin{split}
   \frac{2(\alpha_4+\alpha_4^\prime)}{M_P^4}-&\frac{|\bar{\gamma}_0|^2}{M_P^6}t-g^{T_1}_{4,1}t^2\\
   &=\left\langle F_{4,l}^{++--}(\mu,t) \right\rangle
\end{split}\label{ppmm4}\\
g^{T_1}_{5,0}-g^{T_1}_{4,1}t-g^{T_1}_{5,1}t^2&=\left\langle F_{5,l}^{++--}(\mu,t) \right\rangle\label{ppmm5}\\
\begin{split}
  g^{T_1}_{6,0}-g^{T_1}_{5,1}t+&\(g^{T_1}_{4,2}-g^{T_1}_{6,1}\)t^2\\&=\left\langle F_{6,l}^{++--}(\mu,t) \right\rangle
\end{split}\label{ppmm6}
\end{align}
\end{multicols}
}

\bibliographystyle{JHEP}  
\bibliography{refs}

\end{document}